\documentclass[review, sort&compress, times]{elsarticle}

\usepackage[top = 1 in, bottom = 1 in, left = 1 in, right = 1 in]{geometry}
\usepackage{amsmath, amsthm, amssymb, amsfonts}
\usepackage{mathtools, pifont, fleqn, txfonts, xcolor}
\usepackage{graphicx}
\usepackage{hyperref}
\usepackage{lineno}
\usepackage{pgfplots}
\usepgfplotslibrary{fillbetween, external}
\usepackage{float, array, tabu}
\usepackage{caption, subcaption}
\usepackage[unskipbreak, autobase]{cuted}
\usepackage{tikz}
\usetikzlibrary{shapes, arrows, positioning, plotmarks, patterns}

\definecolor{turquoise}{RGB}{0, 120, 200}

\hypersetup{
	linkcolor = turquoise,
	citecolor = turquoise,
	filecolor = turquoise,
	urlcolor = turquoise,
	colorlinks = true,
	bookmarks = true,
	pdftoolbar = true,
	pdfmenubar = true,
	breaklinks = true,
	pdfstartview = {FitH},
	pdftitle = {A Novel Design and Performance Optimization Methodology for Hydraulic Cross-Flow Turbines using Successive Numerical Simulations},
	pdfauthor = {Goodarz Mehr, Mohammad Durali, Mohammad Hadi Khakrand, Hadi Hoghooghi},
	pdfsubject = {Hydraulic Cross-Flow Turbine Design},
}

\tikzstyle{block} = [rectangle, draw, fill = blue!30, text width = 220 pt, text badly centered, node distance = 1 cm, rounded corners, minimum height = 20 pt]
\tikzstyle{line} = [draw, -latex']
\tikzstyle{communication} = [trapezium, draw, trapezium left angle = 80, trapezium right angle = 100, fill = purple!20, text width = 220 pt, text badly centered, node distance = 1 cm, minimum height = 20 pt]
\tikzstyle{checkpoint} = [ellipse, draw, fill = yellow!30, text width = 20 pt, text badly centered, node distance = 1 cm, minimum height = 20 pt]
\tikzstyle{firststep} = [rectangle, draw, fill = white, dashed = on, text width = 230 pt, node distance = 1 cm, rounded corners, minimum height = 50 pt]
\tikzstyle{secondstep} = [rectangle, draw, fill = white, dashed = on, text width = 230 pt, text badly centered, node distance = 1 cm, rounded corners, minimum height = 160 pt]
\tikzstyle{thirdstep} = [rectangle, draw, fill = white, dashed = on, text width = 230 pt, node distance = 1 cm, rounded corners, minimum height = 40 pt]

\begin{document}
	
	\begin{frontmatter}
		
		\title{
		A Novel Design and Performance Optimization Methodology for Hydraulic Cross-Flow Turbines using Successive Numerical Simulations
		}
		
		\author[SUTME,VT]{Goodarz Mehr\corref{cor1}}
		\ead{goodarzm@alum.sharif.edu,goodarzm@vt.edu}
		
		\author[MD]{Mohammad Durali}
		\ead{durali@sharif.edu}
		
		\author[SUTME]{Mohammad Hadi Khakrand}
		\ead{khakrand\_mohammadhadi@mech.sharif.edu}
		
		\author[SUTME,ETH]{Hadi Hoghooghi}
		\ead{hoghooghihadi@alum.sharif.edu}
		
		\cortext[cor1]{Corresponding author.}
		
		\address[SUTME]{Department of Mechanical Engineering, Sharif University of Technology, Tehran, Iran}
		\address[VT]{Present address: Department of Mechanical Engineering, Virginia Tech, Blacksburg, VA 24061, USA}
		\address[MD]{Professor of Mechanical Engineering, Sharif University of Technology, Azadi Avenue, Tehran 11155-9567, Iran}
		\address[ETH]{Present address: Department of Mechanical and Process Engineering, ETH Zurich, Zurich, Switzerland}
		
		\begin{abstract}
		
			This paper introduces a new methodology for designing and optimizing the performance of hydraulic Cross-Flow turbines for a wide range of operating conditions. The methodology is based on a one-step approach for the system-level design phase and a three-step, successive numerical analysis approach for the detail design phase. Compared to current design methodologies, not only does this approach break down the process into well-defined steps and simplify it, but it also has the advantage that once numerical simulations are conducted for a single turbine, most of the results can be used for an entire class of Cross-Flow turbines. In this paper, after a discussion of the research background, we explain the design process used and the ANSYS\textsuperscript{\textregistered}-based CFD model of the turbine in detail. The design process consists of three steps. First, designing nozzle geometry; second, optimizing runner parameters; and third, enhancing turbine performance by analyzing various load conditions. A turbine designed using this process in a simulation case study achieves a peak hydraulic efficiency of 91\% and peak overall efficiency of 82\% that is maintained for volume flow rates as low as 14\% of the nominal value and water head variations up to 30\% of the nominal value.
			
			\copyright 2021. This manuscript version is made available under the \href{http://creativecommons.org/licenses/by-nc-nd/4.0/}{CC-BY-NC-ND 4.0 license}.
			
		\end{abstract}
		
		
	\end{frontmatter}
	
	
	\section{Introduction} \label{Section1}
	
	Hydropower has been utilized for more than a century and is the most common and most efficient source of renewable electricity generation \cite{Williamson}. In 2018, 4325 TWh of electricity was produced by hydropower and it accounted for 17.3\% of world electricity and 68.7\% of total renewable electricity production. This amount is expected to surpass 4500 TWh by 2020 \cite{IEA}. \par
	
	Hydropower is commonly produced by turbines installed in large dams, but the high initial cost of installation and power transmission as well as environmental challenges associated with dams make local power generation an attractive and efficient option, especially in remote areas \cite{Banos, Paish, Weijermars}. Small turbines can be installed on rivers, fish farms, and water purification facilities to provide electricity for homes, farms, and small plantations \cite{Mohibullah}. However, the turbine used for this purpose should have unique characteristics, including an affordable price, easy repairs, and low maintenance cost. In addition, a high efficiency under varying load conditions that is present in many local water sources is expected. A Cross-Flow turbine satisfies these requirements. \par
	
\begin{figure}[ht!]
		\centering
		\includegraphics[width = 0.6\columnwidth]{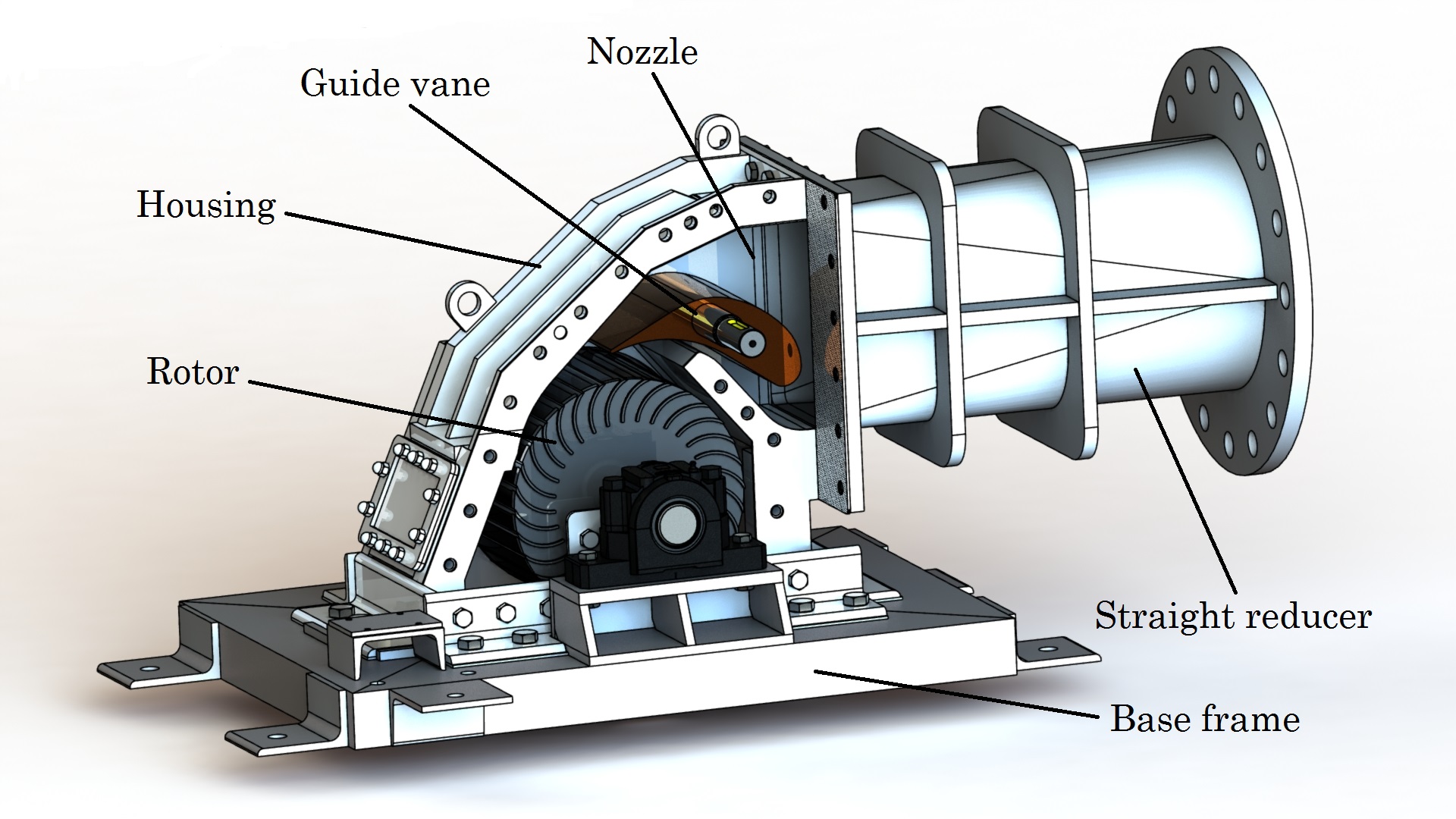}
		\caption{Components of a Cross-Flow turbine. The straight reducer is used to connect a circular discharge line to the rectangular entrance of the turbine. Some parts have been omitted for better visual appearance.} \label{Figure1}
	\end{figure}

	In a Cross-Flow turbine, water enters the runner tangentially, passes through it transversely, and leaves it radially, thereby transmitting the majority of its kinetic energy to the runner. The specific speed of a Cross-Flow turbine is greater than an impulse Pelton turbine but lower than a mixed-flow Francis turbine. A Cross-Flow turbine can achieve efficiencies up to 86\% which is slightly less than other types of hydraulic turbines (Pelton, Francis, and Kaplan) \cite{Pereira1}. However, it has a flat efficiency curve and if built as a multi-cell turbine with a 2:1 division, it can handle low-flow conditions and operate at its optimal efficiency for any water flow from 1/6 to full design flow rate \cite{Sammartano1}. It is suitable for applications where water head is between 2.5 and 200 m and volume flow rate is between 0.025 and 13 m$ ^ {3}$/s \cite{OssbergerCrossFlow}. A Cross-Flow turbine is generally used in pico, micro and small power plants, with output powers ranging from 1 kW to approximately 5 MW \cite{DOE}. Components of a typical Cross-Flow turbine are shown in \autoref{Figure1}.
	
	In this paper, we present a novel methodology for designing and optimizing the performance of Cross-Flow turbines. We use a simple chart and previous research results in the system-level design phase to select initial turbine parameters. After that, the detail design phase consists of three steps. In the first step, flow inside an isolated nozzle is simulated and nozzle geometry is optimized to produce a uniform velocity and angle of attack profile at runner inlet. In the second step, flow inside an isolated runner is simulated and the most important runner parameters and characteristics - rotational speed, admission angle, blade profile, diameter ratio, and the number of blades - are optimized. Finally, flow inside the entire turbine is simulated for various load conditions. These simulations help further modify turbine geometry and enhance its performance. \par
	
	Compared to current design methods, our proposed methodology is structured into well-defined steps that help simplify the design process and allow the designer to evaluate the effect of each parameter on turbine performance. Besides, once simulations are performed for a single turbine, the results can be used to design similar turbines without any new simulations which greatly reduces the time and cost of development and is a major advantage over current design methods. \par
	
	Moving forward, we first provide some background information about Cross-Flow turbines in \autoref{Section2} to familiarize the reader with important concepts and the limits of previous methods. \autoref{Section3} lays out the new design process, the CFD model, and numerical and experimental validation of that model. We implement the new methodology by designing a new Cross-Flow turbine as a simulation case study, presenting the results and discussing them in \autoref{Section4}. Finally, \autoref{Section5} concludes the findings of this paper and discusses the path forward.
	
	\section{Background} \label{Section2}
	
	A Cross-Flow turbine consists of a nozzle and a runner. The runner is made up of two parallel discs that are joined to each other at the rim using several curved blades. The nozzle, which has a rectangular cross-sectional area, guides the water jet towards the runner, which then strikes the blades on the rim, flows inside the channels between the blades for the first time, passes through the empty space in the middle of the runner and flows through the mid-blade channels for a second time, exiting the runner almost radially \cite{Mockmore}. A summary of the turbine's theory of operation is presented in \ref{AppendixA}. \par
	
	Australian engineer A. G. M. Michell, Hungarian inventor Donat Banki, and German entrepreneur Fritz Ossberger are generally credited as the first people to develop a Cross-Flow turbine. Mockmore and Merryfield laid the foundation of research on Cross-Flow turbines by translating Banki's transcripts into English. They also built an experimental model of the turbine and investigated the effect of rotational speed on turbine efficiency under various load conditions, verifying that maximum efficiency occurred near the nominal speed \cite{Mockmore}. Since then, many researchers have performed experimental studies on Cross-Flow turbines. Khosrowpanah et al. attempted to optimize the number of blades, diameter ratio, and admission angle \cite{Khosrowpanah}. Fiuzat et al. experimented on the turbine to determine the contribution of each stage to the output power \cite{Fiuzat}. Aziz et al. performed several experiments to identify the optimal values of angle of attack, diameter ratio, and the number of blades \cite{Aziz, Desai1, Desai2}. Olgun conducted experiments to investigate the effect of diameter ratio and internal guide tubes in the runner on water flow and efficiency \cite{Olgun1, Olgun2}. Kaunda et al. performed experimental and numerical studies to investigate the effect of water head, rotational speed, and guide vane opening on the turbine's performance \cite{Kaunda1, Kaunda2}.
	
	In recent years many researchers have used CFD alone or along with experiments to analyze and optimize Cross-Flow turbine's performance. Choi et al. used CFD extensively to analyze the effect of various parameters as well as nozzle shape, draft tube, and air layer on the turbine's performance \cite{Choi1, Choi2, Choi3, Kokubu1, Chen1}. De Andrade et al. investigated internal flow characteristics of a Cross-Flow turbine numerically \cite{DeAndrade}. Sammartano et al. used a transient model to analyze the behavior of a Cross-Flow turbine, investigated the effect of runner shaft on turbine performance, and optimized various parameters such as rotational speed, diameter ratio, and the number of blades. They further validated their results experimentally and investigated the use of Cross-Flow turbines in hydraulic power plants \cite{Sammartano1, Sammartano2, Sammartano3, Sinagra1, Sinagra2, Sammartano4, Sammartano5}. Yassen and Adhikari et al. used CFD to optimize various turbine parameters and geometry profiles \cite{Yassen, Adhikari1, Adhikari2, Adhikari3, Adhikari4, Adhikari5}, while Woldemariam et al. employed a hybrid numerical simulation-based approach to improve the performance of the turbine \cite{Woldemariam1, Woldemariam2, Woldemariam3}. Finally, Leguizamon and Avellan introduced a design methodology for Cross-Flow turbines under the constraints imposed by the discrete set of dimensions available for commercial steel pipes \cite{Leguizamon}. \par
	
	\autoref{TableB1} in \ref{AppendixB} summarizes the results of previous studies on the Cross-Flow turbine. Most studies have investigated the performance of a low-head micro Cross-Flow turbine, so the value of runner diameter falls mostly between 150 and 350 mm. Moreover, \autoref{TableB1} suggests that the results obtained in these studies are partly inconsistent and even occasionally contradictory. \par

	Most studies agree that an optimal angle of attack is between 12 and 20 degrees, with 16 being the favorable value. Only a few have concluded that an angle of attack greater than 20 degrees leads to higher efficiencies (\cite{Sammartano1, Sammartano2, Totapally}); however, it seems that in these studies the value of hydraulic efficiency is reported instead of turbine efficiency, leading to the erroneous conclusion that 22 degrees is the optimal angle of attack. Furthermore, data in \autoref{TableB1} is mostly consistent for diameter ratio, and its optimal value should fall between 0.65 and 0.69. \par
	
	Inconsistencies become notable when considering the number of blades and admission angle. While the data suggests that the optimal value of admission angle is between 80 and 120 degrees, some researchers reported 90 degrees as the optimal value while others favored 120 degrees. Furthermore, data in \autoref{TableB1} does not reveal any correlation between the number of blades and site-specific values such as water head or volume flow rate, only that the value mostly resides between 15 and 35. \par
	
	Overall, while these studies provide a good insight into the effect of each parameter on the efficiency of the turbine, only a few present a concrete framework for designing a new turbine. Only one study offers details about the system-level design process \cite{Woldemariam1} and only a few (such as \cite{Sammartano1, Leguizamon}) propose methods for the detail design phase.
	
	\section{Methodology} \label{Section3}
	
	\stripsep = 10 pt
	
	Designing a new product comprises of five phases: concept development, system-level design, detail design, prototype testing and refinement, and production ramp-up \cite{Ulrich}. For a new hydraulic turbine, in the concept development phase, it is essential to collect data on the water head and volume flow rate of a potential site for several months (usually one year). This data should be analyzed to determine whether installation of a hydraulic turbine is technically feasible and if so, the kind of turbine(s) that should be used. Also, the project should be examined from a financial standpoint. \par
	
	The next two phases, system-level design and detail design, explore determining product architecture, parameter value selection, and geometry optimization \cite{Ulrich}.
	
	\subsection{System-level design} \label{Section3.1}
	
	The first step in the system-level design phase is defining product architecture. The architecture that we use consists of four major components: nozzle, runner (rotor), housing, and guide vane, as seen in \autoref{Figure1}. The runner is installed at the nozzle's outlet with a small clearance to ensure that efficiency losses due to air entrainment are kept at a minimum. Furthermore, a curved guide vane is used inside the nozzle to guide and regulate water flow. A low-head turbine can feature a draft tube to increase the net water head and an air vent to regulate air pressure inside the housing.
	
	The second stage is determining turbine parameter values, including runner diameter and length, diameter ratio, rotational speed, angle of attack, blade entrance and exit angles, admission angle, and the number of blades. These are only initial values and some are subject to change in the detail design phase. \par
	
	Theoretical relations presented in \ref{AppendixA} are utilized to determine parameter dependency. Specifically, if values of runner diameter, angle of attack, admission angle, blade exit angle, and the number of blades - which we call key turbine parameters - are known, runner length is obtained from \autoref{EquationA7} and rotational speed is calculated using \autoref{EquationA2}. Furthermore, blade entrance angle is computed from \autoref{EquationA4} and diameter ratio is obtained from \autoref{EquationA6}. Finally, nozzle height and blade geometry are determined using \autoref{EquationA5}, \autoref{EquationA8}, and \autoref{EquationA9}, respectively. Therefore, it suffices to determine the value of key turbine parameters and others will be determined accordingly. \par
	
	As discussed in \ref{AppendixA}, the value of blade exit angle must be 90 degrees to avoid efficiency loss. Based on the discussion in \autoref{Section2} we use the accepted value of 16 degrees for the angle of attack, as well as the initial values of 120 degrees for the admission angle and 32 for the number of blades. \par
	
\begin{figure}[ht!]
		\centering
		\includegraphics[width = 0.6\columnwidth]{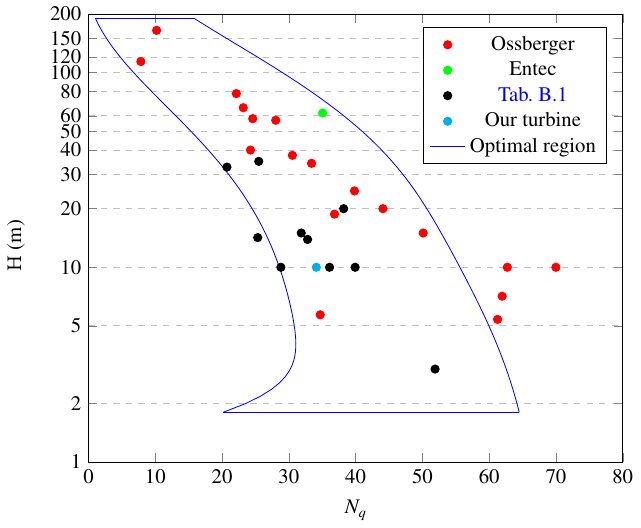}
		\caption{Plot of the optimal region of operation for Cross-Flow turbines along with $N_{q}$ values of our turbine, several operational turbines, and those found in \autoref{TableB1} \cite{OssbergerCrossFlow, Entec, Kaniecki1, Kaniecki2, DeAndrade, Sammartano1, Sammartano3, Sinagra1, Chen1, Acharya, Adhikari4, Adhikari5}.} \label{Figure2}
	\end{figure}

	To determine runner diameter, we use a parameter similar to the specific speed of turbomachinery called $N_{q}$, defined as
	\begin{equation} \label{Equation1}
	 	N_{q} = N(\mathrm{rpm})[H(\mathrm{m})] ^ {-\frac{3}{4}}\sqrt{Q(\mathrm{m} ^ {3}/\mathrm{s})}.
	\end{equation}
	$N_q$ has been calculated for many operational turbines and the results have been plotted against water head to obtain the recommended region of optimal performance for different turbine types \cite{Giesecke}. The optimal region for the Cross-Flow turbine is plotted in \autoref{Figure2} along with values calculated for several operational turbines and those found in \autoref{TableB1} \cite{OssbergerCrossFlow, Entec, Kaniecki1, Kaniecki2, DeAndrade, Sammartano1, Sammartano3, Sinagra1, Chen1, Acharya, Adhikari4, Adhikari5}. We only plotted points corresponding to turbines in \autoref{TableB1} for which: a) values of $H$, $Q$, and $N$ were known, b) $H$ was at least 2 m, and c) the reported efficiency was above 70\%. \autoref{Figure2} shows that the value of $N_{q}$ for most of these turbines falls into the optimal region and the few that do not are low-head, low-speed turbines where other factors (such as gearbox efficiency) affect the choice of $N_{q}$ and runner diameter.

	In addition to \autoref{Figure2}, selecting the proper value of $N_{q}$ should take external design factors, such as generator speed and the availability of a gearbox, into consideration. Using the value of $N_{q}$ and \autoref{Equation1}, we can calculate rotational speed $N$. This can then be used with \autoref{EquationA2} to calculate runner diameter. \par
	
\begin{figure}[ht!]
		\centering	
		\includegraphics[width = 0.6\columnwidth]{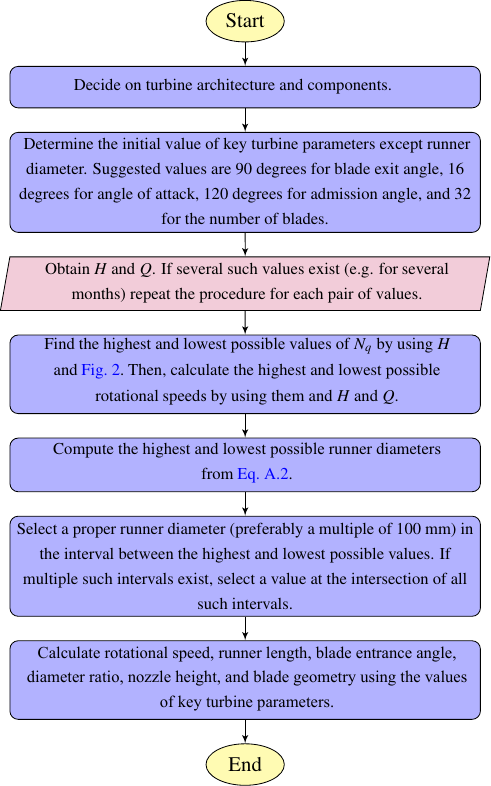}
		\caption{Flowchart of the system-level design process. This procedure leads to initial values of key turbine parameters.} \label{Figure3}	
	\end{figure}

	The procedure used for the system-level design phase can be summarized in the flowchart in \autoref{Figure3}. As explained earlier, this procedure receives water head and volume flow rate values as input and outputs initial values of key turbine parameters.
	
	\subsection{Detail design} \label{Section3.2}
	
	The detail design phase includes multiple steps and much of it, especially the mechanical design of the turbine, is beyond the scope of this paper. Our focus is on the first step of this process for a Cross-Flow turbine which is determining its internal geometry and dimensions. In the following subsections, we first provide a brief overview of the design procedure, then the CFD model and meshing used for numerical simulations, and finally the experimental validation of the CFD model.
	
	\subsubsection{Design procedure} \label{Section3.2.1}
	
\begin{figure}[ht!]
		\centering	
		\includegraphics[width = 0.6\columnwidth]{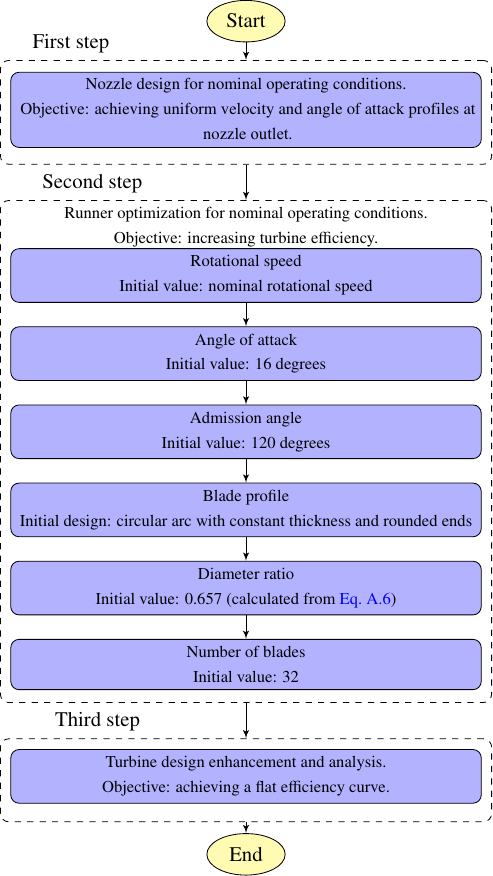}
		\caption{Flowchart of the design process for the detail design phase.} \label{Figure4}
	\end{figure}

	The design process can be broken down into three steps; nozzle design for nominal conditions, runner parameter optimization for nominal conditions, and turbine performance enhancement and analysis under different load conditions. This approach is further illustrated in \autoref{Figure4}.
	
	In the second step of the proposed design process, each parameter is optimized separately, instead of optimizing them all at once. If the latter approach were to be adopted using a general optimization algorithm, the effect of each parameter on the turbine's performance would remain unknown. Alternatively, if an experimental method were to be used, and assuming that at least three values were considered for each parameter, at least $3 ^ {6} = 729$ experimental setups would be required which is both economically unjustifiable and time-consuming. To help reduce computation time and save cost, in our proposed process nozzle design and parameter value optimization are done separately. \par
	
	The proposed methodology is highly iterative at start and several cases must be analyzed, but the results are scalable. Once dimensions have been determined for one case, they can be scaled with respect to runner diameter to get the dimensions of a larger or smaller turbine. This is true for many other parameters as well. Exceptions to this rule are diameter ratio and the number of blades which are related to each other by a parameter called solidity \cite{Durali} and their value depends on water head, so they should be optimized on a case-by-case basis. \par
	
	Our proposed methodology can be implemented both experimentally and numerically. A numerical approach is preferred since an experimental setup is more expensive. In our implementation, we used SolidWorks\textsuperscript{\textregistered} to design CAD models, ANSYS\textsuperscript{\textregistered} Meshing to mesh the geometry, and ANSYS\textsuperscript{\textregistered} CFX for fluid flow simulations.
	
	\subsubsection{CFD model} \label{Section3.2.2}
	
	Water flow in a Cross-Flow turbine is almost transversely uniform (ignoring boundary layer and multi-cell flow division effects), so our analysis was based on 2D simulations instead of computationally-intensive 3D ones. However, since CFX does not allow 2D simulations, a cross-sectional slab with a small width (width of a mesh element) was prepared instead and used with symmetry boundary condition assigned to the sides \cite{Sammartano1}. In our model, we assumed that the flow was steady and that buoyancy forces were negligible due to the small elevation difference in the turbine compared to the water head. Furthermore, since no noticeable temperature change occurs during turbine's operation, we assumed that heat transfer between the turbine and its surroundings was negligible. Our simulations used the Shear Stress Transport (SST) turbulence model with automatic wall function that gives highly accurate predictions of the onset and the amount of flow separation under adverse pressure gradients, which was useful when studying the effect of blade profile on efficiency \cite{CFX}. \par
	
	In the first step, nozzle design, a single-phase model with water as the only phase was used to simulate flow and the convergence target was set to 1.0e-8. The model used for the second step, runner parameter optimization, was similar to the one used by Kaunda et al. \cite{Kaunda2} and more complex since it modeled runner rotation. In our model, the runner was divided into two separate domains, one rotating that consisted of the space surrounding the blades (``rotor''), and the other stationary consisting of the rest of the geometry (``stator''). \par
	
	For the stator domain, we used the homogeneous multiphase flow model with both water and air as continuous fluids, given that a well-defined interface existed between water and air in the turbine \cite{Kaunda2, CFX}. Surface tension between the two phases was modeled with air as the primary fluid. The model used for the rotor domain was the same except that it revolved around the runner's axis of rotation. For accurate results, our simulations used a physical timescale of $\frac{0.1}{\omega}$ and the convergence target was set to 1.0e-06. We also used Rhie-Chow pressure-velocity coupling to achieve a smooth pressure field \cite{CFX}. Finally, we used the Frozen Rotor frame change model for the interface between the stationary and rotating domains. \par
	
	The model used for the third step, turbine design enhancement, was the same as before, except that the geometry was divided into three domains: nozzle, rotor, and stator. \par
	
	For boundary conditions, in the first step, static pressure at nozzle outlet was set to 1 atm. Mass flow rate was assigned to nozzle inlet, which meant inlet total pressure was an implicit result of the simulation. A smooth no-slip wall boundary condition was assigned to the rest of the surfaces and initial conditions were automatically determined by CFX \cite{CFX}.
	
	In the second step, we assumed that water entered the runner with the desired velocity and angle of attack to analyze the effect of the selected parameters on efficiency. Therefore, we defined the components of water velocity in a cylindrical coordinate frame at runner inlet such that the resulting total pressure matched the previously calculated total pressure at the nozzle outlet, and set water's volume fraction to 1 and that of air to 0. At the outlet, an opening boundary condition was used with static pressure set to 1 atm and values of volume fraction and turbulence were assumed to have a zero gradient. The calculated mass-flow average of the total pressure of water leaving the outlet was used for hydraulic efficiency calculations ($P_{e}$). Same as before, a smooth no-slip wall boundary condition was given to the housing and runner blades. A rotating wall boundary condition with the same speed as the runner was assigned to the runner shaft \cite{Kaunda2}. Initial conditions were automatically determined by CFX \cite{CFX}. \par
	
	 Initial and boundary conditions and interfaces remained the same in the third step, except at the nozzle inlet. To get accurate results, simulations were first performed using a mass flow rate boundary condition, then it was switched to total pressure and the result of the previous simulation was used as the initial condition for the new simulation.
	
	\subsubsection{Meshing and numerical accuracy} \label{Section3.2.3}
	
\begin{figure}[ht!]
		\centering
		\includegraphics[width = 0.6\columnwidth]{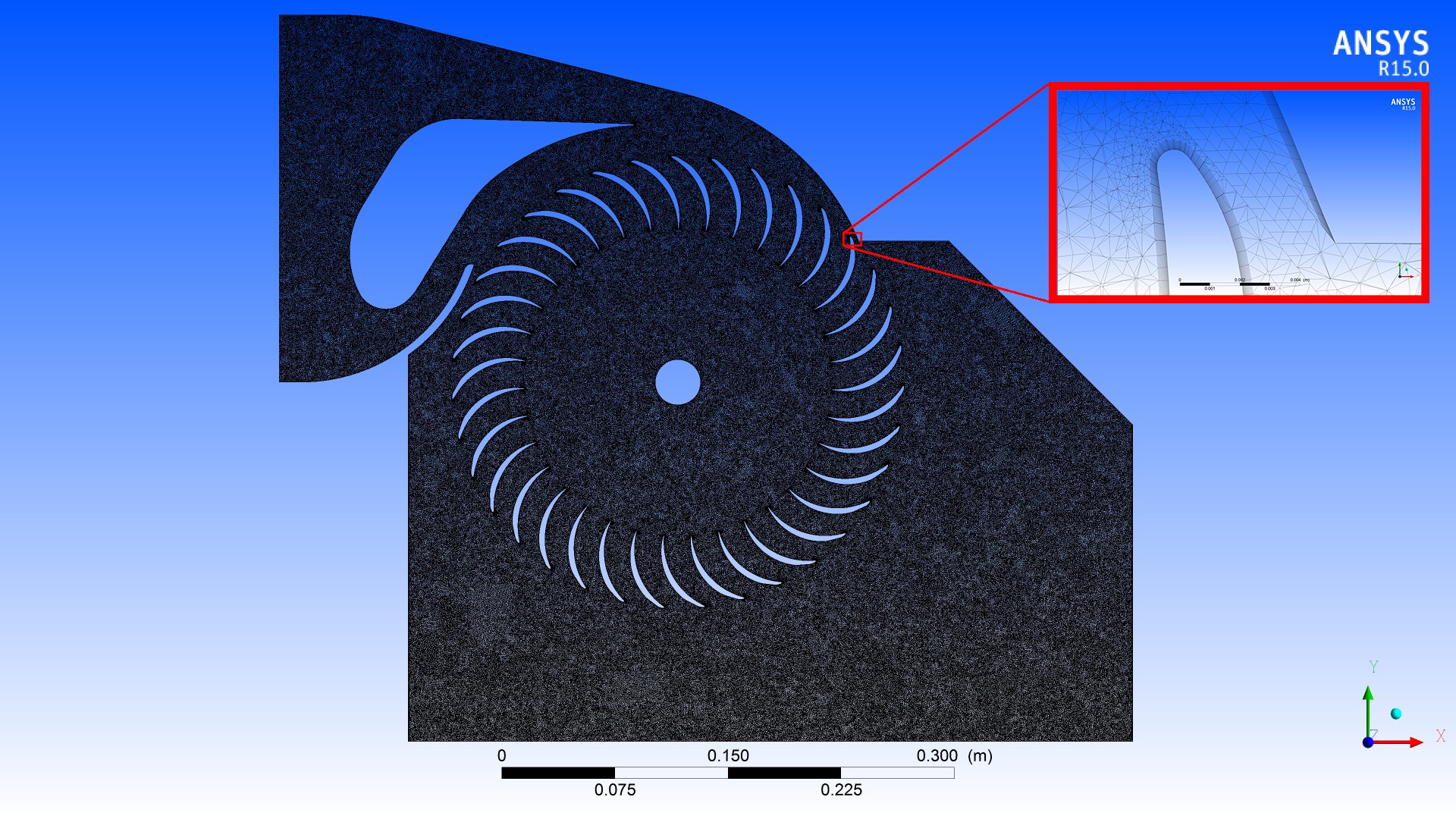}
		\caption{Sample meshing of turbine geometry. Near-wall regions were resolved into several layers of prismatic elements while the rest of the turbine was meshed with mostly tetrahedral elements.} \label{Figure5}
	\end{figure}

	For our simulations, unstructured meshing was favored over a structured one since fluid flow - especially that of air - was largely undetermined. The mesh mostly consisted of tetrahedral elements, the size of each equal to the width of the slab \cite{Sammartano1}. However, near-wall regions were meshed using a pre-meshing inflation algorithm with multiple layers of prismatic elements \cite{Sammartano1, Kaunda2}. The first layer had a height of 10 $\mu$m and layer heights grew geometrically by a factor of 1.05. This resolution was necessary for error reduction when using the SST turbulence model \cite{CFX}. A sample meshing of the turbine is shown in \autoref{Figure5} and the number of elements in each domain is shown in \autoref{Table1}.
	
	\begin{table}[h!]
		\centering
		\caption{Number of mesh elements for a typical turbine simulation.} \label{Table1}
		\begin{tabular}{c c c c c}
			\hline
			Domain & Nozzle & Rotor & Stator & Total \\
			\hline
			Number of elements & 410,046 & 1,073,179 & 1,350,351 & 2,833,586 \\
			\hline
		\end{tabular}
	\end{table}
	
\begin{figure}[ht!]
			\centering
			\includegraphics[width = 0.6\columnwidth]{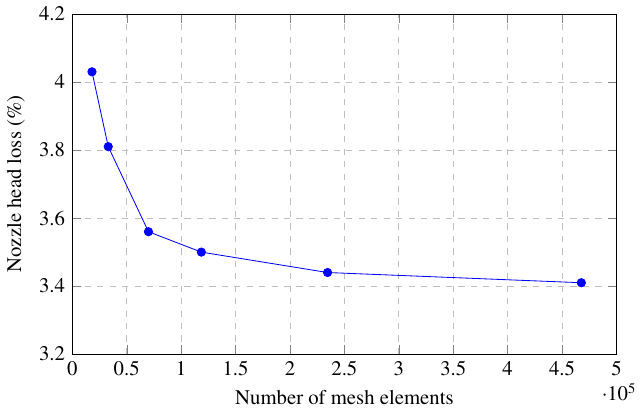}
			\caption{Grid independence of a global variable for simulations of the first step. Nozzle head loss converged when the number of mesh elements surpassed 200,000.} \label{Figure6}
	\end{figure}

	After each simulation, we checked the value of $y ^ {+}$ on each solid surface to ensure it was below the recommended value of 11.06 \cite{CFX}. The average value of $y ^ {+}$ on the runner surface was around 2.4 for simulations of the second step and around 1.8 for simulations of the third step. Moreover, we monitored global and local variables while refining the mesh until all variables converged to a final value for grid (spatial) independence of the results. As an example, this is shown for nozzle head loss in \autoref{Figure6} for simulations of the first step.
	
	\subsubsection{Experimental validation of the CFD model} \label{Section3.2.4}
	
	To validate the proposed CFD model, we used it to simulate fluid flow in a prototype Cross-Flow turbine designed by Durali System Design and Automation (DSDA) which had been tested under various load conditions. Experimental and numerical efficiencies obtained were compared with each other to evaluate the accuracy of our model. The experimental setup is discussed in \ref{AppendixC} and validation results are presented in \autoref{Section4.1}.
	
	\section{Results and discussion} \label{Section4}
	
	In this section, we first discuss the results of experimental validation of the CFD model. Next, we examine a case study of implementing the proposed process to design and optimize the performance of a Cross-Flow turbine for real-world operating conditions.
	
	\subsection{Experimental validation of the CFD model} \label{Section4.1}
	
	The prototype turbine used for our experiments was designed for a nominal water head of 45 m and volume flow rate of 100 l/s, with a rotational speed of 1074 rpm. Five different operating conditions were simulated using the proposed CFD model and the turbine was tested under those same conditions. The operating conditions, numerical and experimental efficiencies, and uncertainty values are shown in \autoref{Table2}.

	\begin{table}[h!]
		\centering
		\caption{Operating conditions and the results of CFD simulations and experimental tests performed on the prototype turbine to validate the CFD model.} \label{Table2}
		\begin{tabular}{c c c c c c c}
			\hline
			Water & Water volume & CFD & Experimental & Experimental & Relative \\
			head (m) & flow rate (l/s) & efficiency (\%) & efficiency (\%) & uncertainty (\%) & error (\%) \\
			\hline
			25.2 & 85.1 & 32.1 & 33.4 & 6.1 & 3.9 \\
			35.0 & 95.2 & 49.9 & 49.9 & 5.4 & 0 \\
			35.1 & 113.3 & 52.6 & 53.7 & 4.8 & 2.0 \\
			45.2 & 119.7 & 59.7 & 58.2 & 4.5 & 2.6 \\
			45.3 & 132.7 & 60.6 & 59.2 & 4.2 & 2.4 \\
			\hline
		\end{tabular}
	\end{table}
	
	As can be seen from \autoref{Table2}, there is good agreement between the experimental and numerical results and the absolute error does not exceed 1.5\% for any of the cases considered (compared to other works that consider an absolute error of 5\% acceptable \cite{Kaunda2, Sammartano2}). It is worth mentioning that the relatively large uncertainty in experimental efficiencies can be attributed to two factors: first, momentary variation of turbine efficiency due to the changing relative position of the runner and nozzle and second, the large relative uncertainty of the flow meter used. \par
	
\begin{figure}[ht!]
		\centering
		\includegraphics[width = 0.6\columnwidth]{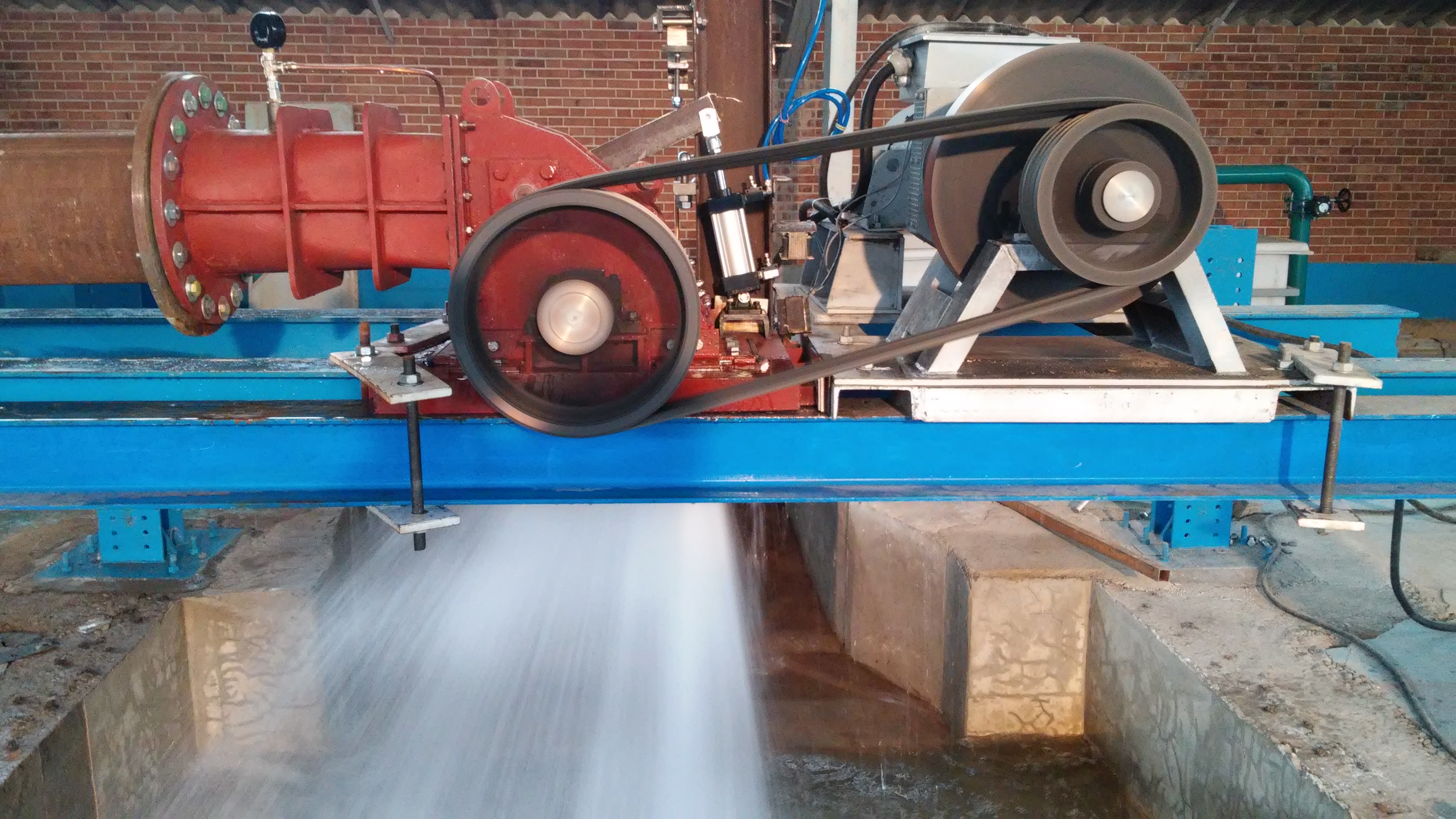}
		\caption{The prototype turbine in operation.} \label{Figure7}
	\end{figure}

\begin{figure}[ht!]
		\centering
		\includegraphics[width = 0.6\columnwidth]{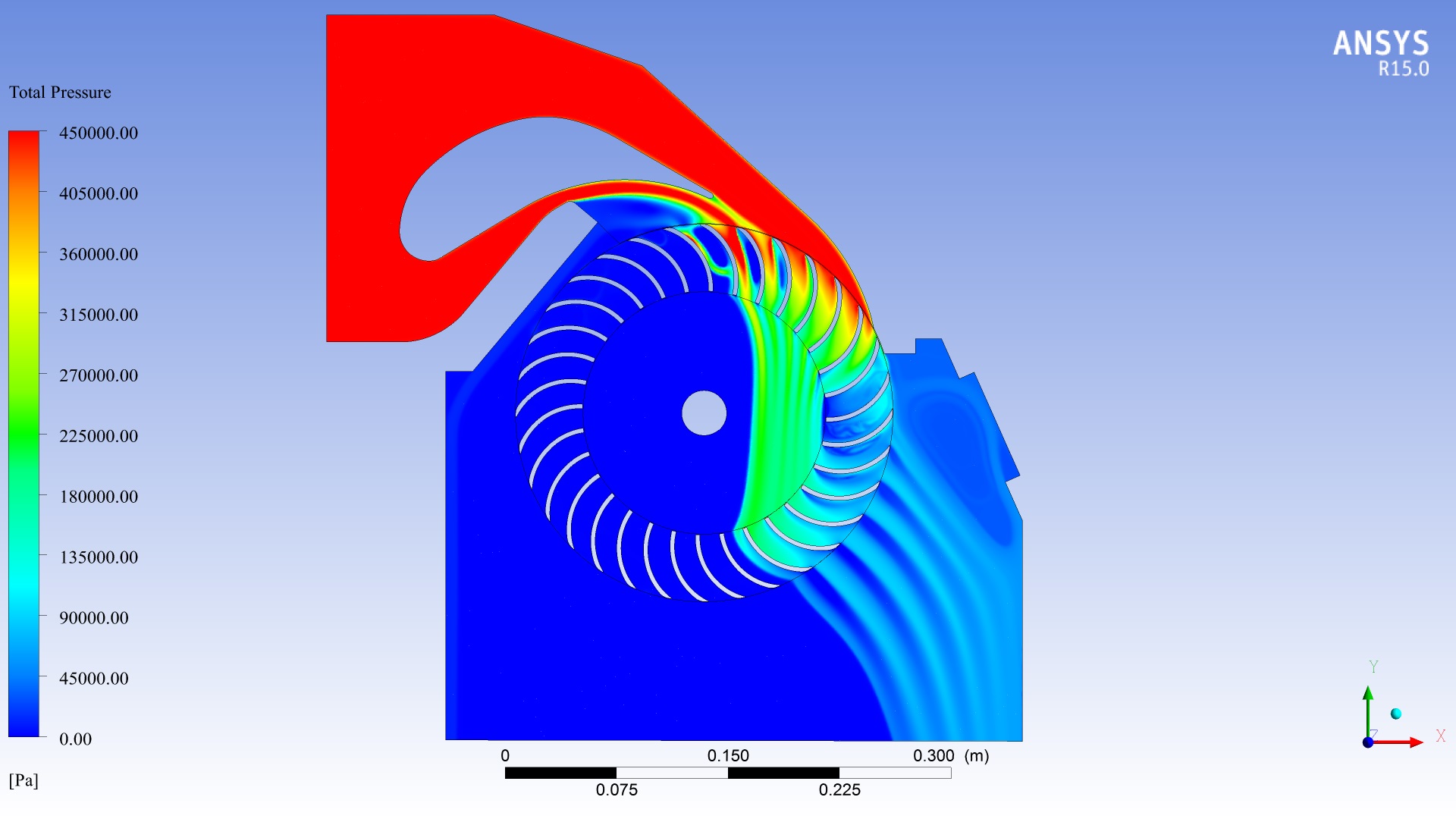}
		\caption{Distribution of total pressure (water head) inside the turbine for a simulation. Water passes through the correct path inside the turbine but exits it in a different direction than the one observed in \autoref{Figure8}. This is the result of using the Frozen Rotor frame change model.} \label{Figure8}
	\end{figure}

	\autoref{Figure7} shows the prototype turbine in operation and \autoref{Figure8} shows the distribution of total pressure (water head) inside the turbine for a simulation. Although water follows through the correct path in \autoref{Figure8}, it exits the turbine in a different direction than the one seen in \autoref{Figure7}. This discrepancy is the result of using the Frozen Rotor frame change model, because when moving from a stationary domain into a rotating one the model only changes the frame of reference and not the relative position of the two domains.
	
	
\begin{figure}[ht!]
		\centering
		\includegraphics[width = 0.6\columnwidth]{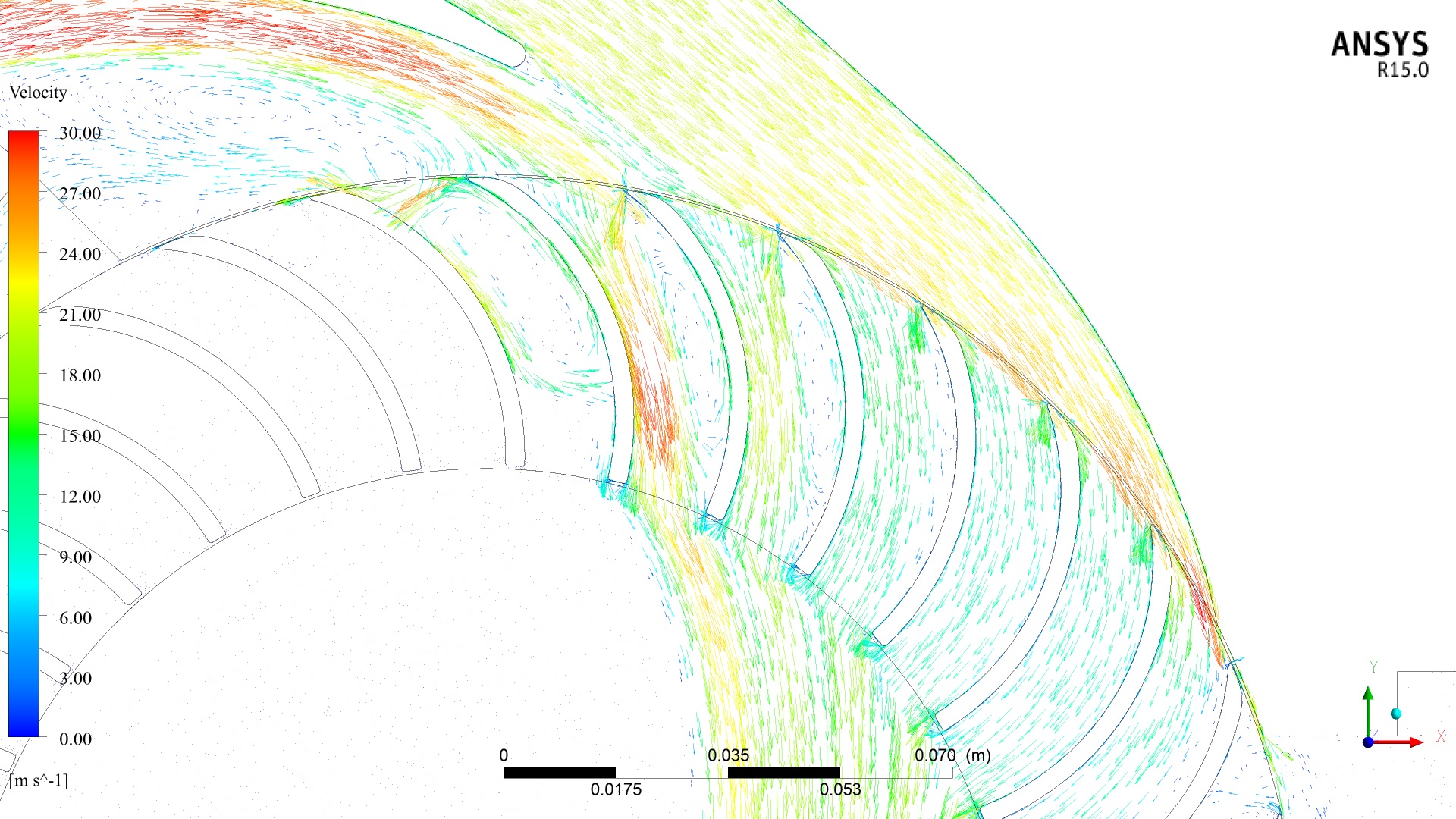}
		\caption{Velocity vectors entering the runner. Poor design of runner blades and high rotational speed means velocity vectors that enter the runner deflect and cause flow separation which ultimately results in the loss of power.} \label{Figure9}
	\end{figure}

	\autoref{Figure9} provides a closer look at velocity vectors entering the runner and shows that the main cause of power loss is the flow separation that occurs at the tip of several blades that prevents water from transferring its energy to the runner. Besides, a portion of the water at the very front and back of the turbine barely enters the runner because of its centrifugal tendency which results in the loss of power.
	
	\subsection{Case study: design of a new turbine} \label{Section4.2}
	
	Because of the low efficiency of the prototype turbine, we used the proposed methodology to design a turbine with higher efficiency. The new turbine was designed for a nominal water head of 10 m, volume flow rate of 200 l/s, resulting in an available hydraulic potential of 19.55 kW which is one of the most prevalent hydro potentials in Iran and around the world. In the following sections, we proceed with system-level design and detail design phases as described in \autoref{Section3}. Numerical simulations for detail design were carried out on a computer with an 8-core Intel\textsuperscript{\textregistered} processor, 6 of which were used during simulations, and 32 GB of RAM. Simulations of the first, second, and third steps took on average 0.5 hr, 6 hr, and 18 hr to complete, respectively. Overall, 123 simulations were conducted across the three steps.
	
	\subsubsection{System-level design} \label{Section4.2.1}
	
	Preliminary values of key turbine parameters, except runner diameter, were selected according to the process described in \autoref{Figure3}. To select the value of runner diameter based on \autoref{Figure2}, for a 10 m water head the minimum and maximum possible values for $N_q$ were 28 and 56, respectively, hence from \autoref{Equation1} the minimum and maximum possible rotational speeds were 352 and 704 rpm, respectively. Therefore, from \autoref{EquationA2} the maximum and minimum possible runner diameters were 365 and 183 mm, respectively.\par
	
	Based on \autoref{Figure3}, two possible runner diameters were 300 and 200 mm. The former was preferred because it resulted in less centrifugal force (which resulted in the loss of power) and reduced the possibility of cavitation \cite{Adhikari2}. Nevertheless, this choice implied the use of a transmission system with a higher step-up ratio which may impact the overall efficiency.\par
	
	Based on the value of runner diameter and using the equations introduced in \ref{AppendixA}, we calculated the remaining turbine parameters, which are shown in \autoref{Table3}. Note that runner aspect ratio (diameter to length ratio) is $\frac{20}{11}$ here which results in a higher efficiency compared to runners with an aspect ratio of approximately 1 \cite{Khosrowpanah}.
	
	\begin{table}[h!]
		\centering
		\caption{Initial values of turbine parameters for the new turbine.} \label{Table3}
		\begin{tabular}{c c c c}
			\hline
			Turbine parameter & Value & Turbine parameter & Value \\
			\hline
			runner diameter (mm) & 300 & rotational speed (rpm) & 429 \\
			runner length (mm) & 165 & blade entrance angle (deg) & 30 \\
			nozzle height (mm) & 86.5 & blade central angle (deg) & 73.8 \\
			blade radius (mm) & 49.1 \\				
			\hline
		\end{tabular}
	\end{table}
	
	\subsubsection{Detail design: first step} \label{Section4.2.2}
	
\begin{figure}[ht!]	
		\centering
		\begin{subfigure}[h]{0.48\columnwidth}
			\centering
			\includegraphics[width = \columnwidth]{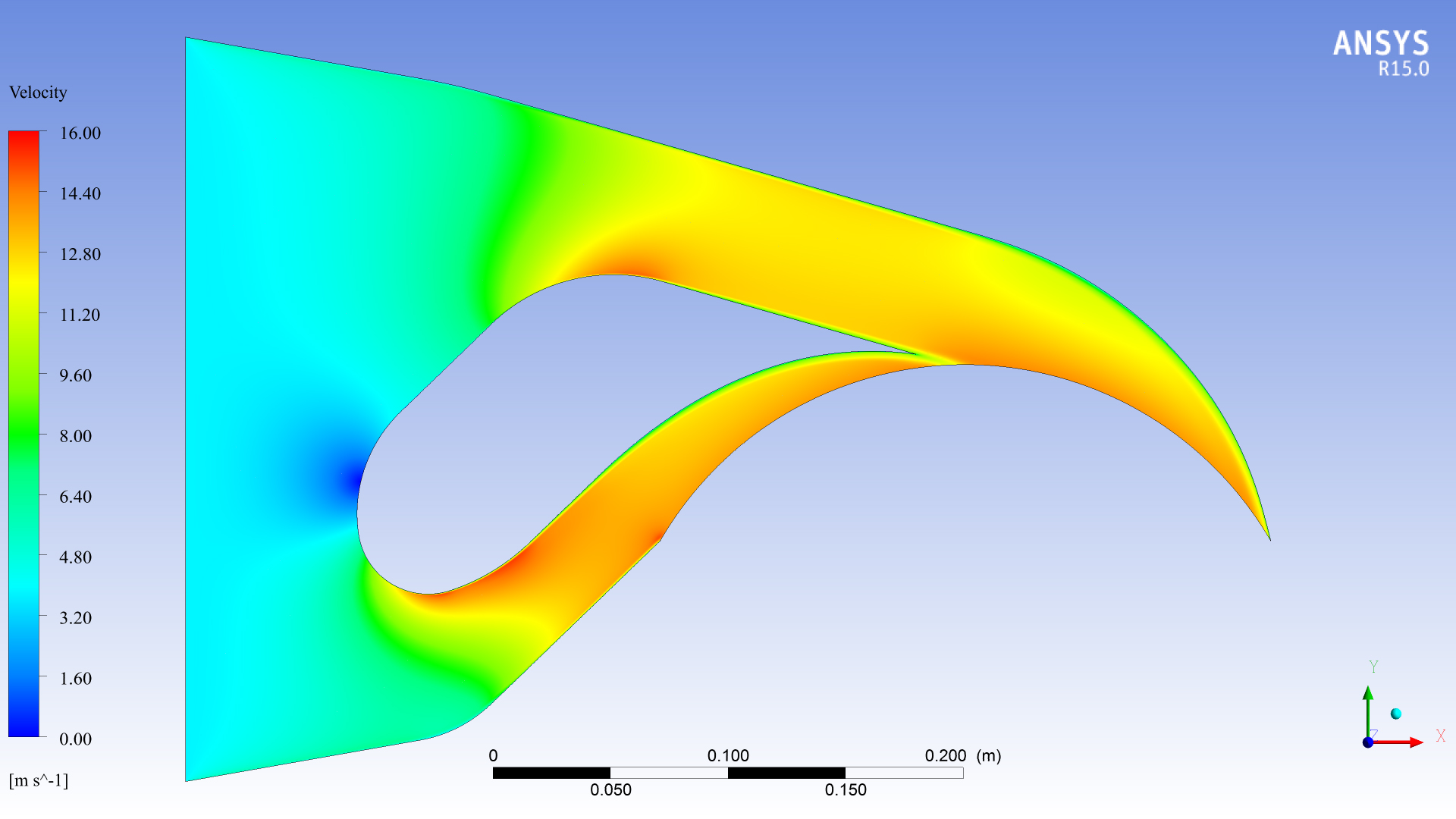}
			\caption{initial nozzle} \label{Figure10a}
		\end{subfigure}
		\begin{subfigure}[h]{0.48\columnwidth}
			\centering
			\includegraphics[width = \columnwidth]{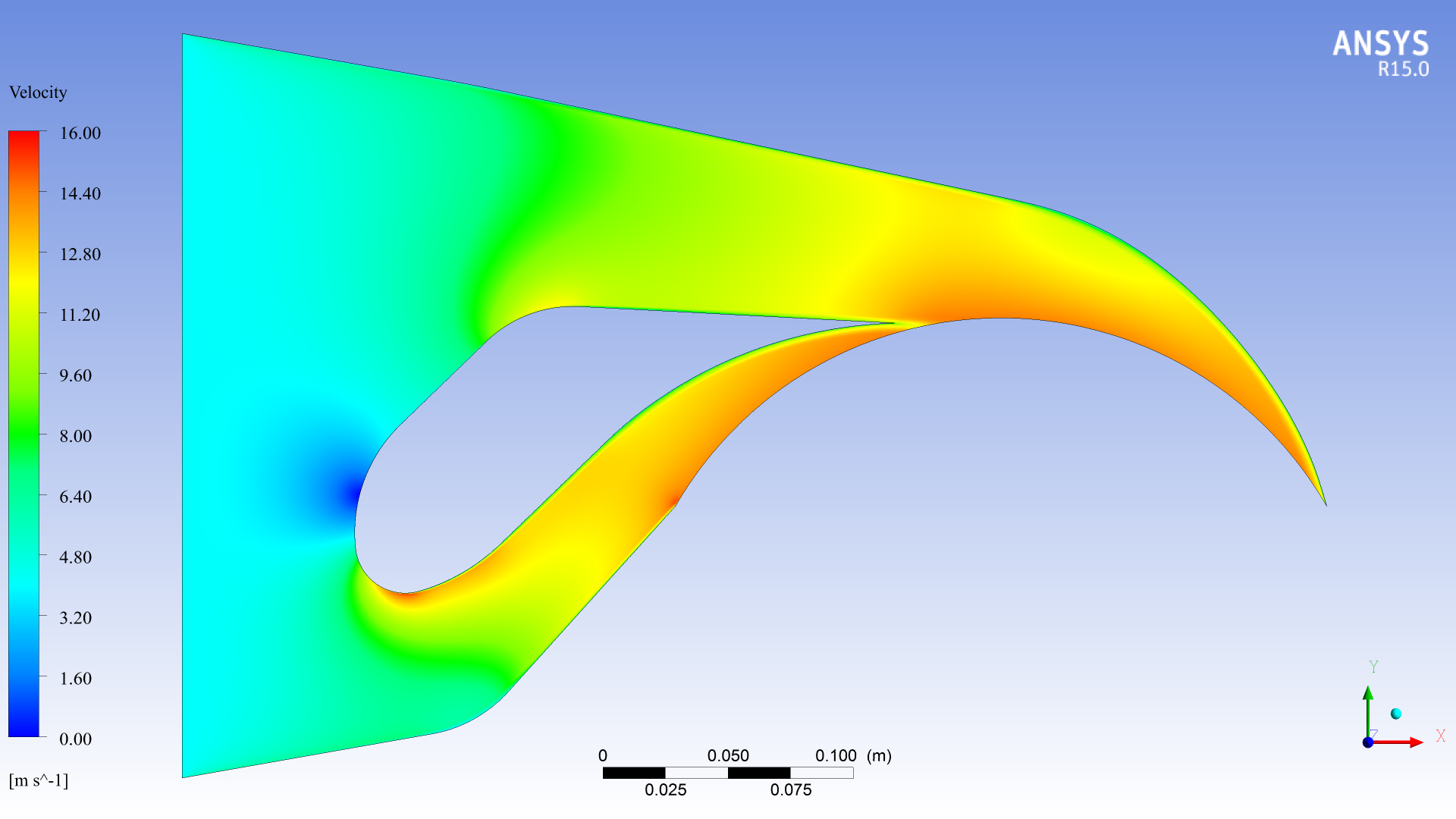}
			\caption{final nozzle} \label{Figure10b}
		\end{subfigure}
		\caption{Water velocity distribution inside the nozzle. Note that changes in velocity in the final nozzle are less abrupt compared to the initial nozzle.} \label{Figure10}
	\end{figure}

	The objective of the first step is to design a nozzle that produces almost uniform water velocity and angle of attack profiles at its outlet to ensure optimal entrance of the water jet into the runner. To that end, we started with an initial sketch of the nozzle and guide vane, seen in \autoref{Figure10a}, adhering to three criteria. First, the guide vane had to be able to close the water passage completely by rotating around its axis. Second, the sum of the heights of the passages above and below the fully-opened guide vane had to equal the value calculated for nozzle height (in this case 86.5 mm). Finally, on both endpoints of the admission arc, the angle between tangents to the nozzle curve and admission arc had to form an angle equal to the angle of attack (here 16 degrees). \par
	
\begin{figure}[ht!]
		\centering
		\begin{subfigure}[h]{0.28\columnwidth}
			\centering
			\includegraphics[width = \columnwidth]{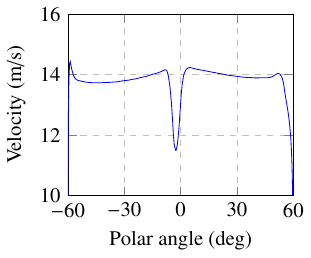}
			\caption{initial nozzle} \label{Figure11a}
		\end{subfigure}
		\begin{subfigure}[h]{0.28\columnwidth}
			\centering
			\includegraphics[width = \columnwidth]{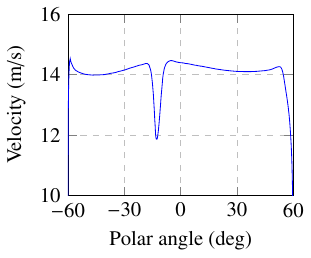}
			\caption{final nozzle} \label{Figure11b}
		\end{subfigure}
		\caption{Water velocity profile at nozzle outlet. The notch in the middle of the profile is due to wake caused by the tip of the guide vane.} \label{Figure11}
	\end{figure}

\begin{figure}[ht!]
		\centering
		\begin{subfigure}[h]{0.28\columnwidth}
			\centering
			\includegraphics[width = \columnwidth]{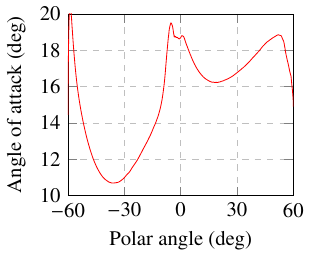}
			\caption{initial nozzle} \label{Figure12a}
		\end{subfigure}
		\begin{subfigure}[h]{0.28\columnwidth}
			\centering
			\includegraphics[width = \columnwidth]{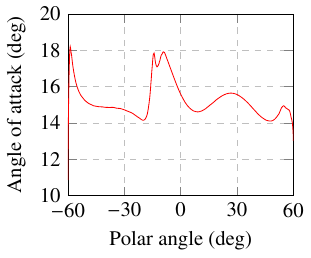}
			\caption{final nozzle} \label{Figure12b}
		\end{subfigure}
		\caption{Angle of attack profile at nozzle outlet. The profile for the final nozzle is more uniform compared to the initial one.} \label{Figure12}
	\end{figure}

	We then used an iterative approach to optimize the shape of the nozzle. Starting from the initial geometry, we simulated water flow through the nozzle and plotted the exiting velocity and angle of attack profiles. These profiles were then used to alter the geometry, through which water flow was simulated again. For example, if the angle of attack was low for a part of the profile, we would change the nozzle (or guide vane) curve corresponding to that profile to increase it. This iterative process was carried out until satisfactory profiles were achieved. The initial and final nozzle geometries are shown in \autoref{Figure10}. \autoref{Figure11} and \autoref{Figure12} show plots of exiting velocity and angle of attack profiles of those nozzles.
	
	
	\autoref{Figure11} illustrates that both nozzles have almost uniform velocity profiles at the outlet because exiting water velocity only depends on nozzle height. However, \autoref{Figure12} shows a considerable difference in angle of attack profiles. This difference stems from the fact that angle of attack is much more dependent on nozzle geometry than velocity and several segments of the geometry play a role in determining it. In the final nozzle, the curve below the guide vane and the curve at the end of the nozzle are linear in terms of polar angle (similar to \cite{Sammartano1}). \par
	
	Calculations show that nozzle loss coefficient for the initial and final nozzles are 0.972 and 0.977, respectively. Using the latter value and \autoref{EquationA1}, and assuming that on average runner loss coefficient is 0.94, the maximum possible efficiency for this design is 85.6\%.
	
	\subsubsection{Detail design: second step} \label{Section4.2.3}
	
\begin{figure}[ht!]
			\centering
			\includegraphics[width = 0.6\columnwidth]{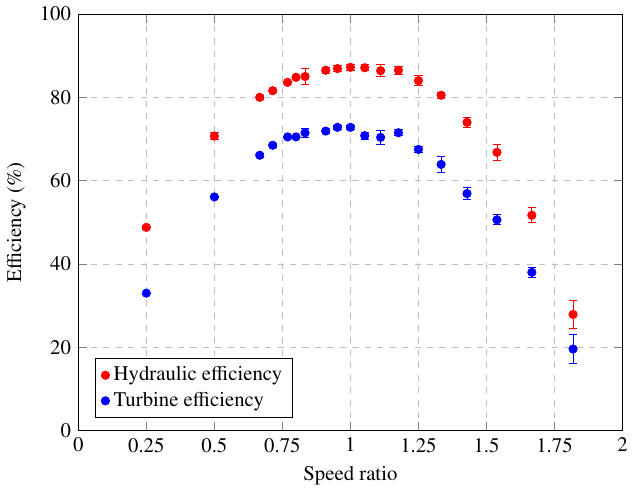}
			\caption{Efficiency as a function of speed ratio. As predicted by \autoref{EquationA1}, the curve is parabolic \cite{Mockmore}.} \label{Figure13}
	\end{figure}

\begin{figure}[h!]
			\centering
			\includegraphics[width = 0.6\columnwidth]{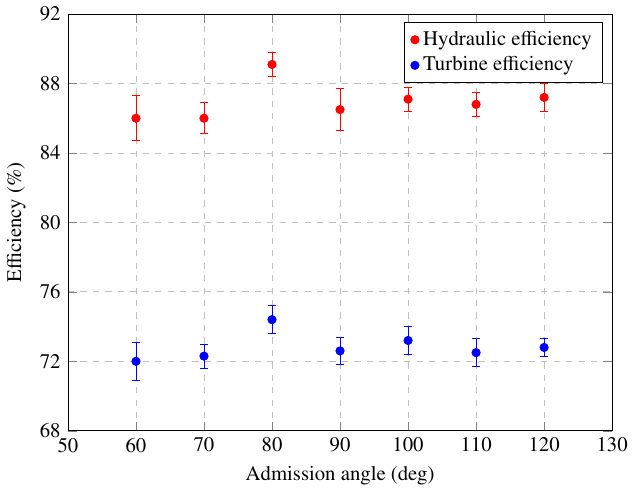}
			\caption{Efficiency as a function of admission angle. The highest efficiency is reached for an admission angle of 80 degrees.} \label{Figure14}
	\end{figure}

\begin{figure}[h!]
			\centering
			\includegraphics[width = 0.6\columnwidth]{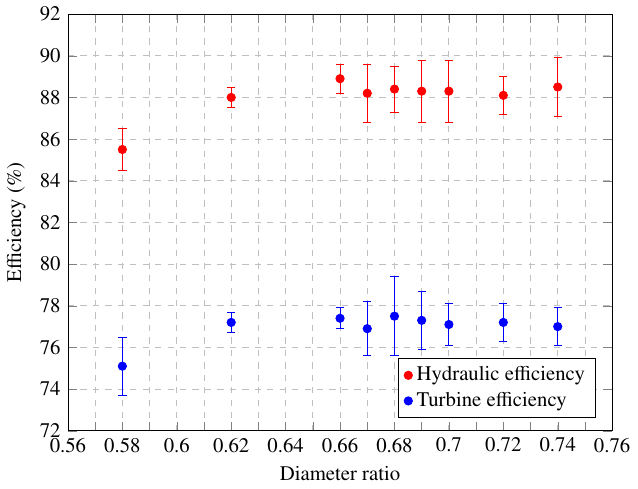}
			\caption{Efficiency as a function of diameter ratio. Turbine efficiency seems constant for diameter ratios between 0.62 and 0.72, with the maximum value occurring at 0.68.} \label{Figure15}
	\end{figure}

\begin{figure}[ht!]
			\centering
			\includegraphics[width = 0.6\columnwidth]{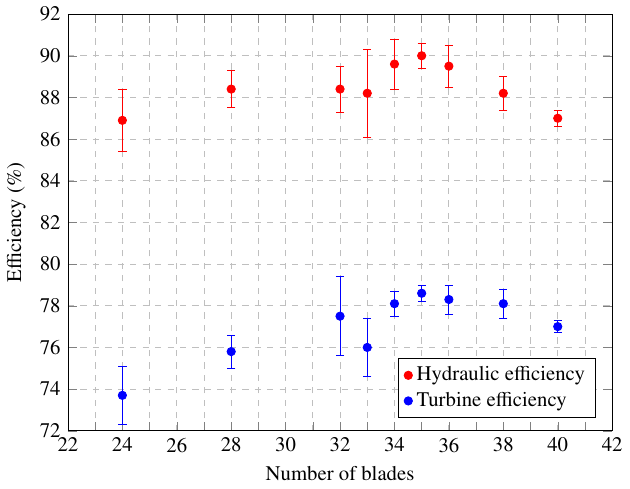}
			\caption{Efficiency as a function of the number of blades. Turbine efficiency seems parabolic in terms of the number of blades and the maximum value is achieved for 35 blades.} \label{Figure16}
	\end{figure}

	The objective of the second step is to optimize and study the effects of the most important turbine parameters on its efficiency. These parameters are rotational speed, angle of attack, admission angle, blade profile, diameter ratio, and the number of blades. For each parameter, we changed its value in a predetermined interval and identified the optimal value (corresponding to the highest efficiency), then used that optimal value in the optimization of the next parameter (a pseudo coordinate ascent method). The initial value, interval of variation, optimal value, and increase in efficiency for each parameter are listed in \autoref{Table4}, and plots of efficiency as a function of each parameter are shown in \autoref{Figure13} to \autoref{Figure16}. In what follows, we discuss the effects of each parameter on the efficiency and flow inside the turbine.
	
	\begin{table}[h!]
		\centering
		\caption{Details of the turbine parameter optimization process in the second step of detail design.} \label{Table4}
		\begin{tabular}{c c c c c}
			\hline
			Turbine parameter & Initial & Variation & Optimal & Efficiency \\
			 & value & interval & value & change (\%) \\
			\hline
			Speed ratio & 1 & 0.25 - $\frac{20}{11}$ & 0.92 & 0 \\
			Angle of attack & 16 & -\footnotemark\addtocounter{footnote}{-1}\addtocounter{Hfootnote}{-1} & 16 & 0 \\
			Admission angle & 120 & 60 - 120 & 80 & 1.6 \\
			Blade profile & see \autoref{Figure4} & other\footnotemark\addtocounter{footnote}{-1}\addtocounter{Hfootnote}{-1} & airfoil\footnotemark & 3.0 \\
			Diameter ratio & 0.657 & 0.58 - 0.74 & 0.68 & 0.1 \\
			Number of blades & 32 & 24 - 40 & 35 & 1.1 \\
			\hline
		\end{tabular}
	\end{table}
	
	\footnotetext{See discussion.}
	
	
	Speed ratio, a dimensionless parameter, is defined as the ratio of rotational speed to the nominal rotational speed ($\frac{N}{N_n}$). \autoref{Figure13} illustrates that turbine efficiency is parabolic in terms of speed ratio (hence rotational speed) and maximum efficiency occurs for speed ratios between 0.9 and 1. This is in accordance with \autoref{EquationA1} and experimental results of Mockmore et al. \cite{Mockmore} and numerous other studies. \autoref{Figure13} also shows that there is a 15\% difference between hydraulic and turbine efficiencies. While the former only takes losses inside the nozzle and runner into account, the latter also accounts for the unused energy of the water that exits the turbine, hence having a lower value. \par
	
\begin{figure}[ht!]
		\centering
		\begin{subfigure}[h]{0.48\columnwidth}
			\centering
			\includegraphics[width = \columnwidth]{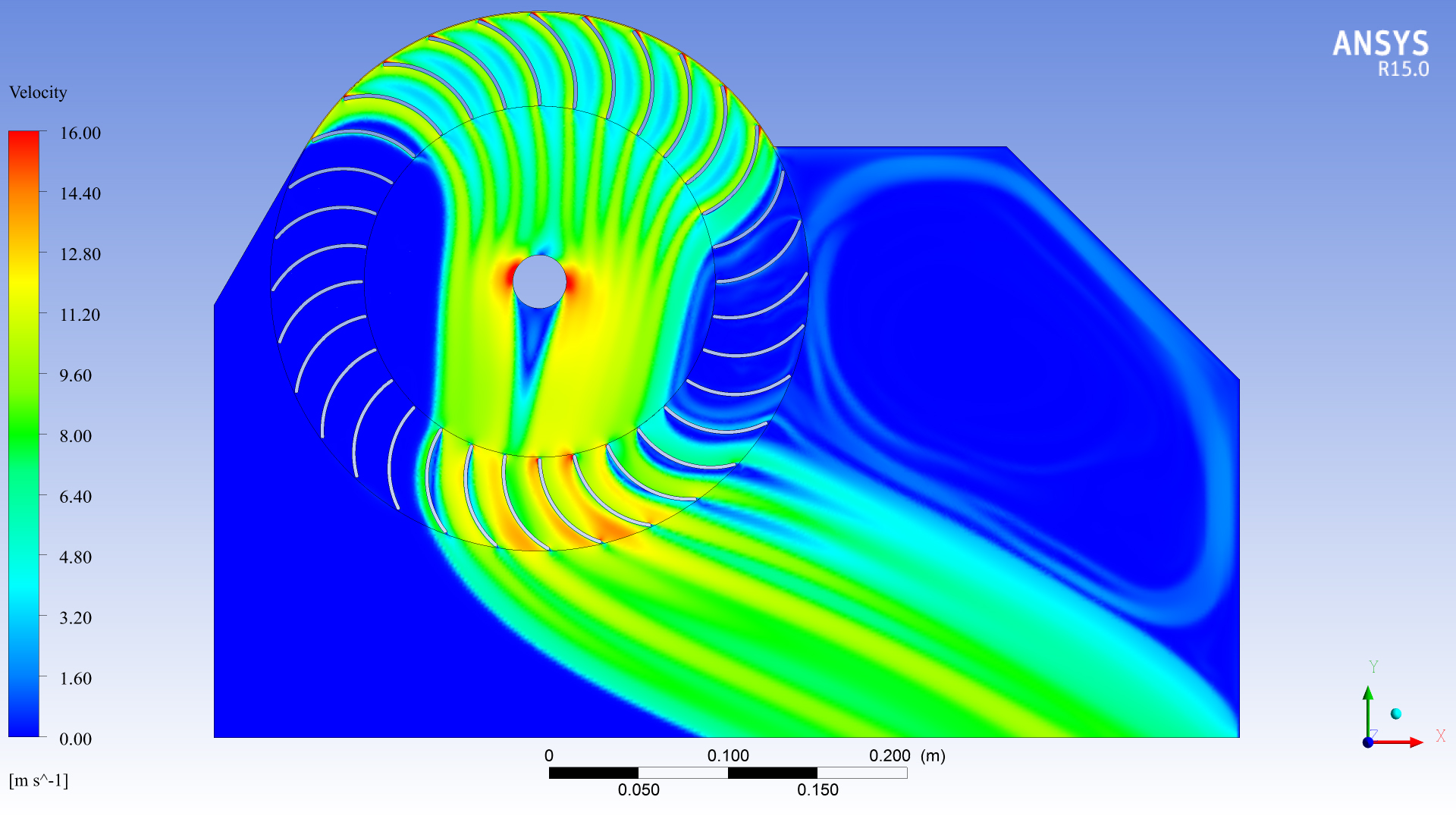}
			\caption{$\frac{N}{N_n} = \frac{1}{4}$} \label{Figure17a}
			\vspace*{4 pt}
		\end{subfigure}
		\begin{subfigure}[h]{0.48\columnwidth}
			\centering
			\includegraphics[width = \columnwidth]{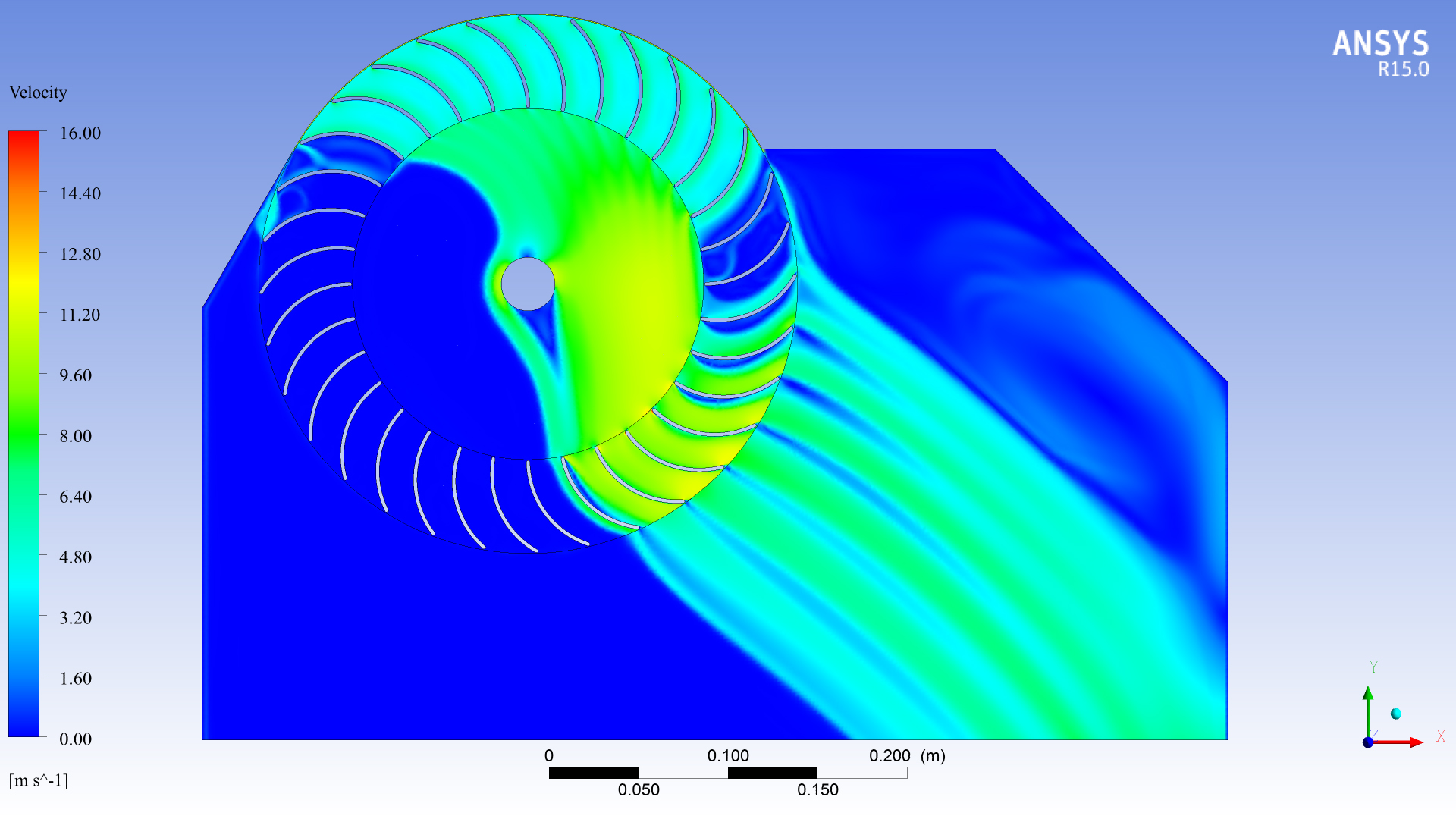}
			\caption{$\frac{N}{N_n} = 1$} \label{Figure17b}
			\vspace*{4 pt}
		\end{subfigure}
		\begin{subfigure}[h]{0.48\columnwidth}
			\centering
			\includegraphics[width = \columnwidth]{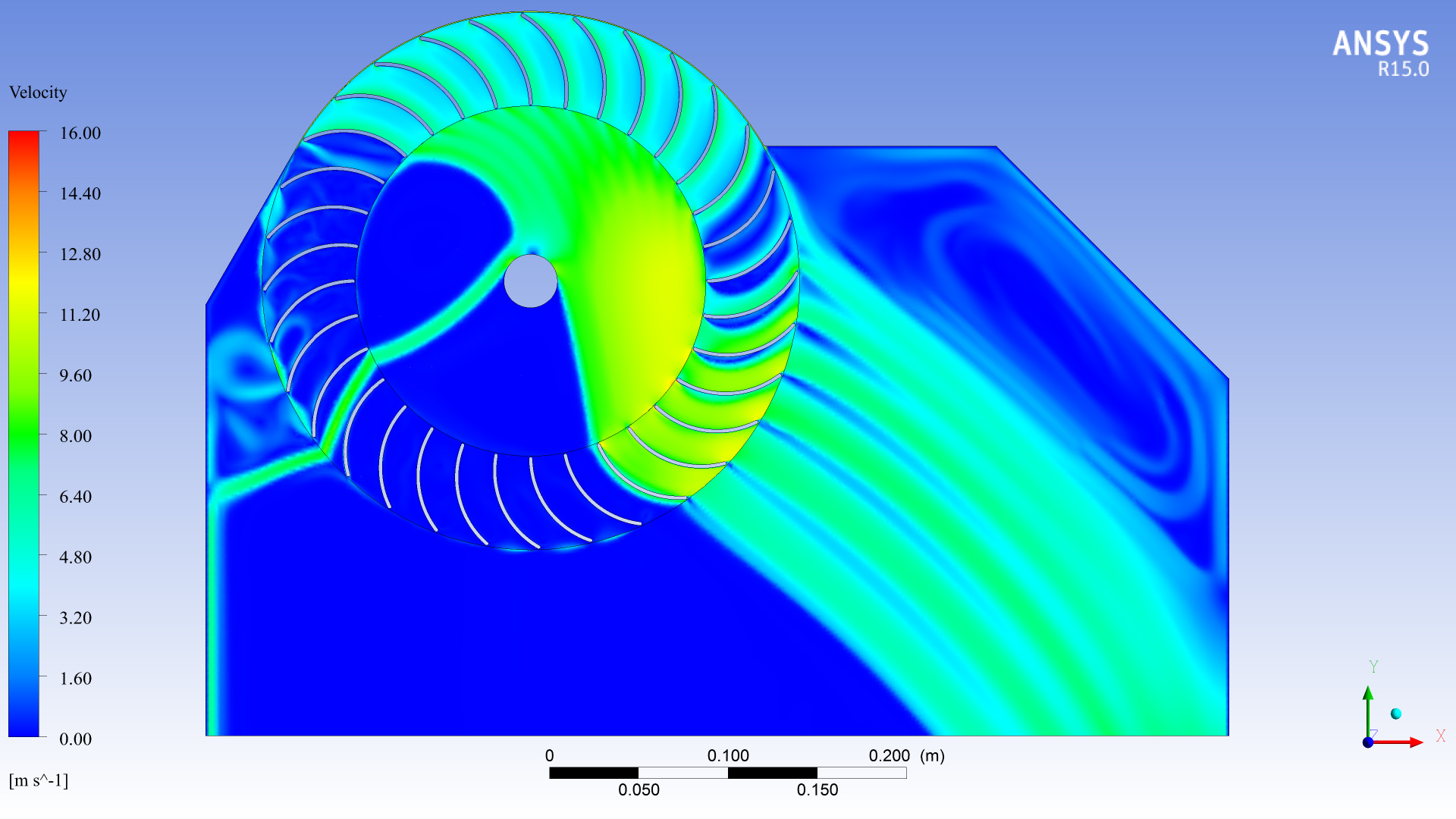}
			\caption{$\frac{N}{N_n} = \frac{10}{9}$} \label{Figure17c}
		\end{subfigure}
		\begin{subfigure}[h]{0.48\columnwidth}
			\centering
			\includegraphics[width = \columnwidth]{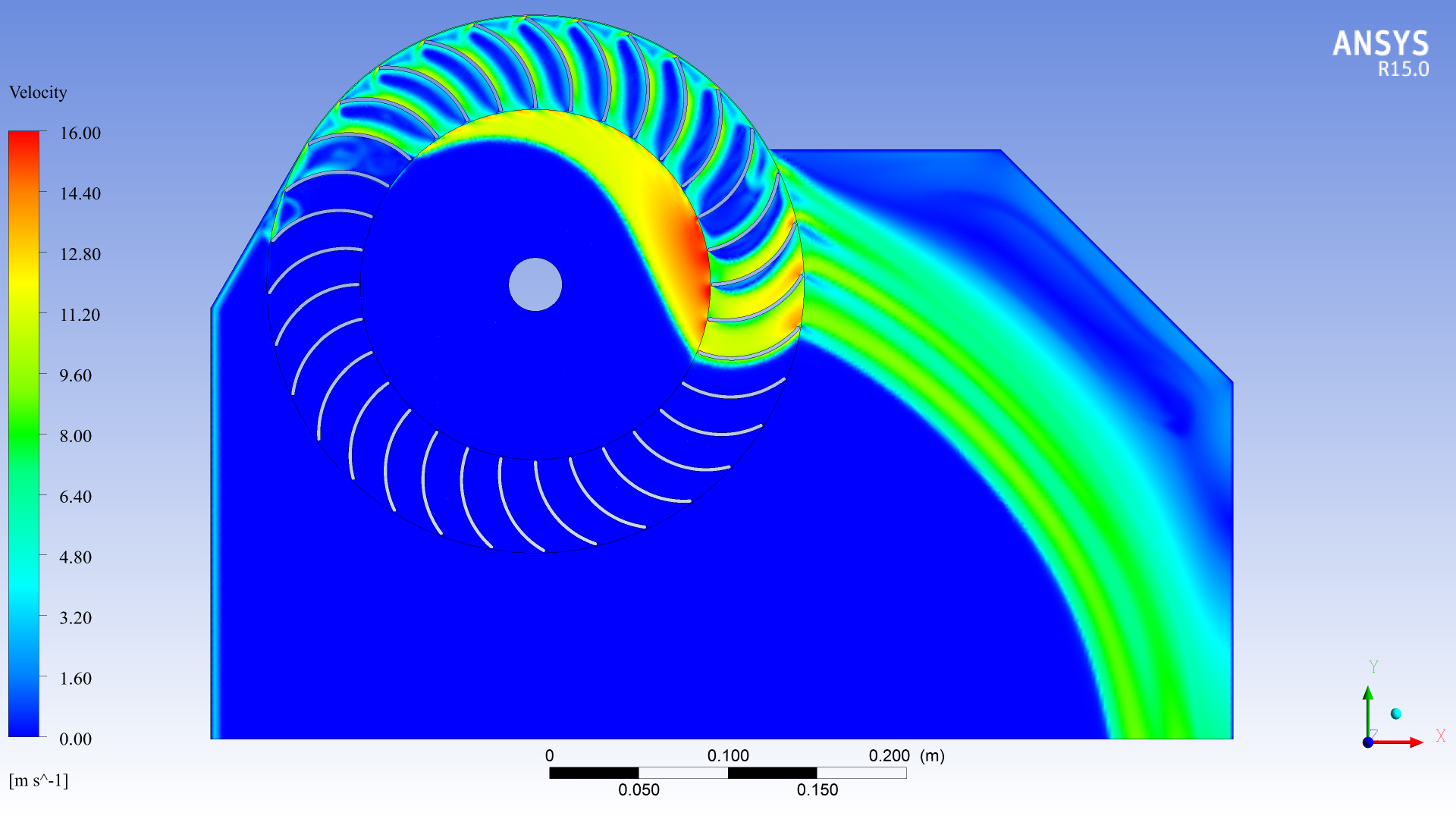}
			\caption{$\frac{N}{N_n} = \frac{20}{11}$} \label{Figure17d}
		\end{subfigure}
		\caption{Water velocity distribution inside the runner for different speed ratios. Very low or very high rotational speeds cause flow separation at the tip of the blades.} \label{Figure17}
	\end{figure}

\begin{figure}[ht!]
		\centering
		\includegraphics[width = 0.6\columnwidth]{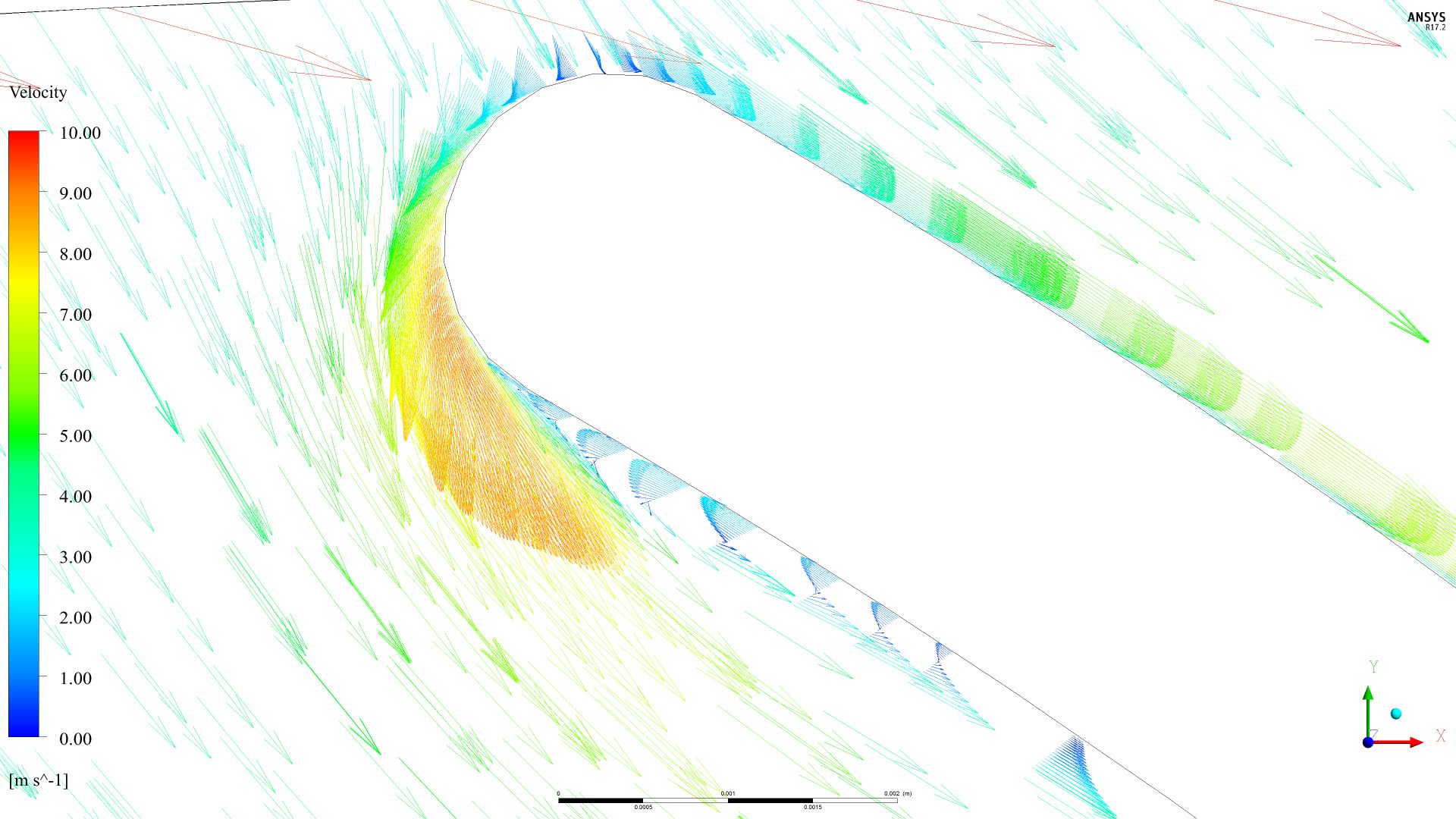}
		\caption{Velocity vectors entering the runner for $\frac{N}{N_n} = \frac{10}{9}$. The vectors are drawn relative to a rotating frame of reference that has a rotational speed equal to that of the runner. Note that even at speed ratios slightly higher than 1, flow separation can be observed on the pressure side of the blades.} \label{Figure18}
	\end{figure}

	At low rotational speeds, the relative angle between the water jet and the runner is less than the blade entrance angle, causing flow separation on the suction side of the blades and reduced efficiency. The same is true about high rotational speeds, only this time flow separation occurs on the pressure side of the blades. This behavior is shown in \autoref{Figure17} and illustrated in detail in \autoref{Figure18} for $\frac{N}{N_n} = \frac{10}{9}$. Note that even when speed ratio is slightly higher than 1, flow separation occurs at the tip of the blades and the adverse pressure gradient can result in undesirable effects such as cavitation \cite{Adhikari2}.
	
	
\begin{figure}[ht!]
		\centering
		\begin{subfigure}[h]{0.48\columnwidth}
			\centering
			\includegraphics[width = \columnwidth]{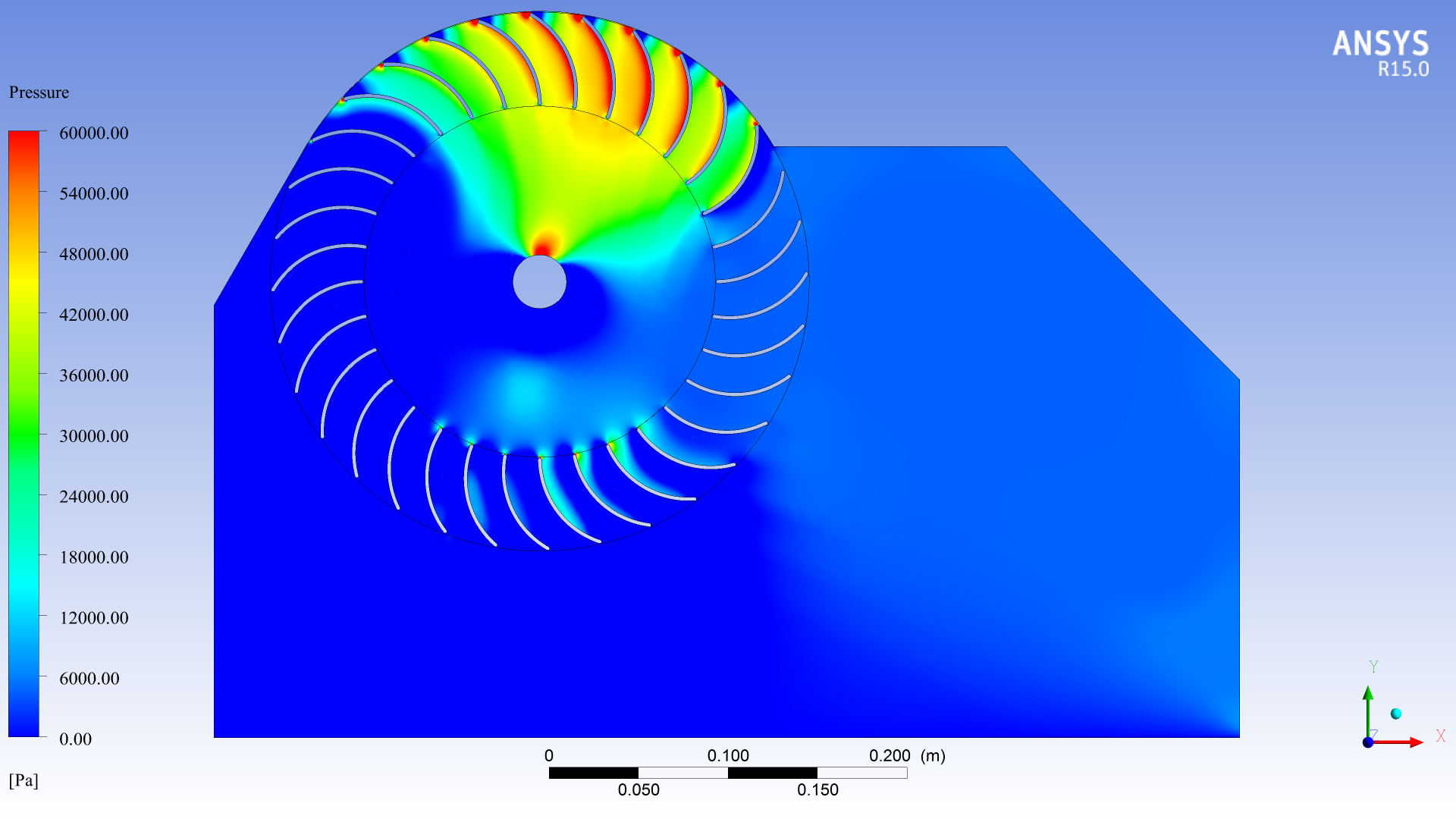}
			\caption{$\frac{N}{N_n} = \frac{1}{4}$} \label{Figure19a}
			\vspace*{4 pt}
		\end{subfigure}
		\begin{subfigure}[h]{0.48\columnwidth}
			\centering
			\includegraphics[width = \columnwidth]{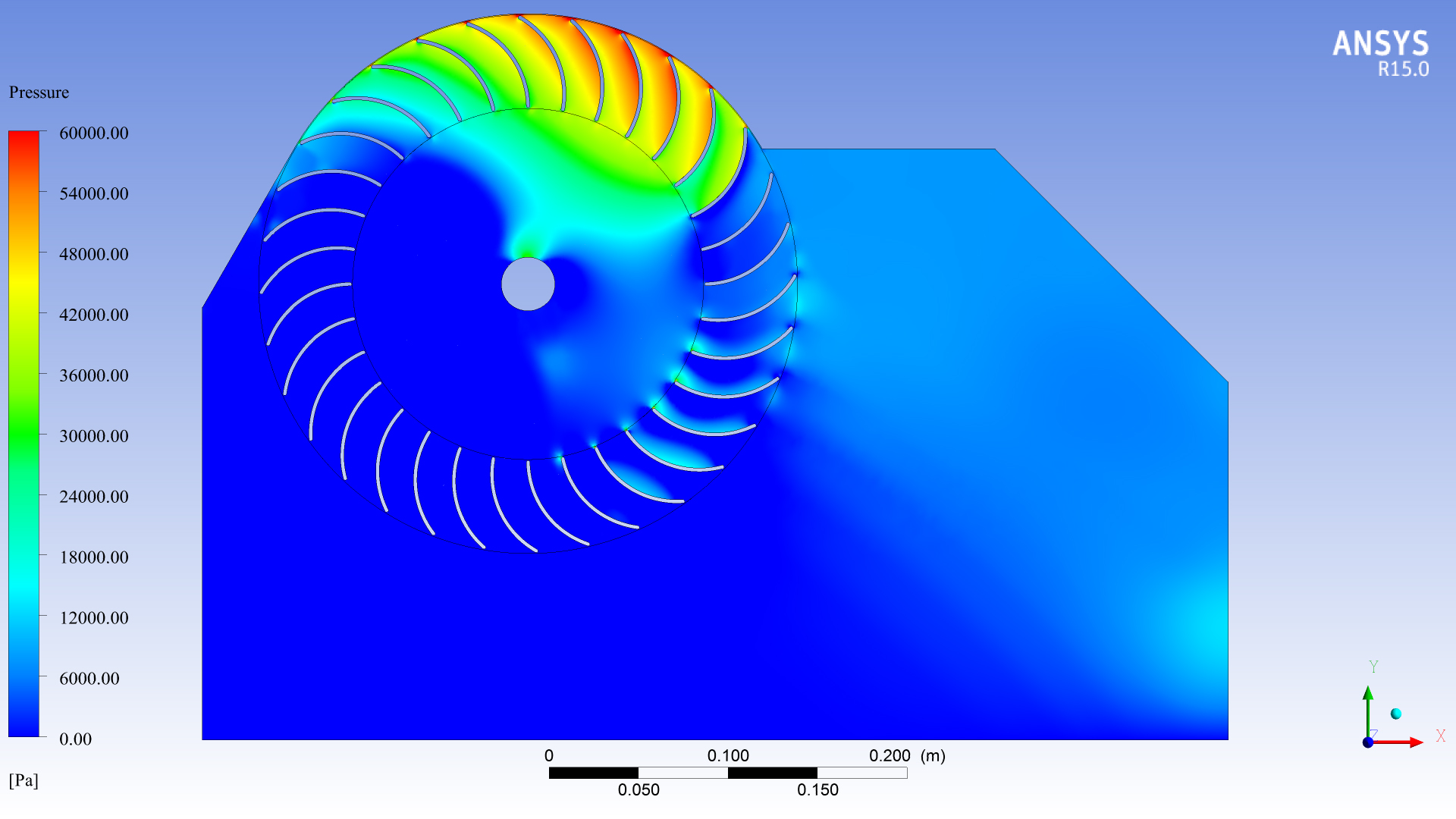}
			\caption{$\frac{N}{N_n} = 1$} \label{Figure19b}
			\vspace*{4 pt}
		\end{subfigure}
		\begin{subfigure}[h]{0.48\columnwidth}
			\centering
			\includegraphics[width = \columnwidth]{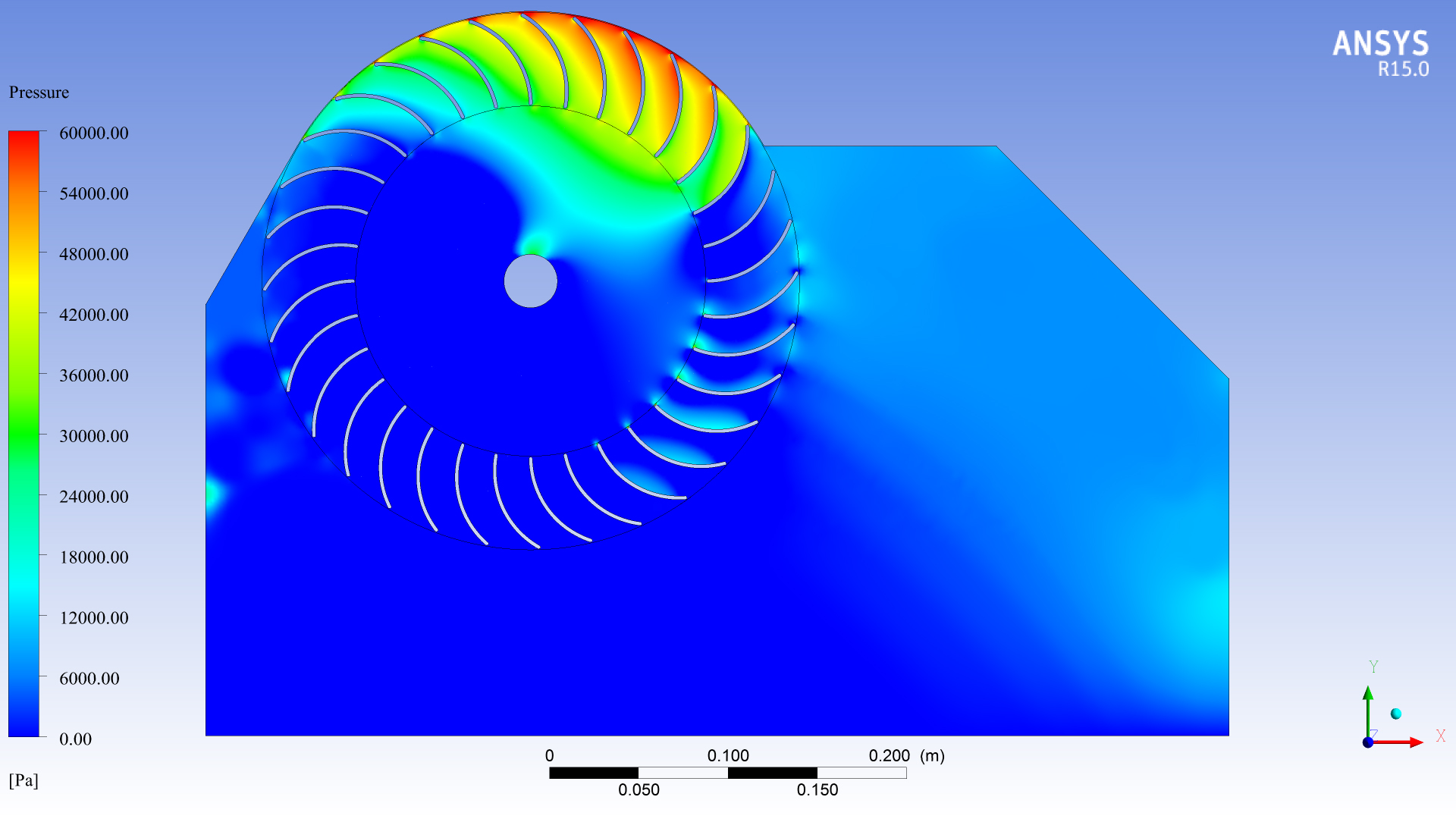}
			\caption{$\frac{N}{N_n} = \frac{10}{9}$} \label{Figure19c}
		\end{subfigure}
		\begin{subfigure}[h]{0.48\columnwidth}
			\centering
			\includegraphics[width = \columnwidth]{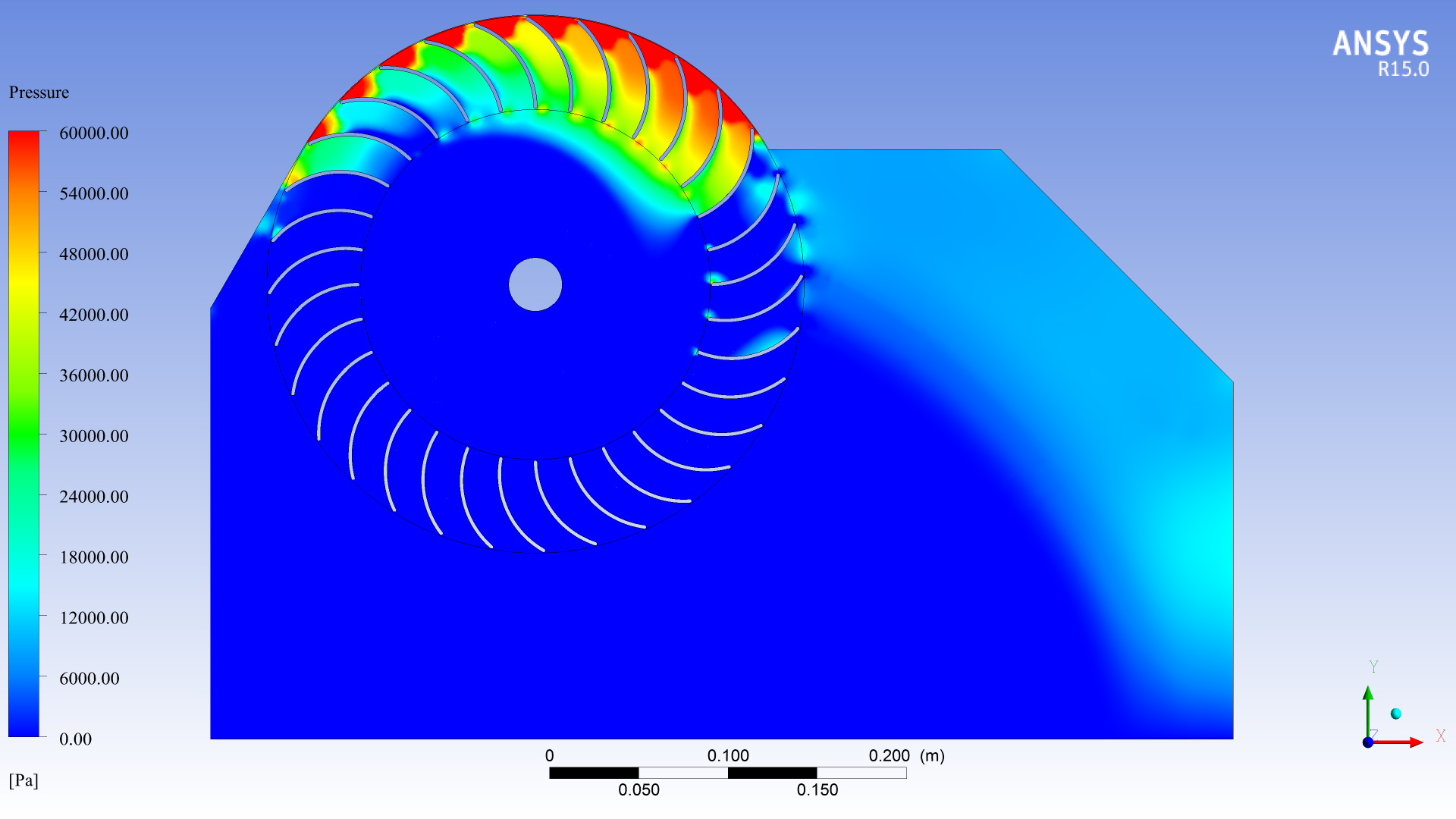}
			\caption{$\frac{N}{N_n} = \frac{20}{11}$} \label{Figure19d}
		\end{subfigure}
		\caption{Relative static pressure distribution inside the runner for different speed ratios. Relative static pressure at the core of the water jet is higher than the sides which are in contact with atmospheric pressure.} \label{Figure19}
	\end{figure}

	Cross-Flow turbines are historically classified as impulse-type turbines. However, \autoref{Figure19} shows that the relative static pressure of the water that enters the runner is significantly higher than zero, leading to the conclusion that this design must be classified as a mixed-flow turbine \cite{Sammartano1, Kaunda2}. This static pressure is caused by the centrifugal force on the water imposed by the rotation of the runner \cite{Sammartano2}. \par
	
	Turbine efficiency reached a maximum value of 72.8\% for an optimal speed ratio of 0.92 (after interpolation), which was used in the next steps to optimize other turbine parameters. \par
	
	The next parameter for optimization was angle of attack. However, it was not optimized in this study because there is a consensus that the 16-degree value, which was used in our simulations, is optimal and no further optimization is required (see the discussion in \autoref{Section2} and \autoref{TableB1} for more details). \par
	
\begin{figure}[ht!]
		\centering
		\begin{subfigure}[h]{0.48\columnwidth}
			\centering
			\includegraphics[width = \columnwidth]{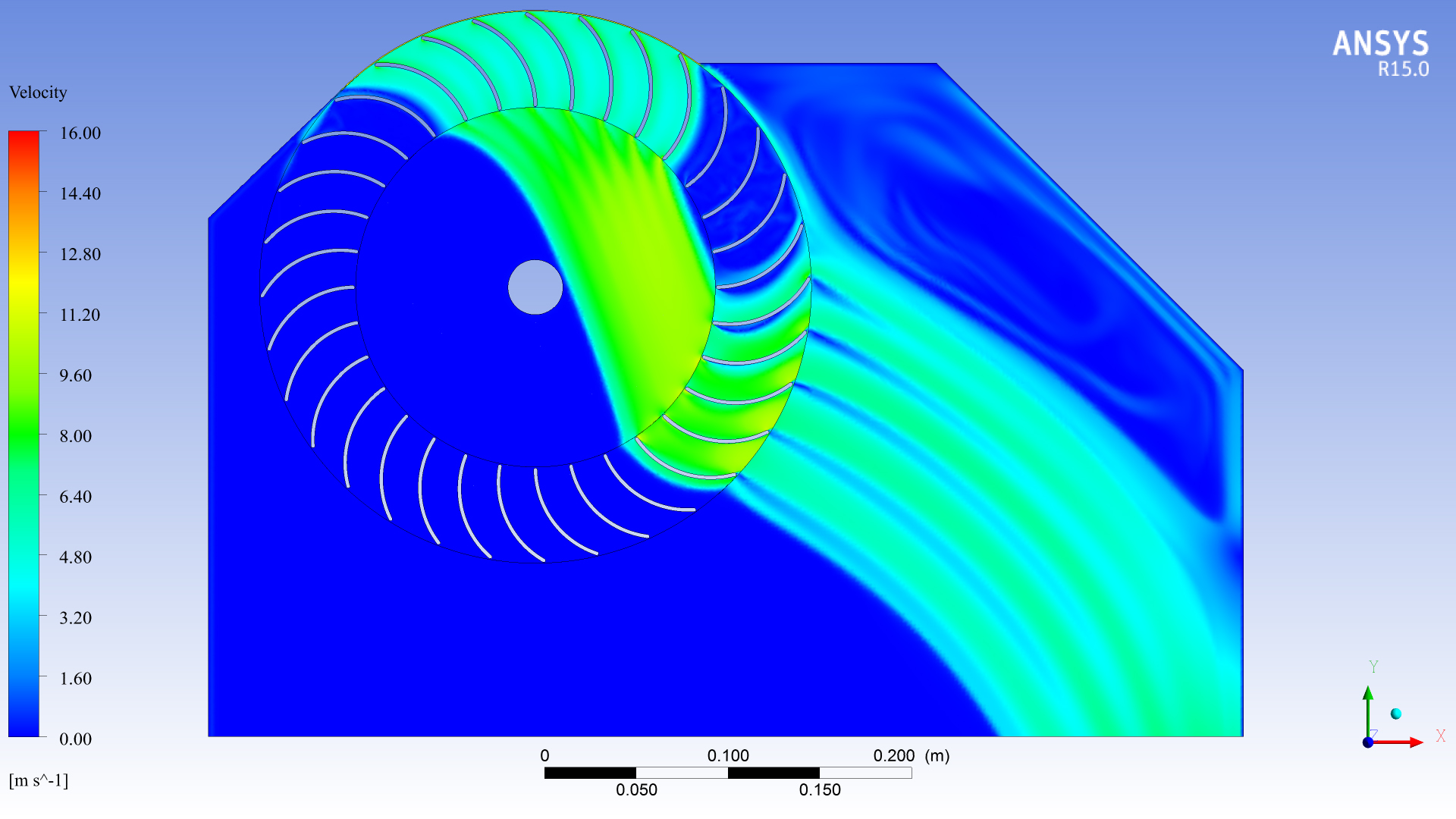}
			\caption{$\lambda = 80^{\circ}$} \label{Figure20a}
		\end{subfigure}
		\begin{subfigure}[h]{0.48\columnwidth}
			\centering
			\includegraphics[width = \columnwidth]{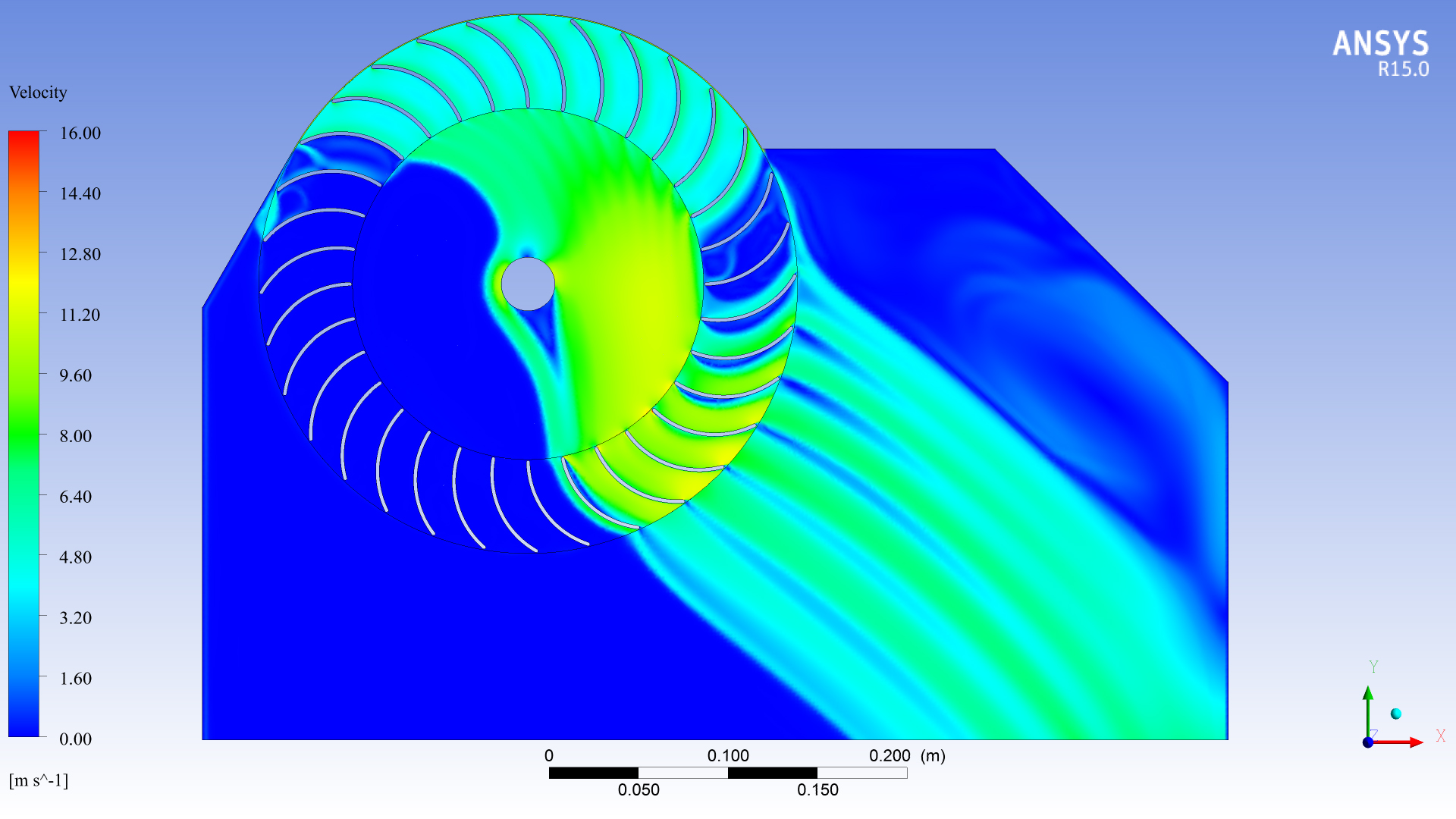}
			\caption{$\lambda = 120^{\circ}$} \label{Figure20b}
		\end{subfigure}
		\caption{Water velocity distribution inside the runner for different admission angles.} \label{Figure20}
	\end{figure}

\begin{figure}[ht!]
		\centering
		\begin{subfigure}[h]{0.48\columnwidth}
			\centering
			\includegraphics[width = \columnwidth]{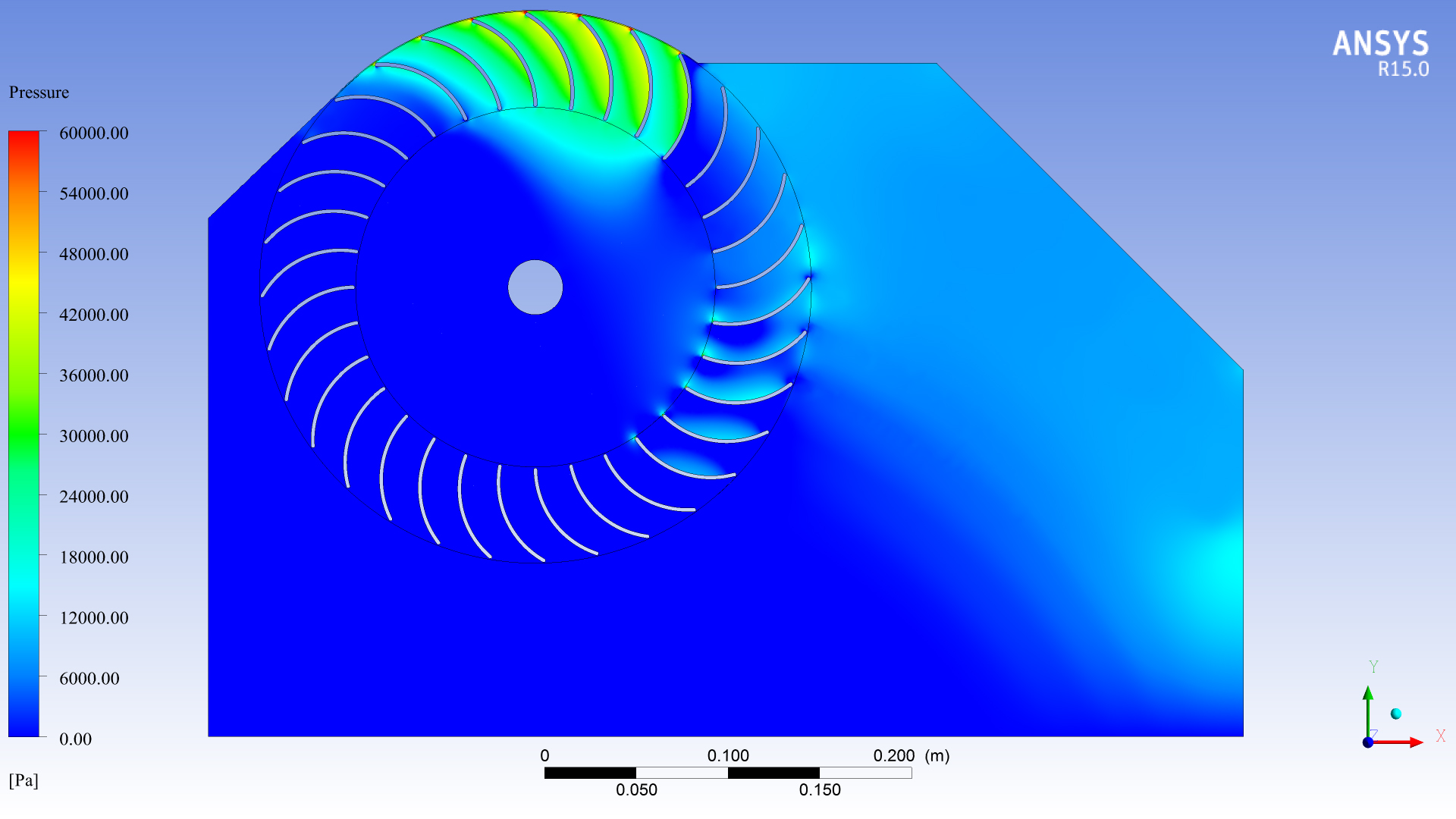}
			\caption{$\lambda = 80^{\circ}$} \label{Figure21a}
		\end{subfigure}
		\begin{subfigure}[h]{0.48\columnwidth}
			\centering
			\includegraphics[width = \columnwidth]{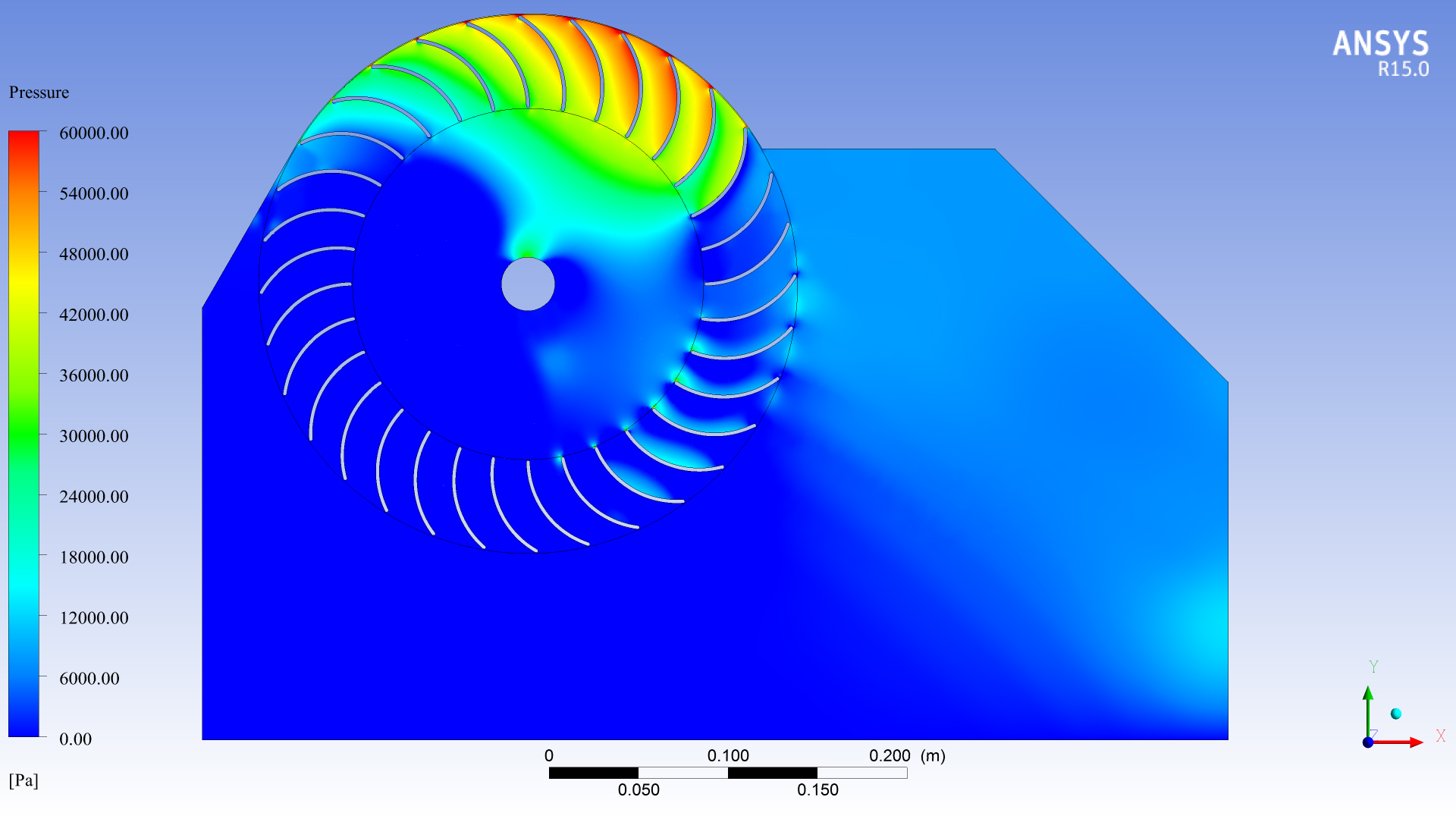}
			\caption{$\lambda = 120^{\circ}$} \label{Figure21b}
		\end{subfigure}
		\caption{Relative static pressure distribution inside the runner for different admission angles. As the admission angle increases, relative static pressure at the core of the water jet increases and so does the average at runner inlet.} \label{Figure21}
	\end{figure}

	The next step was optimizing the admission angle for which the results are shown in \autoref{Figure14}. While turbine efficiency is mostly constant in \autoref{Figure14}, it is notably higher for an 80-degree admission angle which can be attributed to several factors. As can be seen in \autoref{Figure20}, for larger angles a portion of the water that exits the first stage hits the runner shaft, decreasing turbine efficiency. Another factor is the relative static pressure at runner inlet. \autoref{Figure21} shows that a static pressure gradient exists in the water jet that passes through the first stage of the turbine. This pressure is highest at the core of the jet and equal to the atmospheric pressure on the sides. Decreasing the admission angle narrows the entering water jet, reducing its relative static pressure and allowing the turbine to operate more closely to its ideal conditions. On the other hand, decreasing the admission angle increases runner length, which in turn increases water entrainment by the air that is trapped inside the housing, resulting in the loss of power \cite{Durgin}.
	
	
	For an admission angle of 80 degrees, turbine efficiency reached a maximum of 74.4\% and there was a balance between the phenomena described above. This optimal admission angle was used in the next steps. \par
	
\begin{figure}[ht!]
		\centering
		\begin{subfigure}[h]{0.48\columnwidth}
			\centering
			\includegraphics[width = \columnwidth]{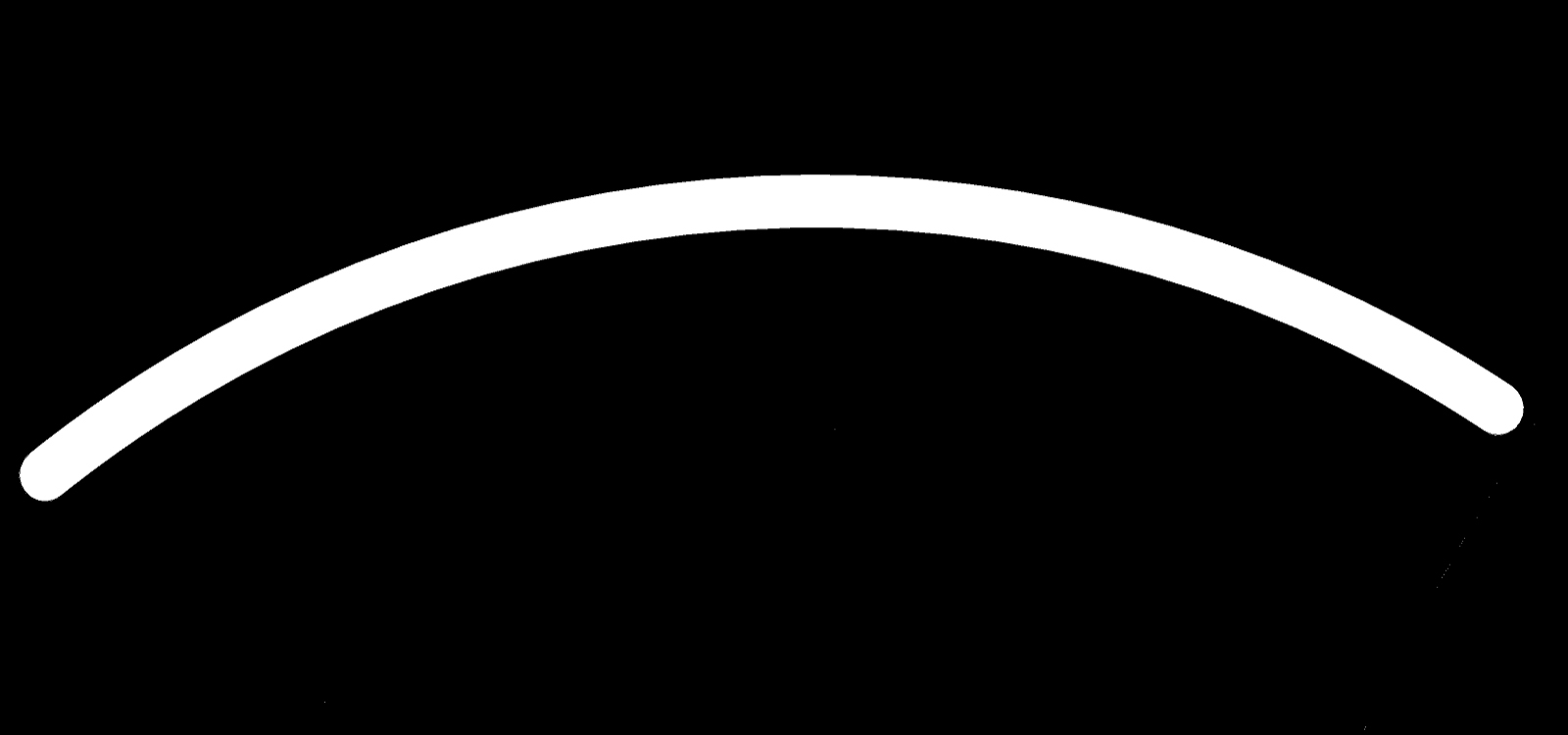}
			\caption{initial design} \label{Figure22a}
		\end{subfigure}
		\begin{subfigure}[h]{0.48\columnwidth}
			\centering
			\includegraphics[width = \columnwidth]{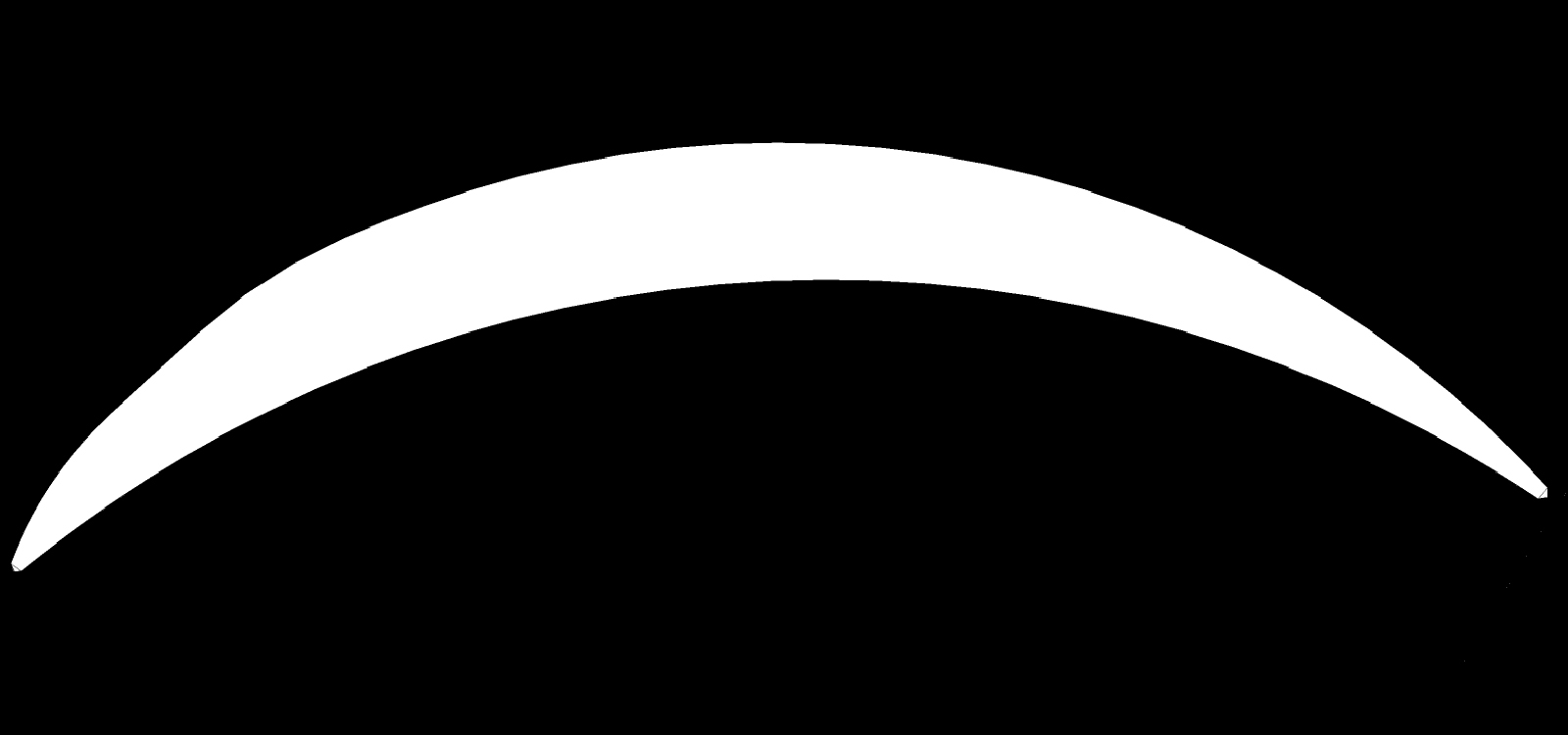}
			\caption{final design} \label{Figure22b}
		\end{subfigure}
		\caption{The initial and final designs of the blade profile. The final profile utilizes the above-zero relative static pressure at the runner inlet and has the minimum amount of resistance to the incoming and outgoing flows.} \label{Figure22}
	\end{figure}

	We continued the optimization process with the blade profile. The initial design of the blades was simply a circular arc with constant thickness and rounded ends, as shown in \autoref{Figure22a}. This profile was then optimized according to the following guidelines. First, the blade tip facing the incoming water had to be as small as possible to avoid efficiency losses due to friction. We chose curved tips over wedge-shaped ones because the latter caused flow separation in off-design load conditions. Second, the blade had to be shaped like an airfoil to utilize the above-zero relative static pressure at the runner inlet. Finally, the other end of the blade had to be small to limit the wake in the exiting flow from the first stage and friction losses of the incoming water jet to the second stage.
	
	Following these guidelines, we modified the blade profile iteratively (testing various characteristics such as blade thickness, airfoil shape, blade tip curvature radius, etc.) and examined the resulting efficiency. The final blade profile is shown in \autoref{Figure22b} which increased turbine efficiency to 77.4\% and was used in the next optimization steps.
	
\begin{figure}[ht!]
		\centering
		\begin{subfigure}[h]{0.48\columnwidth}
			\centering
			\includegraphics[width = \columnwidth]{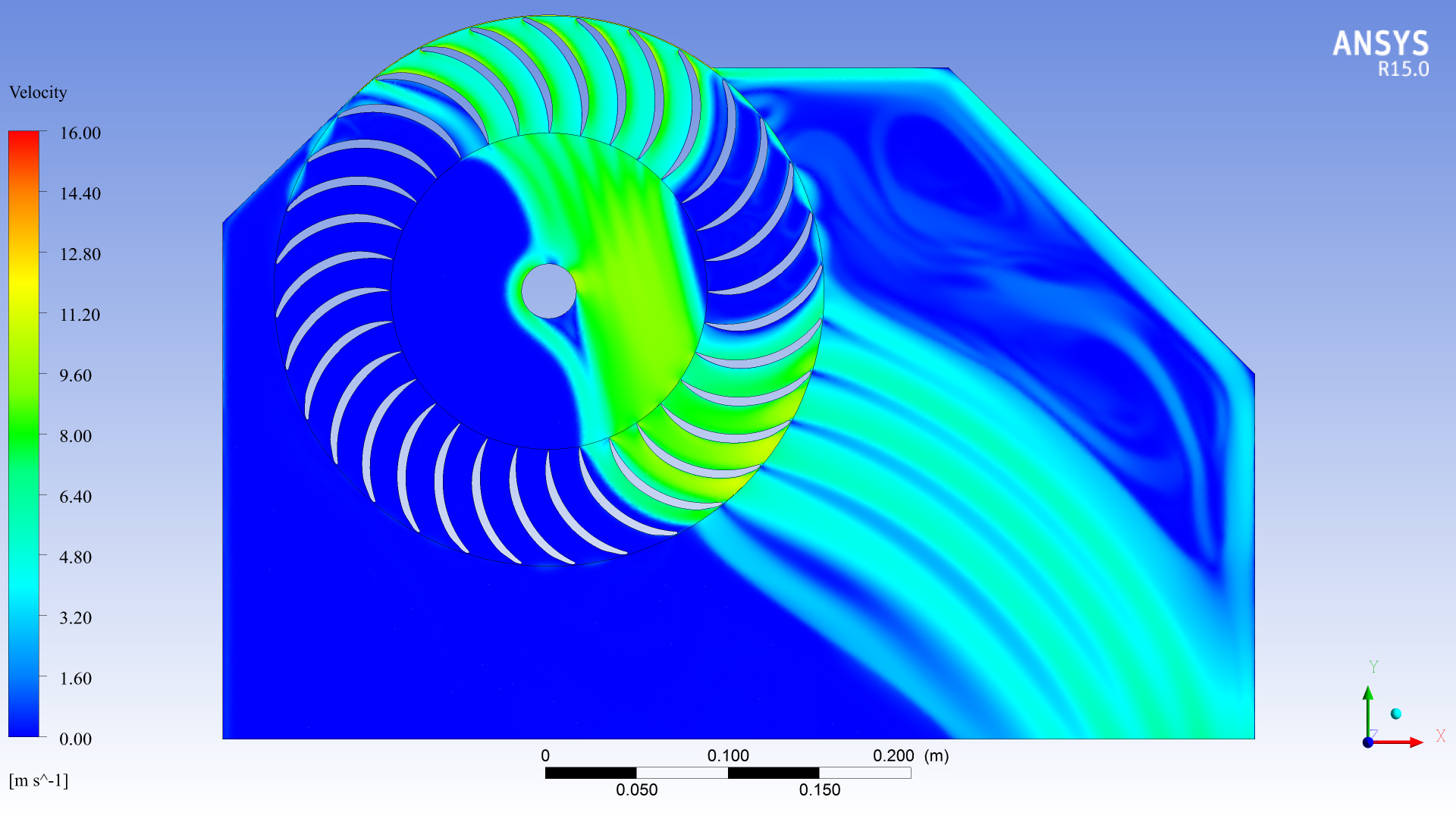}
			\caption{0.58 diameter ratio} \label{Figure23a}
		\end{subfigure}
		\begin{subfigure}[h]{0.48\columnwidth}
			\centering
			\includegraphics[width = \columnwidth]{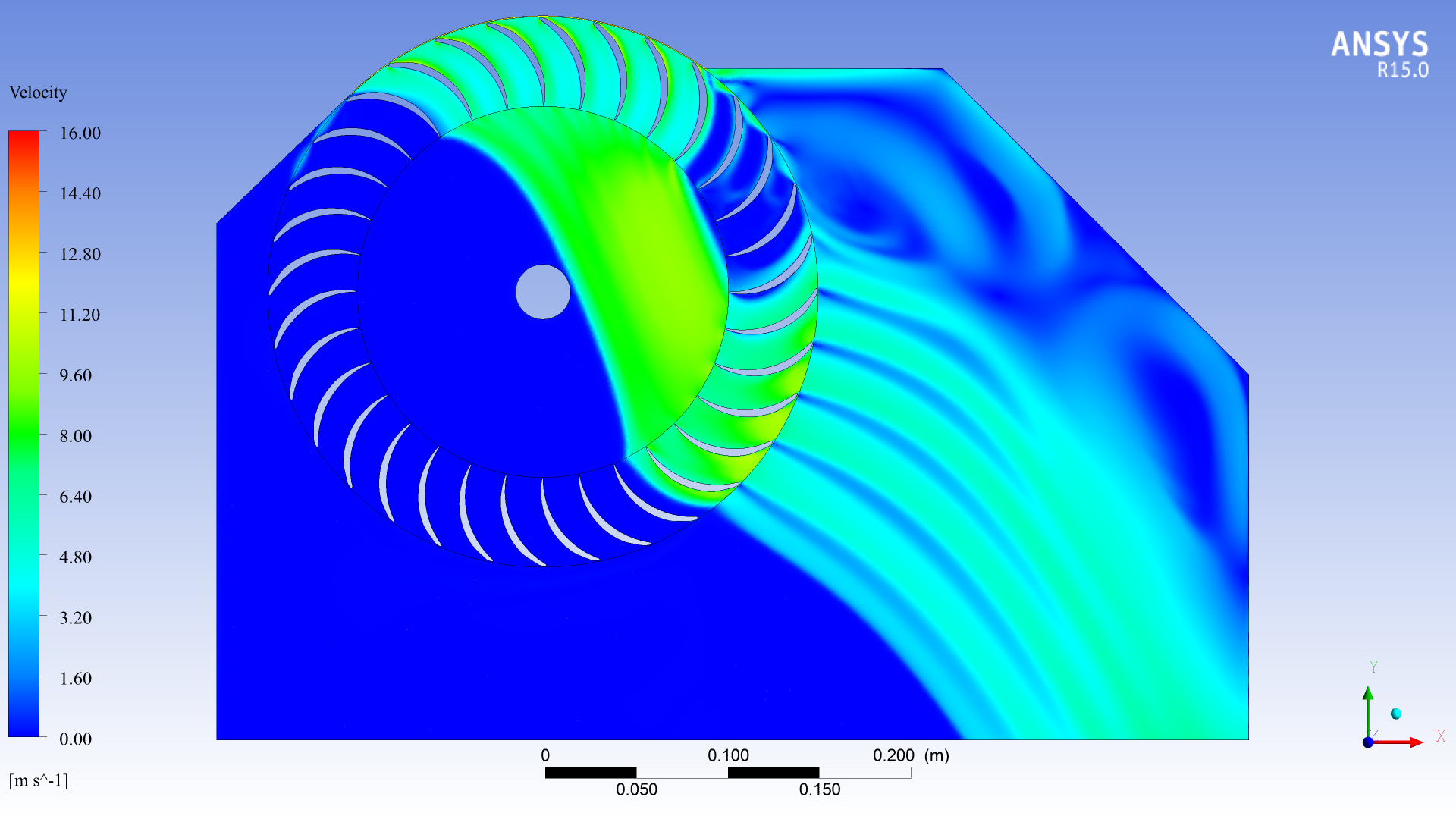}
			\caption{0.68 diameter ratio} \label{Figure23c}
		\end{subfigure}
		\caption{Water velocity distribution inside the runner for different diameter ratios. Increasing the diameter ratio changes the tangential component of the velocity of water that exits the first stage of the turbine and moves the flow sideways.} \label{Figure23}
	\end{figure}

	The next step was optimizing the diameter ratio, and the results are shown in \autoref{Figure15}. \autoref{Figure15} illustrates that turbine efficiency seems to be parabolic in terms of diameter ratio \cite{Yassen} and has a low variation for diameter ratios from 0.62 up to 0.74. Maximum turbine efficiency was reached when diameter ratio was 0.68 and can be attributed to several factors. First, increasing the diameter ratio increases the tangential component of water velocity at each blade's inner tip, which deflects velocity vectors away from the shaft, as shown in \autoref{Figure23}. Second, increasing the diameter ratio reduces the amount of built-up back-pressure at the runner inlet that helps increase turbine efficiency \cite{Sammartano2}. Finally, increasing the diameter ratio reduces the length of the blade arc. Aside from reducing the time that water transfers its energy to the blade, this increased blade curvature amplifies associated hydraulic curvature losses.
	
	For a diameter ratio of 0.68, turbine efficiency reached a maximum of 77.5\%. This diameter ratio maintained a good balance between the factors discussed above and was also close to the theoretical value of 0.657 and the consensus of \autoref{TableB1}. It was used to optimize the number of blades. \par
	
	Results of the final step, optimizing the number of blades, is shown in \autoref{Figure16}. Increasing the number of blades increases the total area of energy transfer to the runner and reduces the volume of semi-filled blade channels on the sides of the admission arc \cite{Sammartano2}. On the other hand, increasing the number of blades increases the total solid thickness that blocks the incoming water jet and reduces the equivalent hydraulic diameter of each blade channel, increasing friction and curvature loss. As illustrated in \autoref{Figure16}, a balance between these factors was achieved for 35 blades and the efficiency reached a maximum of 78.6\%. \par
	
	\autoref{Figure16} shows that turbine efficiency is parabolic in terms of the number of blades \cite{Yassen}; however, the value calculated for 33 blades seems an outlier to the rest of the data. As explained in \autoref{Section4.1}, this is caused by the Frozen Rotor frame change model that does not account for the relative position of runner blades and the admission arc.
	
	\subsubsection{Detail design: third step} \label{Section4.2.4}
	
\begin{figure}[ht!]
			\centering
			\includegraphics[width = 0.6\columnwidth]{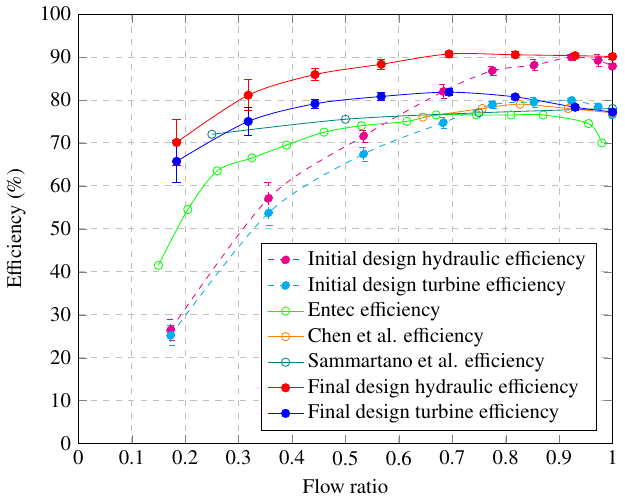}
			\caption{Efficiency as a function of flow ratio for different turbine designs. The initial design operates well for flow ratios between 0.775 and 1, but the efficiency drops significantly for lower flow ratios due to the detachment of water from the lower surface of the guide vane. In the final design the efficiency is significantly improved for low flow ratios and is approximately constant. Maximum efficiency is reached for an opening of 15 degrees which corresponds to a flow ratio of 0.69. The Entec turbine efficiency curve was recorded for a 15 kW nominal power of the 29 kW model turbine \cite{Entec}. Chen et al. and Sammartano et al. efficiency curves are reproduced from \cite{Chen1} and \cite{Sammartano5}, respectively.} \label{Figure24}
	\end{figure}

\begin{figure}[ht!]
		\centering
		\begin{subfigure}[h]{0.48\columnwidth}
			\centering
			\includegraphics[width = \columnwidth]{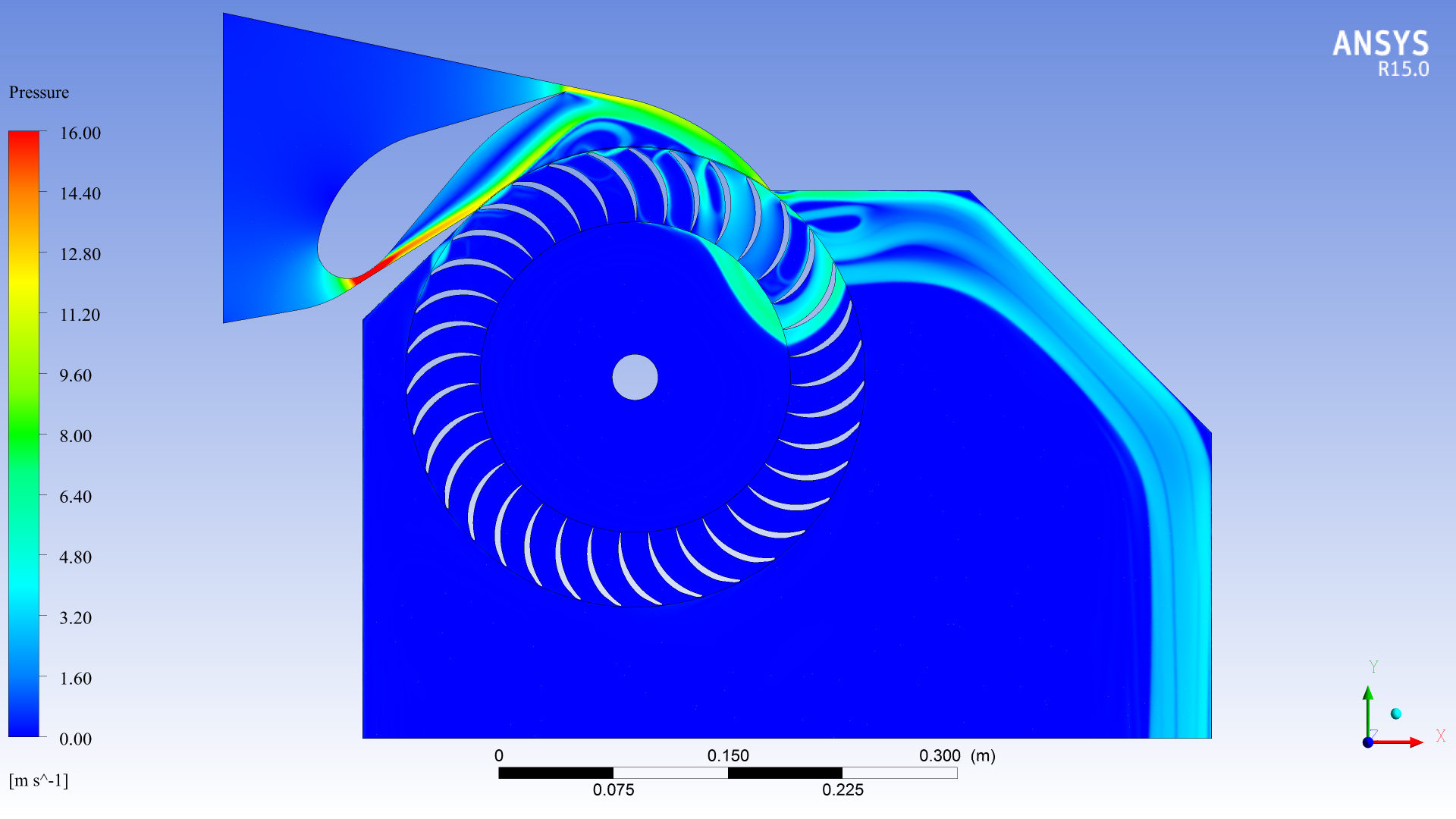}
			\caption{3 degrees open} \label{Figure25a}
			\vspace*{4 pt}
		\end{subfigure}
		\begin{subfigure}[h]{0.48\columnwidth}
			\centering
			\includegraphics[width = \columnwidth]{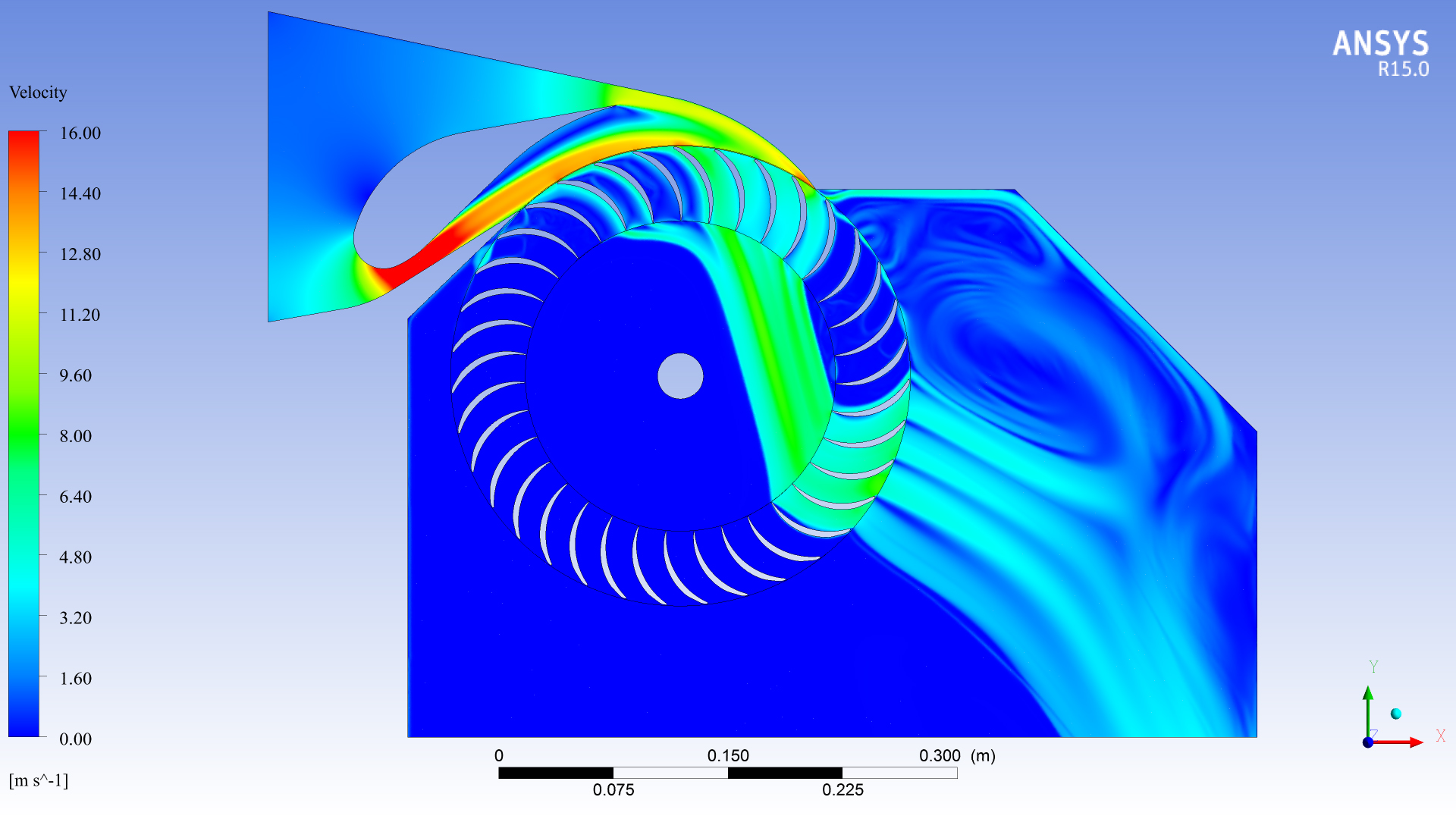}
			\caption{9 degrees open} \label{Figure25b}
			\vspace*{4 pt}
		\end{subfigure}
		\begin{subfigure}[h]{0.48\columnwidth}
			\centering
			\includegraphics[width = \columnwidth]{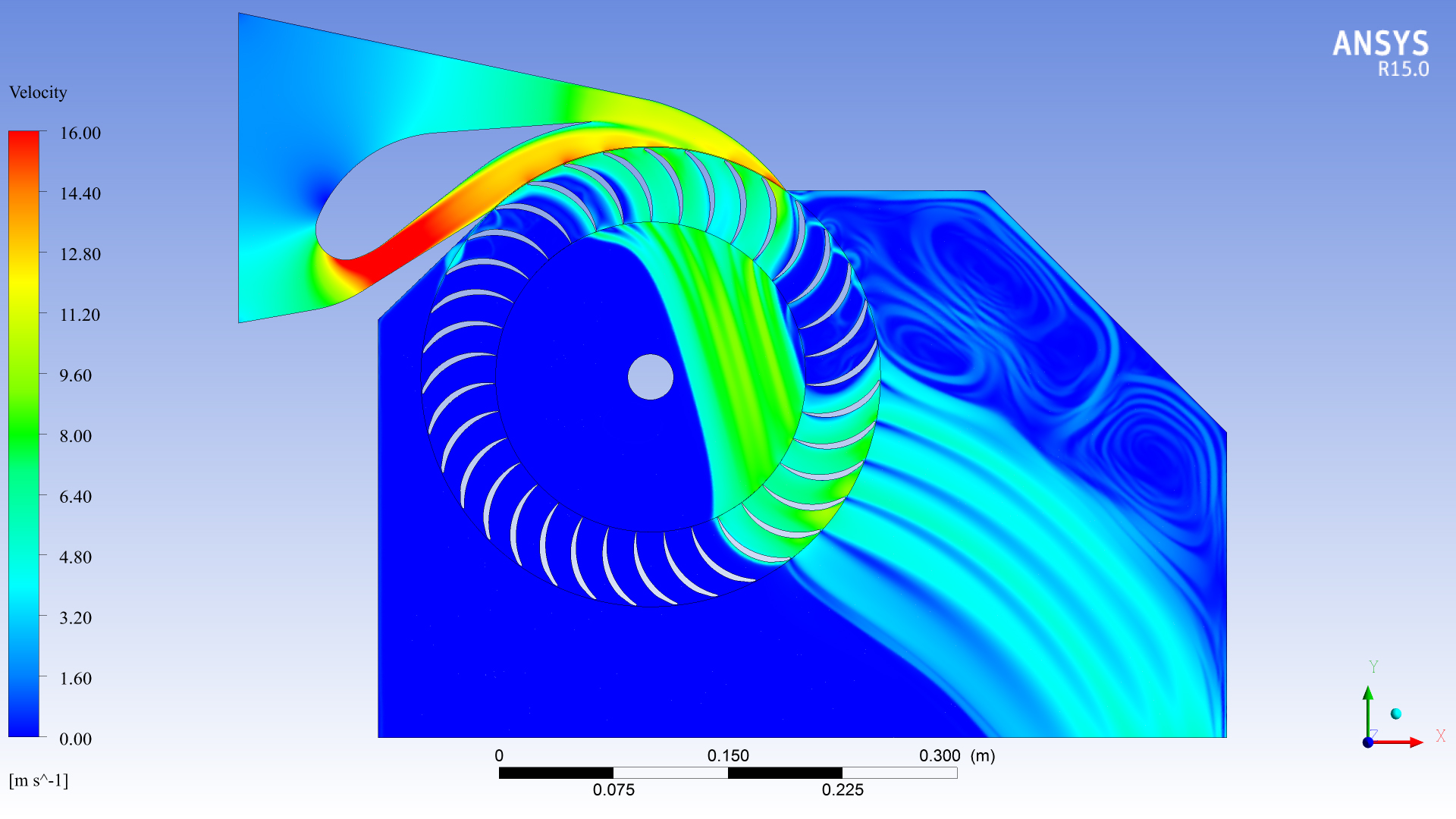}
			\caption{15 degrees open} \label{Figure25c}
		\end{subfigure}
		\begin{subfigure}[h]{0.48\columnwidth}
			\centering
			\includegraphics[width = \columnwidth]{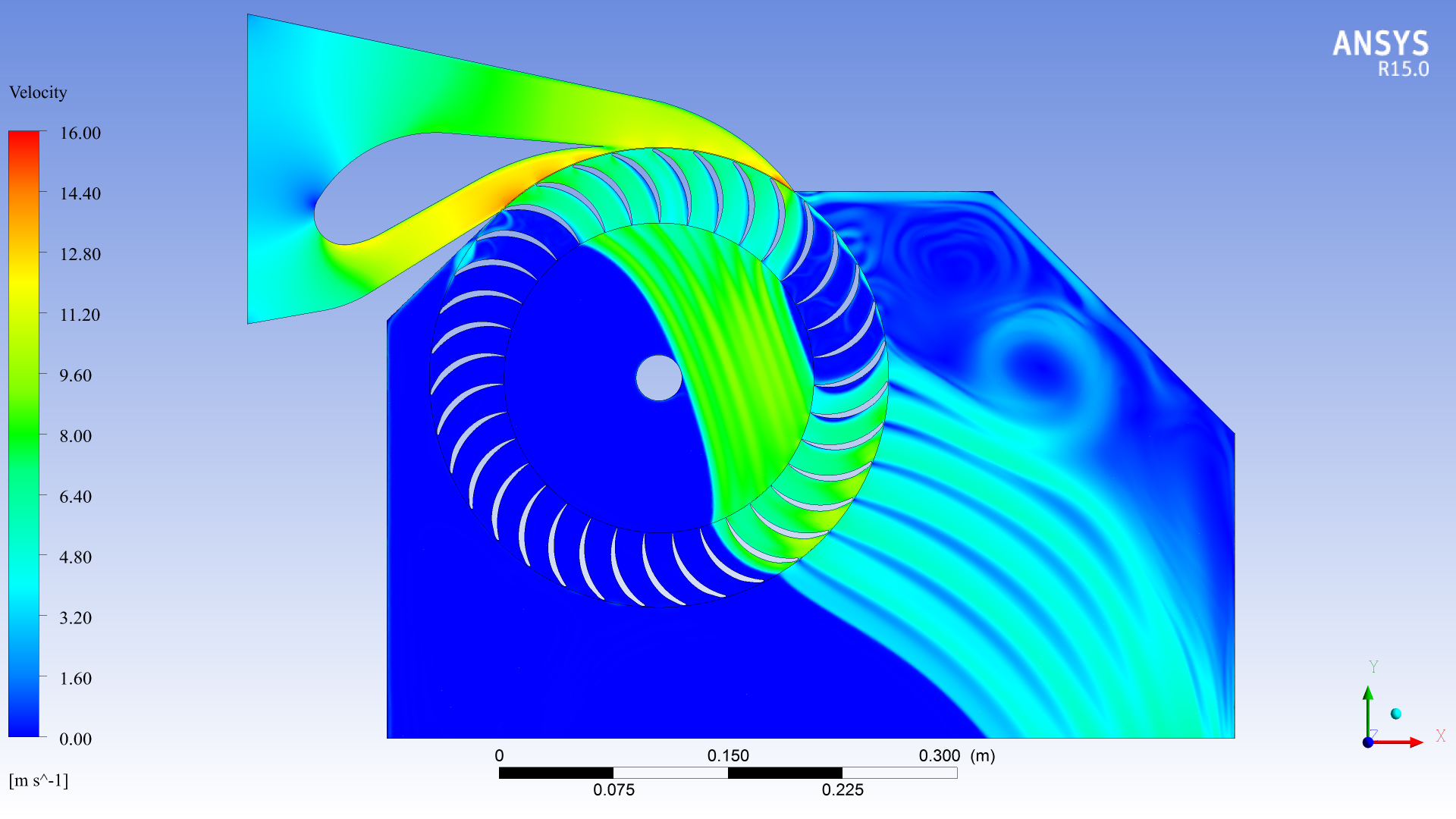}
			\caption{24 degrees open} \label{Figure25d}
		\end{subfigure}
		\caption{Water velocity distribution inside the initial turbine design for different guide vane angles. For small angles, there is an abrupt increase in the velocity of the stream that passes below the guide vane and it gets detached from the guide vane's lower surface. Together with multiple half-filled blade channels, this results in a significant loss of power.} \label{Figure25}
	\end{figure}

	In the final step, the designed nozzle and runner were assembled and flow through the resulting turbine was simulated under various load conditions. The turbine was modified as necessary to reach a relatively flat efficiency curve over a wide range of operating conditions (nozzle geometry was modified to reflect the new 80-degree admission angle). For these simulations, the guide vane angle was varied from fully closed to the fully open position in 3-degree increments to evaluate the performance of the turbine under constant head, variable flow conditions. The results are plotted in terms of flow ratio (the ratio of volume flow rate to the nominal volume flow rate) in \autoref{Figure24} along with a few efficiency curves from the literature \cite{Chen1, Sammartano5}. Water velocity distribution for some cases is shown in \autoref{Figure25}.
	
	
	As can be seen from \autoref{Figure24}, although turbine efficiency is approximately constant for flow ratios higher than 0.775, it drops significantly for those ratios below that limit. This is caused by an increase in nozzle loss and poor entrance of the water jet to the runner. As the guide vane angle decreases, the shape of the channel below the guide vane transforms from a nozzle into a diffuser, causing the lower water jet to detach from the guide vane's lower surface. This water jet partially enters the runner, increasing the number of half-filled blade channels, with the rest of the water traveling in the space above the runner and merging with the upper water jet. \par
	
	Note that there is a small uptick in maximum turbine efficiency compared to the previous simulations, from 78.6\% to 79.9\%. This can be attributed to the off-design conditions (the maximum occurring at a flow ratio of 0.923), as well as difference in boundary conditions. In previous simulations, the water that entered the runner was constrained by the theoretical values and uniformity of velocity and angle of attack. However, since only water head is specified at nozzle inlet for the simulations of the third step, the water jet that reaches the runner has different, non-uniform velocity and angle of attack profiles, slightly improving the efficiency. \par
	
	Though decoupling the nozzle and runner domains and optimizing them individually simplifies and speeds up the design process, the discussion above and the fact that the two domains are intricately interrelated shows that this approach is not without its consequences. It is important to note that the simplifications made in the process of defining the boundary conditions for individual domains can distort the optimization process. For example, we assigned a uniform atmospheric static pressure boundary condition to the nozzle outlet in the first step, but as the results of the second step showed that assumption does not completely reflect the reality and may have had a small impact on the optimization process. Similarly, we assumed uniform water velocity and angle of attack profiles at the runner inlet for simulations of the second step, but the real velocity and angle of attack profiles were different and this may have slightly distorted the parameter optimization process. Nevertheless, we think the benefits gained in having a faster and simpler design process outweigh these errors, as they can be accounted for in the third step. Furthermore, by studying the presented results for the third step, these errors can be remedied in future simulations by defining boundary conditions that more accurately reflect reality, for example a non-uniform, above-atmospheric static pressure profile at the nozzle outlet. \par
	
\begin{figure}[ht!]
		\centering
		\begin{subfigure}[h]{0.48\columnwidth}
			\centering
			\includegraphics[width = \columnwidth]{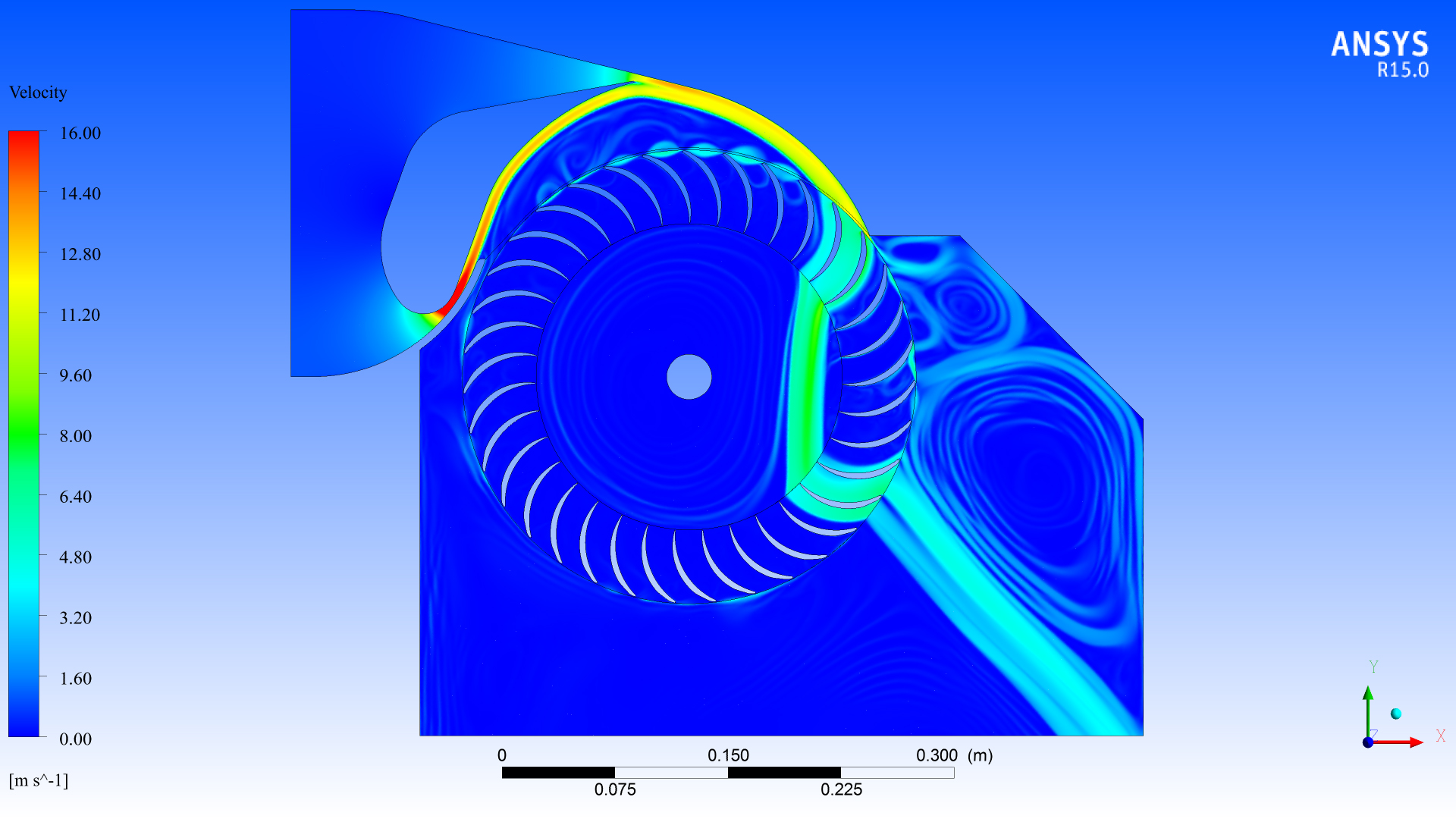}
			\caption{3 degrees open} \label{Figure26a}
			\vspace*{4 pt}
		\end{subfigure}
		\begin{subfigure}[h]{0.48\columnwidth}
			\centering
			\includegraphics[width = \columnwidth]{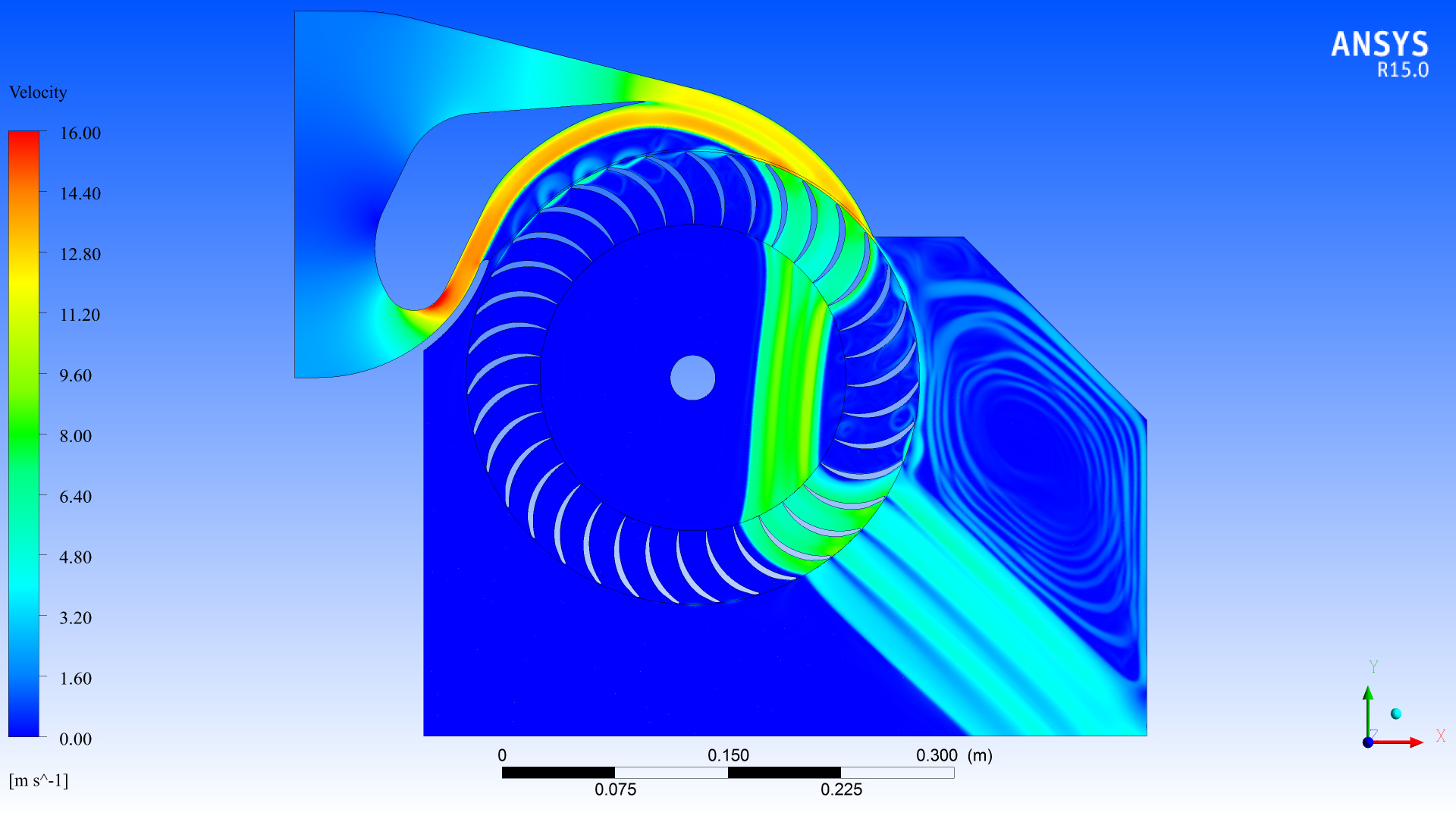}
			\caption{9 degrees open} \label{Figure26b}
			\vspace*{4 pt}
		\end{subfigure}
		\begin{subfigure}[h]{0.48\columnwidth}
			\centering
			\includegraphics[width = \columnwidth]{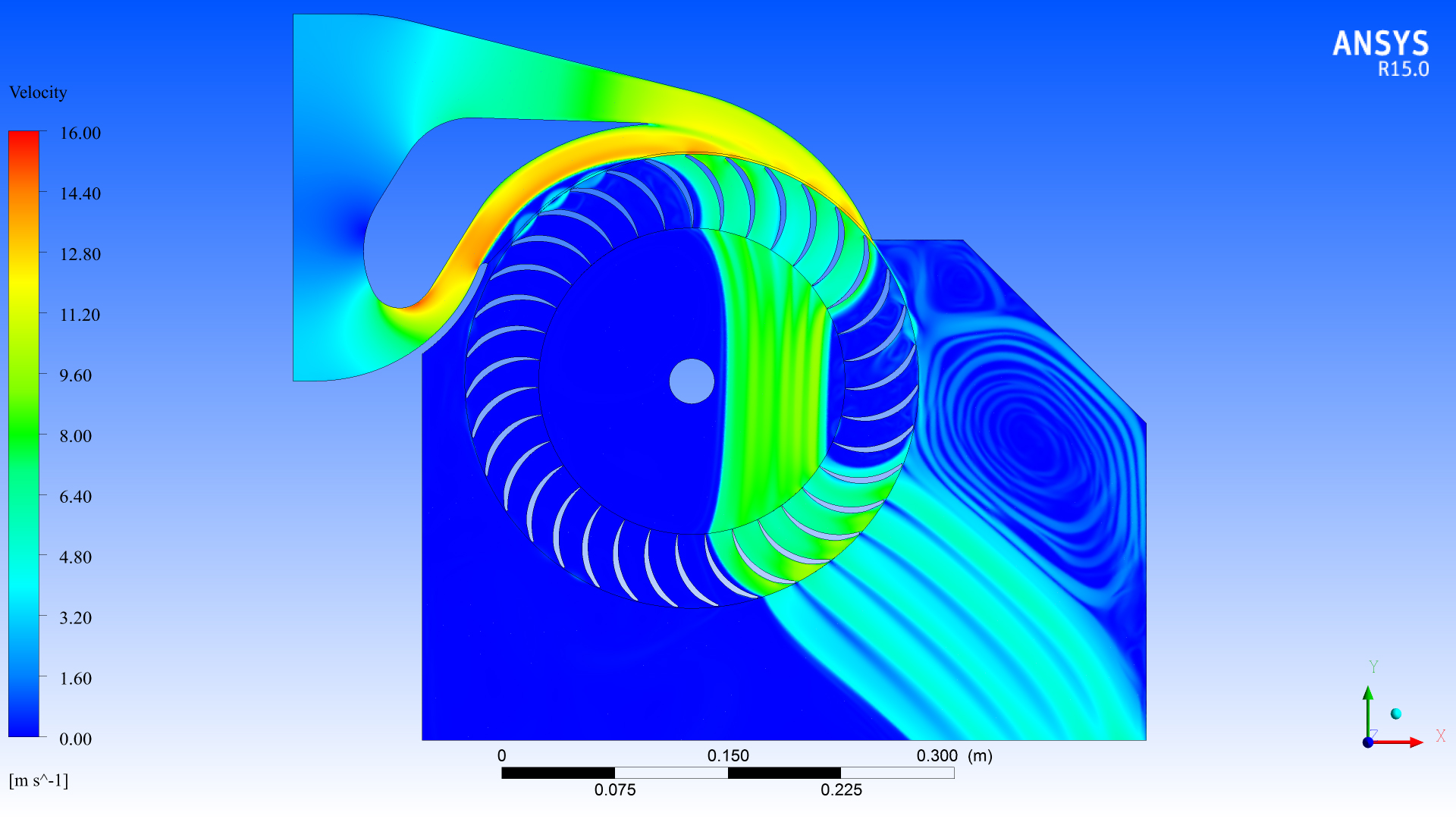}
			\caption{15 degrees open} \label{Figure26c}
		\end{subfigure}
		\begin{subfigure}[h]{0.48\columnwidth}
			\centering
			\includegraphics[width = \columnwidth]{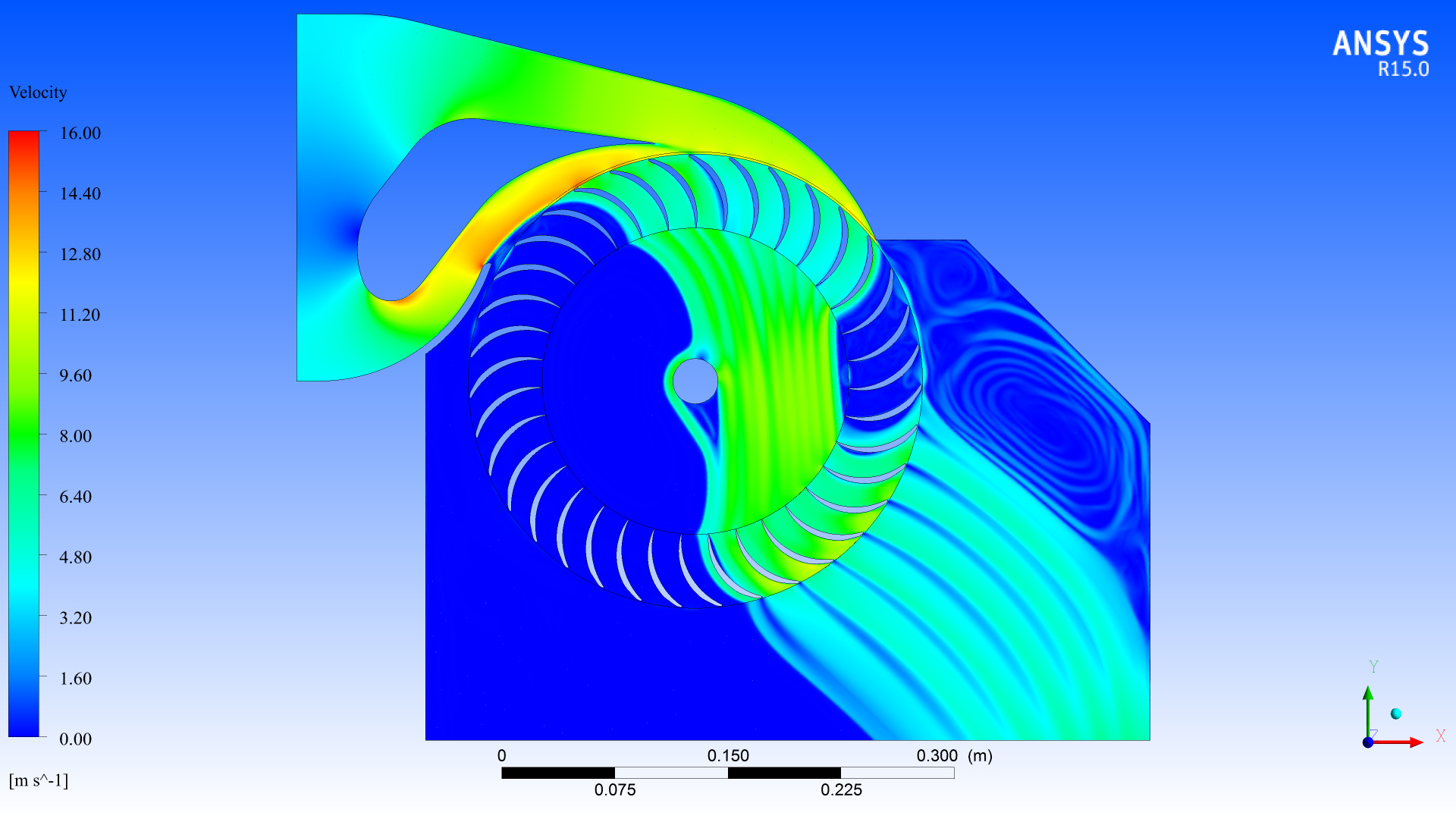}
			\caption{21 degrees open} \label{Figure26d}
		\end{subfigure}
		\caption{Water velocity distribution inside the final turbine design for different guide vane angles. Note that the lower water jet is attached to the lower surface of the guide vane even for small angles.} \label{Figure26}
	\end{figure}

	In order to improve the efficiency curve, we redesigned the channel below the guide vane so that it could maintain a nozzle shape even for small guide vane angles; however, the new geometric constraints forced us to increase the admission angle to 110 degrees as a trade-off. The efficiency curve for the final turbine is shown in \autoref{Figure24} and water velocity distribution for some guide vane angles is shown in \autoref{Figure26}.
	
\begin{figure}[ht!]
			\centering
			\includegraphics[width = 0.6\columnwidth]{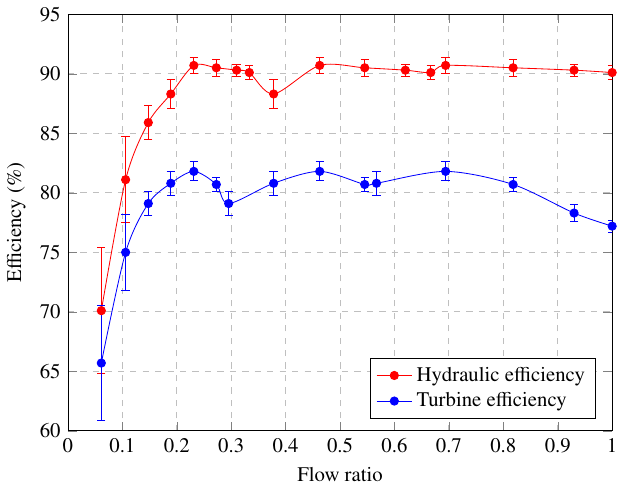}
			\caption{Efficiency as a function of flow ratio for a 2:1 multi-cell turbine design.} \label{Figure27}
	\end{figure}

	As can be seen from \autoref{Figure26}, in the final design the lower water jet remains attached to the guide vane surface even for small angles. This results in a significant efficiency improvement for low flow ratios, as shown in \autoref{Figure24}. \autoref{Figure24} also shows that due to the increase in admission angle, the maximum efficiency point has shifted from a flow ratio of 0.92 to a flow ratio of 0.69 with a value of 81.8\%. Because of this, the efficiency curve is more uniform, which in turn helps a multi-cell design maintain an efficiency value of about 78\% for flow ratios from 0.14 to 1, as shown in \autoref{Figure27}. In this study we assume that a multi-cell turbine has a 2:1 split, i.e. nozzle inlet is broken span-wise into two parts with a 2:1 ratio. The efficiency curves in \autoref{Figure27} were obtained by properly scaling the results shown in \autoref{Figure24}, accounting for individual guide vane angles on the two cells. \autoref{Figure27} assumes that flow division into the two cells does not break span-wise symmetry, allowing the use of 2D simulation results. For our future designs, nominal conditions will correspond to the 0.69 flow ratio as this allows the turbine to operate at maximum efficiency for nominal conditions and suitably handle overload situations.
	
\begin{figure}[ht!]
			\centering
			\includegraphics[width = 0.6\columnwidth]{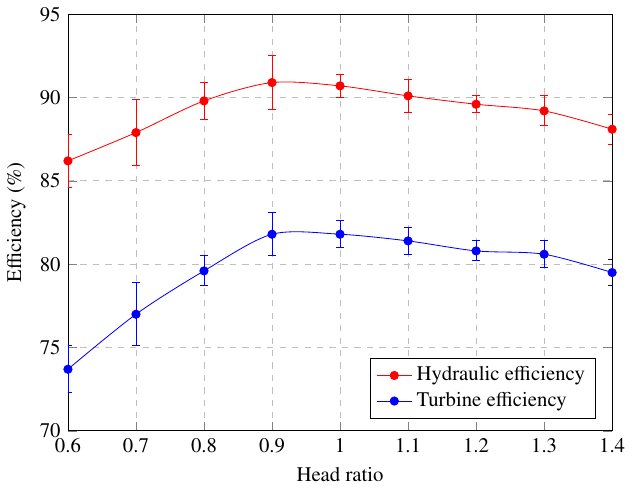}
			\caption{Efficiency as a function of head ratio for the final turbine design.} \label{Figure28}
	\end{figure}

	In the next step, the guide vane angle was held constant at 15 degrees (corresponding to the maximum turbine efficiency) and, without changing the rotational speed, water head at nozzle inlet was changed by as much as 40\% to examine the turbine's performance. The efficiency curve is plotted in terms of head ratio (ratio of water head to the nominal water head) in \autoref{Figure28}. It shows that turbine efficiency is approximately constant for above unity head ratios, but drops for head ratios below 0.9. This can be attributed to the designed blade profile which can handle a range of incoming flow angles without causing significant flow separation and a drop in efficiency.
	
\begin{figure}[ht!]
			\centering
			\includegraphics[width = 0.6\columnwidth]{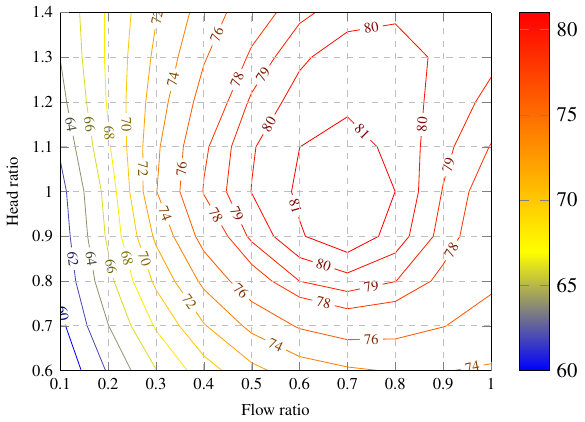}
			\caption{Normalized hill chart of the turbine. For a wide range of operating conditions, the turbine can guarantee an efficiency of 78\% with a maximum of 81.8\% and this range can be expanded if a multi-cell turbine is used.} \label{Figure29}
	\end{figure}

	To get an overview of the performance of the designed turbine, data from \autoref{Figure24} and \autoref{Figure28} were combined (efficiencies computed for a set of simulations at different flow ratios and constant head in the horizontal direction and those computed for a set of simulations at different heads and constant flow ratio in the vertical direction) to produce the normalized flow-head hill chart of the turbine shown in \autoref{Figure29}. It shows that for a wide range of operating conditions, the turbine can guarantee an efficiency of 78\% and a maximum efficiency of 81.8\%. This range can be significantly expanded (as shown in \autoref{Figure27}) if a multi-cell turbine is used. Note that in plotting the hill chart we have implicitly assumed that the nominal volume flow rate increases proportional to the nominal water head such that for different cases, identical flow ratios correspond to identical guide vane openings.
	
\begin{figure}[t!]
			\centering
			\includegraphics[width = 0.6\columnwidth]{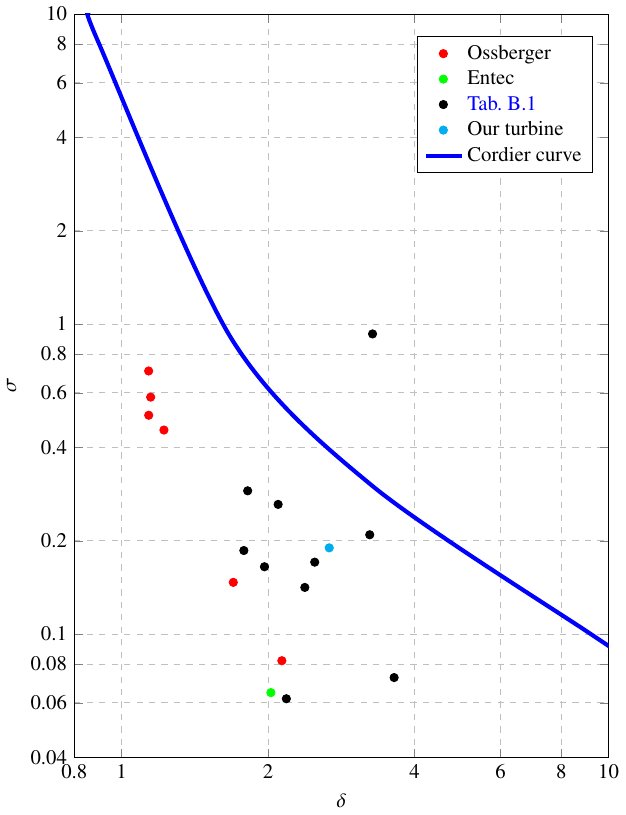}
			\caption{Cordier diagram showing the Cordier curve along with points representing our turbine, operational turbines with known diameter, and those found in \autoref{TableB1} \cite{OssbergerCrossFlow, Entec, Kaniecki1, Kaniecki2, DeAndrade, Sammartano1, Sammartano3, Sinagra1, Chen1, Acharya, Adhikari4, Adhikari5}.} \label{Figure30}
	\end{figure}

	Finally, knowledge of the optimum operating conditions facilitates a comparison between our turbine and others using the Cordier diagram \cite{Cordier}. Based on two dimensionless parameters $\delta$ and $\sigma$ defined as
	\begin{equation} \label{Equation2}
		\delta = 2N \frac{\sqrt{\pi Q}}{(2gH) ^ {1.5}},
	\end{equation}
	\begin{equation} \label{Equation3}
		\sigma = \frac{D}{2}\sqrt[4]{\frac{2\pi ^ {2}gH}{Q ^ {2}}},
	\end{equation}
	the Cordier curve in the $\delta$-$\sigma$ plane reflects the locus of high-efficiency turbomachinery. \autoref{Figure30} shows the Cordier curve along with points representing our turbine, operational turbines with known diameter, and those found in \autoref{TableB1} (based on the criteria established for \autoref{Figure2}) \cite{OssbergerCrossFlow, Entec, Kaniecki1, Kaniecki2, DeAndrade, Sammartano1, Sammartano3, Sinagra1, Chen1, Acharya, Adhikari4, Adhikari5}. It shows that our turbine is closer to the Cordier curve than most other turbines, and along with \autoref{Figure29}, demonstrates that the proposed methodology can be implemented successfully to design a high-performance Cross-Flow turbine.
	
	\section{Conclusions} \label{Section5}
	
	This paper introduced a new methodology for designing and optimizing the performance of high-efficiency hydraulic Cross-Flow turbines, and implemented that methodology in a simulation case study. In the system-level design phase water head and volume flow rate were used to determine runner diameter and select initial turbine parameters. We started the detail design phase by optimizing nozzle geometry to reach uniform velocity and angle of attack profiles at the runner inlet. This was followed by successively optimizing turbine parameters to both understand their effect on the performance of the turbine and select the optimal value to achieve the highest efficiency. Finally, the performance of the designed turbine was evaluated under various load conditions and the geometry was modified as necessary to obtain a nearly flat efficiency curve for different flow and head ratios. \par
	
	The case study showed that the proposed methodology provides a fast and computationally efficient process for designing a large number of high-efficiency turbines. In addition, we made several observations regarding different turbine characteristics throughout the case study, as follows:
	
	\begin{enumerate}
		
		\item In a high performance nozzle, the curve below the guide vane and the curve at the end of the nozzle are linear in terms of polar angle.
		
		\item Turbine efficiency is parabolic in terms of speed ratio and maximum efficiency occurs for speed ratios between 0.9 and 1. Furthermore, the relative static pressure of the water that enters the runner is significantly higher than zero, and it is caused by the centrifugal force on the water imposed by the rotation of the runner.
		
		\item The optimal admission angle has to provide a good balance between factors such as water entrainment and relative static pressure.
		
		\item A blade profile shaped like an airfoil with small ends utilizes the above-zero relative static pressure of water at the runner inlet and has the minimum amount of resistance to the incoming and outgoing flows.
		
		\item The optimal diameter ratio has to maintain a good balance between the built-up back-pressure at the runner inlet, time that water transfers energy to the blade, and hydraulic curvature losses.
		
		\item The optimal number of blades should provide a good balance between factors such as the total area of energy transfer, volume of semi-filled blade channels, and the total solid thickness blocking the incoming water jet.
		
		\item To maintain high efficiency at low flow ratios, nozzle design has to ensure that the lower water stream remains attached to the guide vane.
		
	\end{enumerate}
	
	Future research will look beyond the limitations of this work. In this paper, we used 2D approximations to simulate flow inside the turbine, but further 3D analysis can provide insight into multi-cell turbine design as well as the effect of sidewalls on turbine performance, especially for high-head, low-flow cases where runner length is relatively small. Future work will also analyze interaction between the water jet and the runner from a structural standpoint to help improve blade design and runner lifespan and reduce service needs. Transient behavior of the turbine will also be a topic of study, as this paper only focused on the steady-state behavior.
	
	\section*{Acknowledgment}
	
	This research was fully funded by DSDA, a research center at Sharif University of Technology (SUT). The experimental data and information on the manufactured prototype turbine used in this paper were from research conducted by DSDA under a contract between SUT and Water Research Institute of Iran (WRI) funded through Iran Water and Power Resources Development Company (IWPCO). The authors wish to express their gratitude to all participating parties for their positive role in the completion of this research and to the anonymous reviewers for their constructive feedback on this paper.
	
	\appendix
	\section{Theory of operation} \label{AppendixA}
	\setcounter{figure}{0}
	
\begin{figure}[ht!]
		\centering
		\includegraphics[width = 0.6\columnwidth]{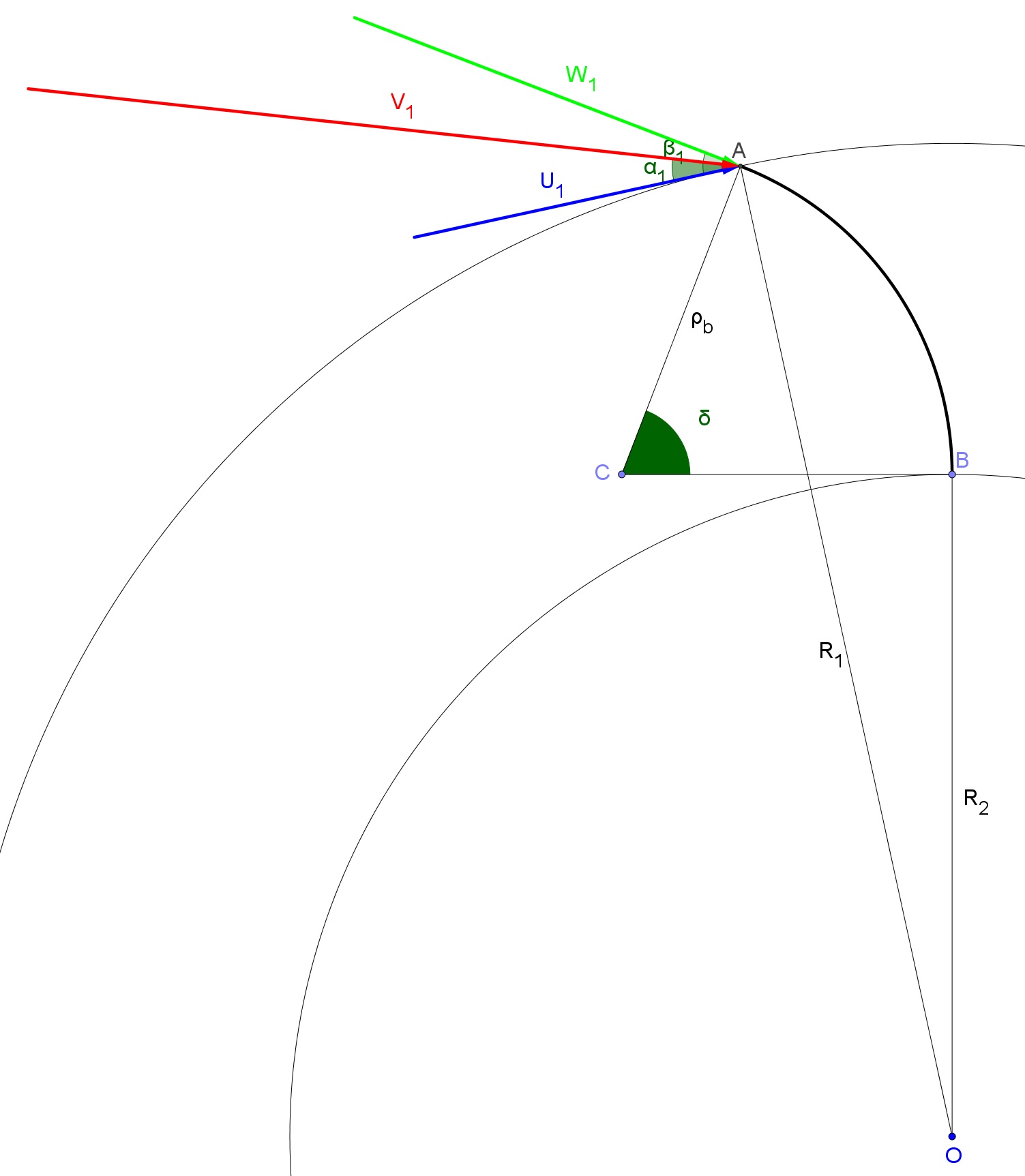}
		\caption{The nomenclature of turbine geometry and water entering the first stage. $R_{1}$ and $R_{2}$ denote the outer and inner runner radii, $\rho_{b}$ denotes the blade radius, $\delta$ denotes the blade central angle, $\alpha_{1}$ and $\beta_{1}$ denote the attack and blade entrance angles, and $U_{1}$, $V_{1}$, and $W_{1}$ denote the runner peripheral velocity, absolute water flow velocity, and water flow velocity relative to the blade, respectively.} \label{FigureA1}
	\end{figure}

	\autoref{FigureA1} shows the nomenclature of turbine geometry and water entering the first stage. \par
	
	Using the Reynolds Transport Theorem, Mockmore and Merryfield \cite{Mockmore} showed that the maximum theoretical efficiency is
	\begin{equation} \label{EquationA1}
		\eta_{t, \max} = \frac{1}{2}C_{l, n} ^ {2}(1 + \psi)\cos ^ {2}(\alpha_{1}),
	\end{equation}
and it is achieved if
	\begin{equation} \label{EquationA2}
		\frac{U_{1}}{V_{1}} = \frac{\cos(\alpha_{1})}{2},
	\end{equation}
	where $U_{1} = \frac{D_{1}\omega}{2}$ and $V_{1} = \sqrt{2gH}$.

	It is worth noting that $\eta_t$ denotes turbine efficiency, which is the ratio of output power $P_{o}$ (mechanical power at the turbine shaft) to input power $P_{i} = \rho gQH$, and should not be confused with the turbine's hydraulic efficiency, which is defined as,
	\begin{equation} \label{EquationA3}
		\eta_{h} = \frac{P_{o}}{P_{i} - P_{e}}.
	\end{equation}
$P_{e}$ is the amount of water energy (total pressure) leaving the turbine. Hydraulic efficiency merely accounts for the amount of energy loss in the turbine, and ideally has a value of 1. \par

	Using \autoref{EquationA2}, blade entrance angle is calculated by \autoref{EquationA4}:
	\begin{equation} \label{EquationA4}
		\tan(\beta_{1}) = 2\tan(\alpha_{1}).
	\end{equation}
Blade exit angle $\beta_{2}$ is selected similarly, and its value must be equal to 90 degrees to avoid loss of efficiency \cite{Durali}. \par

	Nozzle height can be determined using \autoref{EquationA5}:
	\begin{equation} \label{EquationA5}
		S_{n} = \sin(\alpha_{1})R_{1}\lambda.
	\end{equation}
Furthermore, Mockmore and Merryfield showed that, accounting for the centrifugal force of the runner, the optimal runner inner-to-outer diameter ratio (in this paper called ``diameter ratio'') can be calculated using \autoref{EquationA6} \cite{Mockmore}:
	\begin{equation} \label{EquationA6}
		\left(\frac{D_{2}}{D_{1}}\right) ^ {2} - \left[1 - \frac{1}{\cos ^ {2}(\beta_{1})}\right]\left(\frac{D_{2}}{D_{1}}\right) - \tan ^ {2}(\beta_{1}) = 0,
	\end{equation}
and runner length is given by,
	\begin{equation} \label{EquationA7}
		B = \frac{Q}{V_{1}S_{n}} = \frac{Q}{V_{1}\sin(\alpha_{1})R_{1}\lambda} = \frac{Q}{C_{l, n}\sqrt{2gH}\sin(\alpha_{1})R_{1}\lambda}.
	\end{equation}

	To reduce manufacturing costs, turbine blades are assumed to be circular arcs having a constant thickness. Blade radius $\rho_{b}$ is calculated using \autoref{EquationA8}:
	\begin{equation} \label{EquationA8}
		\rho_{b} = \frac{R_{1} ^ {2} - R_{2} ^ {2}}{2(R_{1}\cos(\beta_{1}) - R_{2}\cos(\beta_{2}))}.
	\end{equation}
Similarly, for blade central angle $\delta$:
	\begin{equation} \label{EquationA9}
		\delta = 2\tan^{-1}\left(\frac{\cos(\beta_{1}) - \frac{R_{2}}{R_{1}}\cos(\beta_{2})}{\sin(\beta_{1}) + \frac{R_{2}}{R_{1}}\sin(\beta_{2})}\right).
	\end{equation}
	
	\section{Previous studies} \label{AppendixB}
	\setcounter{table}{0}
	\setlength{\tabcolsep}{3 pt}
	
	\autoref{TableB1} summarizes the results of previous studies on the Cross-Flow turbine and contains turbine parameter values used or optimized in each study, i.e. the values shown in \autoref{TableB1} describe the final, optimized turbine. A - mark indicates the information was not mentioned in the respective research. Furthermore, ``theo'' indicates a theoretical, ``con'' a conceptual, ``num'' a numerical, and ``exp'' an experimental study. For studies that the reported efficiency is greater than 85\%, the reported value is most likely the hydraulic efficiency and not the turbine efficiency.
	
	\begin{table}[h!]
		\centering
		\vspace*{4 pt}
		\caption{Summary of turbine parameters used or optimized in previous studies. A - mark indicates a lack of information in that study.} \label{TableB1}
		\resizebox{\textwidth}{!}{
		\begin{tabular}{c c c c c c c c c c c c c}
			\hline
			\multicolumn{3}{c}{Study details} & \multicolumn{9}{c}{Used or optimized parameters} \\
			\hline
			Researcher(s) & Study & Study & Water & Water volume & Rotational & Runner & Runner & Number & Diameter & Angle of & Admission & Efficiency \\
			 & type & year & head & flow rate & speed & diameter & length & of & ratio & attack & angle & (\%)\\
			 &  &  & (m) & (l/s) & (rpm) & (mm) & (mm) & blades &  & (deg) & (deg) &  \\
			\hline
			Mockmore and Merryfield \cite{Mockmore} & theo/exp & 1949 & 2.7 - 5.5 & 62.9 & 270 & 333 & 305 & 20 & 0.66 & 16 & - & 68 \\
			Durali \cite{Durali} & theo & 1976 & 10 & 86.7 & 300 & 380 & 165 & 24 & 0.6 & 16 & 30 & 60 \\
			Durgin and Fey \cite{Durgin} & theo/exp & 1984 & - & - & - & - & - & 18 & - & 16 & 63 & 61 \\
			Fukutomi et al. \cite{Fukutomi} & theo/exp & 1985 & - & - & - & - & - & - & - & - & 30 - 120 & - \\
			Khosrowpanah et al. \cite{Khosrowpanah} & exp & 1988 & 0.335 - 2.60 & 20.4 - 44.2 & 87 - 356 & 304.8 & 152.4 & 15 & 0.68 & 16 & 90 & 80 \\
			Fiuzat and Akerkar \cite{Fiuzat} & exp & 1991 & - & - & - & 304.8 & 152.4 & - & 0.667 & - & 90 & 78.8 \\
			Aziz and Desai \cite{Aziz, Desai1} & exp & 1994 & 0.340 - 0.343 & 25.6 - 25.8 & - & 304.8 & 152.4 & 25 & 0.68 & 24 & 90 & 84.49 \\
			Desai and Aziz \cite{Desai2} & exp & 1994 & - & - & - & 304.8 & 101.6 & 30 & 0.6 & 22 & 90 & 88 \\
			Totapally and Aziz \cite{Totapally} & exp & 1994 & - & - & - & - & 101.6, 152.4 & 35 & 0.68 & 22, 24 & 90 & 92 \\
			Joshi et al. \cite{Joshi} & exp & 1995 & 2 - 9 & - & - & 300 & 325 & 24 & 0.66 & 16 & 36 & 64.8 \\
			Costa Pereira and Borges \cite{Pereira1} & exp & 1996 & 0.89 - 5.2 & - & - & 300 & 215 & 25 & 0.667 & 15 & 80 & 73.8 \\
			Reddy et al. \cite{Reddy} & exp & 1996 & 3 - 9 & 90 & 360 & 300 & 325 & 24 & 0.66 & 16 & 36 & 67.58 \\
			Olgun \cite{Olgun1} & exp & 1998 & 4 - 30 & 14 - 55 & 400 - 1300 & 170 & 114 & 28 & 0.67 & 16 & 46 & 72 \\
			Olgun \cite{Olgun2} & exp & 2000 & 4 - 30 & 14 - 55 & 968 - 1107 & 170 & 114 & 24 & 0.54 & 16 & 46 & 70 \\
			Kaniecki \cite{Kaniecki1} & exp/num & 2002 & 10, 15 & 300, 195 & 410, 550 & 300 & 300, 150 & 30 & 0.667 & 16 & - & 79, 81 \\
			Kaniecki and Steller \cite{Kaniecki2} & exp/num & 2003 & 10, 15 & 300, 195 & 410, 550 & 300 & 296, 150 & 30 & 0.667 & 16 & 82, 105 & 74.3, 78 \\
			Choi et al. \cite{Choi1} & num & 2008 & 2.9 & - & - & 250 & 150 & 26 & - & - & 120 & 65.7 \\
			Walseth \cite{Walseth} & exp & 2009 & - & - & 350 & 270 & - & 24 & 0.696 & 16 & 120 & 77.5 \\
			Choi et al. \cite{Choi2} & exp/num & 2010 & 20 & 516 & 530 & - & 500 & 30 & - & - & - & - \\
			De Andrade et al. \cite{DeAndrade} & exp/num & 2011 & 35 & 135 & 1000 & 294 & 150 & 24 & 0.68 & 16 & 70 & 75 \\
			Haurissa et al. \cite{Haurissa} & exp & 2012 & - & 0.234 & 300 & 200 & - & 20 & 0.65 & 16 & - & 72.569 \\
			Choi and Son \cite{Choi3} & num & 2012 & 46 & 176 & 1000 & 280 & 135 & 23 & - & 19 & - & - \\
			Kokubu et al. \cite{Kokubu1} & exp/num & 2012 & 3.15 & 6.6, 44.4 & 280 & 250 & 17, 100 & 30 & - & 16 & - & 62.9 \\
			Kokubu et al. \cite{Kokubu2} & exp/num & 2013 & - & - & - & 250 & - & 30 & 0.668 & 17 & 109 & - \\
			Sammartano et al. \cite{Sammartano1} & num & 2013 & 14.2 & 60 & 757 & 161 & 93 & 35 & 0.646 & 22 & 90 & 85.6 \\
			Sammartano et al. \cite{Sammartano3} & con & 2014 & - & - & 380 & 735 & 130 & 60 & 0.75 & 15 & 120 & 89 \\
			 &  &  & 13.9 & 620 & 300 & 390 & 616 & 60 & 0.75 & 15 & 120 & 84 \\
			Sinagra et al. \cite{Sinagra1} & con & 2014 & 32.7 & 620 & 360 & 385 & 530 & 50 & 0.65 & 15 & 120 & 89 \\
			Kaunda et al. \cite{Kaunda1, Kaunda2} & exp/num & 2014 & 3 - 10 & - & 350 & 268 & - & 24 & 0.693 & 16 & 90 & 79 \\
			 &  &  & 5 & - & 350 & - & - & 24 & 0.693 & 16 & 90 & 78 \\
			\hline
		\end{tabular}}
	\end{table}
	\begin{table}[h!]
		\centering
		\resizebox{\textwidth}{!}{
		\begin{tabular}{c c c c c c c c c c c c c}
			\hline
			\multicolumn{3}{c}{Study details} & \multicolumn{9}{c}{Used or optimized parameters} \\
			\hline
			Researcher(s) & Study & Study & Water & Water volume & Rotational & Runner & Runner & Number & Diameter & Angle of & Admission & Efficiency \\
			 & type & year & head & flow rate & speed & diameter & length & of & ratio & attack & angle & (\%)\\
			 &  &  & (m) & (l/s) & (rpm) & (mm) & (mm) & blades &  & (deg) & (deg) &  \\
			\hline
			Reihani et al. \cite{Reihani} & num & 2014 & 60 & 200 & 1000 & 300 & 152 & 19 & - & 16 & - & 63.73 \\
			Yassen \cite{Yassen} & theo/num & 2014 & - & - & - & 300 & 300 & 30 & 0.65 & 15 & 90 & 77.810 \\
			Sinagra et al. \cite{Sinagra2, Sammartano2} & exp/num & 2015 & - & 60 & 750 & 161 & 93 & 35 & 0.646 & 22 & 90 & 82.1 \\
			Chen and Choi \cite{Chen1} & exp/num & 2015 & 20 & 465.0 & 530 & 340 & 500 & 30 & - & - & 120 & 81.3 \\
			Acharya et al. \cite{Acharya} & num & 2015 & 10 & 100 & 642 & 200 & - & 22 & - & - & - & 80.76 \\
			Elbatran et al. \cite{Elbatran1, Elbatran2} & con & 2015 & - & - & - & 200 & 400 & 26 & 0.7 & - & - & 49 \\
			 &  &  & - & - & - & 200 & 380 & 26 & 0.684 & 16 & - & 52 \\
			Katayama et al. \cite{Katayama} & exp/num & 2015 & 0.71 & 7 & - & 200 & 100 & 20 & - & 16 & - & 49.8 \\
			Adhikari et al. \cite{Adhikari1, Adhikari2, Adhikari3, Adhikari4, Adhikari5} & theo/num & 2016 & 1.17, 1.337 & 42, 46 & 199.1& 304.8 & 101.6 & 35 & 0.68 & 22 & 90 & 90.61 \\
			 &  &  & 3 - 10 & 56 - 105 & 500 & 316 & 94.34 & 35 & 0.67 & 16 & 80 & 91 \\
			Sammartano et al. \cite{Sammartano4} & exp/num & 2016 & - & - & 757 & - & - & 35 - 80 & 0.75 - 0.8 & 22 & 90 & 80.74 \\
			Sammartano et al. \cite{Sammartano5} & con/num & 2016 & 18.6 - 34.9 & 200 - 800 & 400 - 625 & 392 & 450 & 50 & 0.75 & 15 & 27.7 - 120 & - \\
			Dragomirescu and Schiaua \cite{Dragomirescu} & exp/num & 2017 & 0.72 - 0.92 & 67.2 - 68.3 & 174.9 - 268.4 & 250 & 300 & 19 & 0.66 & - & - & 55 \\
			Costa Pereira and Borges \cite{Pereira2} & theo/exp & 2017 & 5.5 & 100 & - & 300 & 215 & 20 & 0.667 & 13 & 80 & 84.8 \\
			Tiwari and Shrestha \cite{Tiwari} & num & 2017 & 88 & 308 & 1500 & 250 & 150 & 24 & 0.7 & 16 & - & 67.3 \\
			Woldemariam et al. \cite{Woldemariam1} & theo/num & 2018 & 10 & 1000 & 350 & 300.9 & - & - & 0.71 & 15 & - & - \\
			Woldemariam et al. \cite{Woldemariam2} & num & 2018 & 5 - 22.2 & - & 360 & 302 & - & 30 & 0.676 & - & - & - \\
			Zaffar et al. \cite{Zaffar} & num & 2018 & 9.3 & 450 & 386 & 300 & 420 & 35 & 0.68 & 22 & 90 & 58.9 \\
			Siswantara et al. \cite{Siswantara} & num & 2018 & - & - & - & 322 & - & 35 & 0.752 & 22 & 90 & 78.5 \\
			Ranjan et al. \cite{Ranjan} & num & 2019 & 10 & 105 & 450 - 600 & 316 & 150 & 20 & 0.67 & - & 65 & 97.8 \\
			Woldemariam et al. \cite{Woldemariam3} & num & 2019 & 5 & 49.70, 44.93 & 250, 350 & 270 & 204 & 24 & 0.694 & - & - & 84.49, 61.80 \\
			Leguizamon and Avellan \cite{Leguizamon} & num & 2020 & 3.2 & 50 & 300 & 250 & 190 & 32 & 0.664 & 13 & 80 & 75.2 \\
			\hline
		\end{tabular}}
	\end{table}
	
	\section{Experimental setup} \label{AppendixC}
	\setcounter{figure}{0}
	\setcounter{table}{0}
	\setlength{\tabcolsep}{6 pt}
	
\begin{figure}[ht!]
		\centering
		\includegraphics[width = 0.84\textwidth]{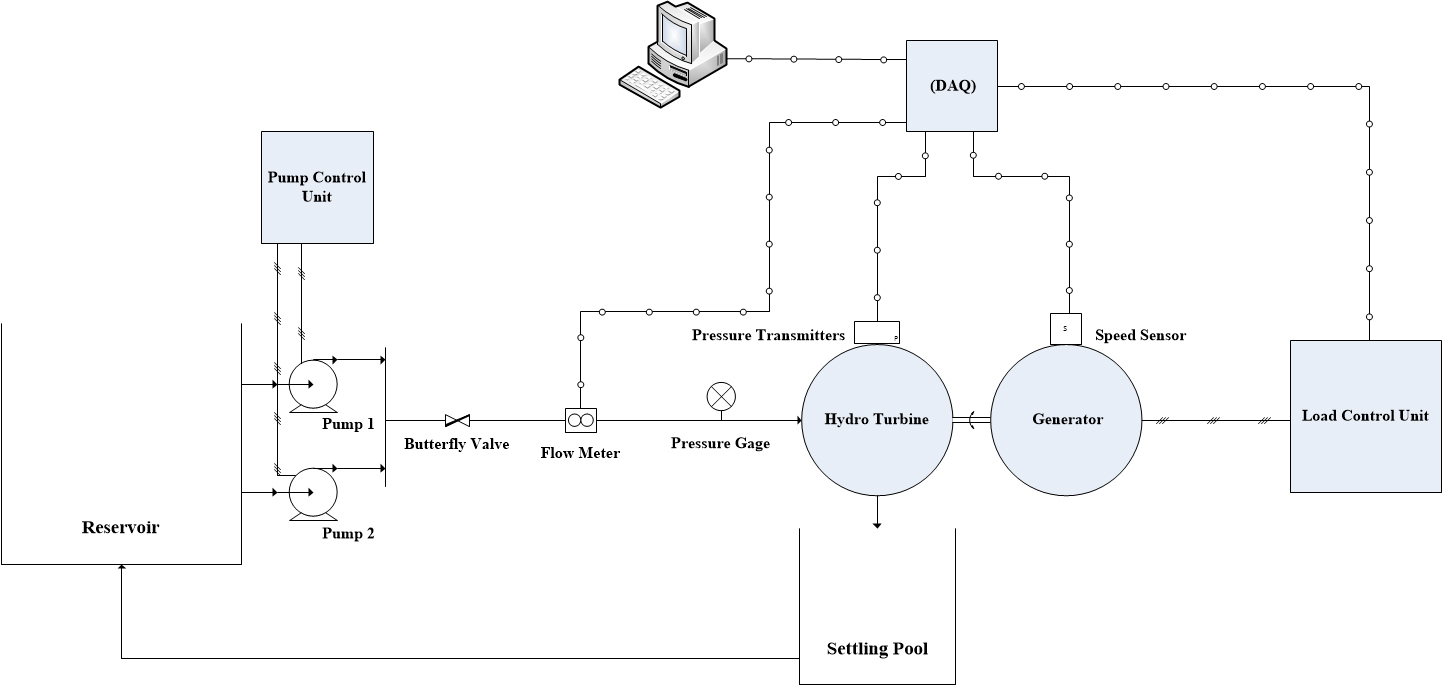}
		\caption{Experimental setup for testing a micro hydraulic turbine \cite{Hoghooghi}.} \label{FigureC1}
	\end{figure}

	Experimental tests were carried out at the testing facility of WRI located in Tehran which is 1300 m above sea level. The facility consisted of a water pumping system and a test stand on which the prototype Cross-Flow turbine was coupled to a synchronous generator. The experimental setup was similar to \cite{Sammartano2} and is shown in \autoref{FigureC1}.
	
	The experimental setup operated in a closed-loop mode and comprised of an open reservoir, a suction pipe, four centrifugal pumps, and a discharge pipe that was connected to the test stand. The generator had four poles and a speed of 1500 rpm when used with 50 Hz electricity. A summary of the specifications of instruments used for measurements is available in \autoref{TableC1}. Measurements were made according to the IEC41 standard for hydraulic turbine testing. Each test case was repeated 5 times and the average value was reported.
	
	\begin{table}[h!]
		\centering
		\caption{Specifications of measurement instruments used for testing the prototype Cross-Flow turbine developed by DSDA.} \label{TableC1}
		\begin{tabular}{c c c c}
			\hline
			Measurement & Uncertainty & Measured & Full-Scale \\
			instrument & amount & value & Output (FSO) \\
			\hline
			pressure transducer & 0.5\% FSO & static inlet pressure & 10 bar \\
			flow meter & 0.2\% FSO & discharge flow & 2222.2 l/s \\
			power analyzer & 1.0\% & generator output & 9.999 MVA \\
			tachometer & 0.1\% & turbine speed & 100000 rpm \\
			\hline
		\end{tabular}
	\end{table}
	
	Turbine efficiency was calculated using \autoref{EquationC1},
	\begin{equation} \label{EquationC1}
		\eta_{t} = \frac{VI}{\eta_{g}\eta_{m}\rho gQH}.
	\end{equation}
	$V$ and $I$ were read from the power analyzer and $Q$ was read from the flow meter. Inlet static pressure was read from the pressure transducer, which was then added to the dynamic pressure (calculated using $Q$ and a pipe diameter of 490 mm) and the elevation difference of the pipe midline and the turbine outlet (34 mm) to obtain the value of $H$, though the latter term could be considered negligible compared to the nominal head of 45 m ($< 1\%$). \par
	
	The reported average generator efficiency was 91.3\% which, after multiplication by a derating factor of 0.982 due to the altitude of the testing site, reduced to 89.7\% \cite{Mahniroo}. The mechanical transmission system had an overall efficiency of 94.1\% (accounting for the belts, pulleys, and bearings) \cite{Carlisle, SKF}, which lowered the overall transmission efficiency ($\eta_{g}\times\eta_{m}$) to 84.4\%. These calculations were made based on data provided by the manufacturers. \par
	
	The uncertainty in $\eta_{t}$ was determined by the Root of the Sum of the Squares (RSS) rule for uncertainty propagation using individual uncertainties of the variables involved \cite{Figliola}. Based on \autoref{TableC1}, the uncertainty in generated power was 1.0\%, in water volume flow rate was 4.44 l/s, and in water head was 51.1 cm. The uncertainty in generator efficiency was 0.5\% and in the mechanical transmission system efficiency was 2\%.
	
	\section*{List of symbols}
	
	\subsection*{Greek symbols}

		\begin{table}[ht!]
		\flushleft
			\begin{tabular}{l p{180 pt} l l}
				$\alpha$ & Angle of attack (deg, rad) & $\beta$ & Blade tip angle (deg, rad) \\
				$\delta$ & Blade central angle (deg, rad) &  & Cordier specific diameter \\
				$\eta$ & Efficiency & $\lambda$ & Admission angle (deg, rad)
			\end{tabular}
		\end{table}
		\begin{table}[ht!]
		\flushleft
			\begin{tabular}{l p{180 pt} l l}
				$\rho$ & Density (kg/m$^3$) &  & Radius (m) \\
				$\sigma$ & Cordier specific speed & $\psi$ & runner loss coefficient \\
				$\omega$ & Rotational speed (rad/s)
			\end{tabular}
		\end{table}
	
	\subsection*{Latin symbols}

		\begin{table}[h!]
		\flushleft
			\begin{tabular}{l p{180 pt} l l}
				B & runner length (m) & C & Coefficient \\
				D & runner diameter (m) & g & Gravitational acceleration (m/s$^2$) \\
				H & Water head (m) & N & Rotational speed (rpm) \\
				N$^{\prime}$ & Specific speed & P & Power (W) \\
				Q & Water volume flow rate (m$^3$/s) & R & runner radius (m) \\
				S & Nozzle height (m) & U & Blade tip linear velocity (m/s) \\
				V & Absolute water velocity (m/s)
			\end{tabular}
		\end{table}
	
	\subsection*{Subscripts}
	
		\begin{table}[h!]
		\flushleft
			\begin{tabular}{l p{180 pt} l l}
				$b$ & Blade & $e$ & Exit \\
				$g$ & Generator & $h$ & Hydraulic \\
				$i$ & Input & $l$ & Loss \\
				$m$ & Mechanical transmission system & $n$ & Nozzle \\
				$o$ & Output & $q$ & Modified \\
				$t$ & Turbine
			\end{tabular}
		\end{table}
		
	\bibliographystyle{elsarticle-num}
	\bibliography{CFBib}

\end{document}